\documentclass[usenatbib]{mnras}
\usepackage[utf8]{inputenc}

\title[Galaxy Zoo: Clump Scout - a two-dimensional aggregation tool for citizen science]{Galaxy Zoo: Clump Scout - Design and first application of a two-dimensional aggregation tool for citizen science}
\author[Dickinson et al.]{Hugh Dickinson$^{1}$, Dominic Adams$^{2}$, Vihang Mehta$^{2,3}$, Claudia Scarlata$^{2}$, Lucy Fortson$^{2}$,
\newauthor
Stephen Serjeant$^{1}$, Coleman Krawczyk$^{4}$, Sandor Kruk$^{5}$, Chris Lintott$^{6}$, \newauthor
Kameswara Mantha$^{2}$, Brooke D. Simmons$^{7}$, Mike Walmsley$^{8}$\\
$^{1}$ School of Physical Sciences, The Open University, Milton Keynes, MK7 6AA, UK \\
$^{2}$ School of Physics and Astronomy, University of Minnesota, 116 Church Street SE, Minneapolis, MN 55455, USA \\
$^{3}$ IPAC, Mail Code 314-6, California Institute of Technology, 1200 E. California Blvd., Pasadena, CA, 91125, USA\\
$^{4}$ Institute of Cosmology \& Gravitation, University of Portsmouth, Dennis Sciama building, Portsmouth, PO1 SFX, UK\\
$^{5}$ Max-Planck-Institut f\"ur extraterrestrische Physik (MPE), Giessenbachstrasse 1, D-85748 Garching bei M\"unchen, Germany\\
$^{6}$ Oxford Astrophysics, University of Oxford, Denys Wilkinson Building, Keble Road, Oxford, OX1 3RH, UK\\
$^{7}$ Physics Department, Lancaster University, Lancaster, LA1 4YB, UK \\
$^{8}$ Jodrell Bank Centre for Astrophysics, Department of Physics \& Astronomy, University of Manchester, Manchester M13 9PL, UK
}

\usepackage{graphicx, url, amsmath, amssymb, aas_macros, xspace, multirow}

\newcommand{\gzcs}{\textit{Galaxy Zoo: Clump Scout}\xspace}
\newcommand{\bvp}{BVP17\xspace}
\newcommand{\nicospaper}{2022ApJ...931...16A}

\usepackage{fontawesome5}
\usepackage{hyperref}

\makeatletter
\newcommand{\github}[1]{%
   \href{#1}{\faGithubSquare}%
}
\makeatother

\begin{document}

\maketitle

\begin{abstract}
    \gzcs is a web-based citizen science project designed to identify and spatially locate giant star forming clumps in galaxies that were imaged by the Sloan Digital Sky Survey Legacy Survey. We present a statistically driven software framework that is designed to aggregate two-dimensional annotations of clump locations provided by multiple independent \gzcs volunteers and generate a consensus label that identifies the locations of probable clumps within each galaxy. The statistical model our framework is based on allows us to assign false-positive probabilities to each of the clumps we identify, to estimate the skill levels of each of the volunteers who contribute to \gzcs and also to quantitatively assess the reliability of the consensus labels that are derived for each subject. We apply our framework to a dataset containing 3,561,454 two-dimensional points, which constitute 1,739,259 annotations of 85,286 distinct subjects provided by 20,999 volunteers. Using this dataset, we identify 128,100 potential clumps distributed among 44,126 galaxies. This dataset can be used to study the prevalence and demographics of giant star forming clumps in low-redshift galaxies. The code for our aggregation software framework is publicly available at: \url{https://github.com/ou-astrophysics/BoxAggregator} \github{https://github.com/ou-astrophysics/BoxAggregator}
\end{abstract}
\begin{keywords}
galaxies: structure -- methods: data analysis -- methods: statistical -- software: data analysis -- software: public release
\end{keywords}
\section{Introduction}

One of the main goals for modern observational cosmology is to discover and understand how galaxies and their constituent substructures have assembled and evolved throughout cosmic history. 

During the last two decades, a large number of observational data have been assembled, which show strong evidence for a substantial evolution in the dominant mode of star formation in galaxies between $z\sim3$ and $z\sim0.2$ \citep[e.g.][]{2014ARA&A..52..415M, Murata+2014,Guo+2015, Shibuya+2016, Guo_2018}. 

Early observations using the Hubble Space Telescope (HST) revealed that typical massive galaxies ($M\gtrsim10^{10} \mathrm{M}_{\odot}$), populating the $z\sim2$ star forming main sequence \citep{2007ApJ...660L..47N}, exhibit thick, gas-rich, clumpy disks with star formation rates  $\dot{M}_{\star}\sim100\,\mathrm{M}_{\odot}\,\mathrm{yr}^{-1}$ \citep[e.g.][]{2011ApJ...733..101G, ElmegreenElmegreenHirst2004a, ElmegreenElmegreenSheets2004b}. Many of these $z\sim2$ galaxies were found to exhibit discrete, sub-galactic regions of enhanced star formation (hereafter referred to as ``clumps'') with apparent radii $\lesssim 1\,$kpc and stellar masses $M_{\star} \gtrsim 10^7\,\mathrm{M}_{\odot}$ \citep{Elmegreen2007}. More recent evidence suggests that these clumps may in fact be aggregations of smaller substructures that could not be resolved by HST \citep[e.g.][]{Wuyts+2014, Fisher+2017, DessuagesZavadskyAdamo2018}, but this remains to be confirmed. The prevalence of giant star-forming clumps at high redshift and the overall characteristics of their host galaxies are in stark contrast with the thin, uniform and generally quiescent ($\dot{M}_{\star}\sim1 \mathrm{yr}^{-1}$) disk morphologies that prevail among star-forming galaxies in the local Universe \citep[e.g.][]{2011ApJS..196...11S, 2013MNRAS.435.2835W}.

The mechanisms that drove this evolution of star formation activity, their onset epochs and the timescales over which they operated, remain to be fully established. If they can be accurately determined, the abundances of clumps within galaxies at different redshifts, together with their spatial distributions and intrinsic properties, provide obvious diagnostics for the transition from clumpy to more diffuse star formation. Historically, the most extensive surveys of clumpy star formation have relied on HST imaging and focused on intermediate and high redshift galaxies \citep[e.g.][]{Murata+2014, Guo+2015, Guo_2018}.
A common conclusion of these studies is that the overall fraction of massive ($M_{\star}\gtrsim10^{9.5}\, \mathrm{M}_{\odot}$), clumpy star forming galaxies decreases rapidly for $z\lesssim2$ and falls below $\sim5\%$ by $z\sim0.2$. 

The scarcity of clumpy galaxies in the local Universe makes the task of identifying them in large numbers much more challenging and related studies at low redshift have entailed focused investigations of small samples containing $\sim50$ galaxies, or fewer \citep[See, however][]{2021ApJ...912...49M}. Identifying enough low-redshift clumpy galaxies to enable accurate inference of their overall population demographics and characteristics requires wide-field imaging surveys that encompass a large fraction of the sky and a reliable method for discovering candidate systems. In recent years, extensive ground-based surveys like the Sloan Digital Sky Survey Legacy Survey \citep[SDSS;][]{York+2000_SDSS_legacy} and the Dark Energy Camera Legacy Survey \citep[DECaLS;][]{2019AJ....157..168D} have delivered publicly available wide field imaging data that make systematic searches for large numbers of low-redshift clumpy galaxies possible.
\gzcs \citep{\nicospaper} is a citizen science project that used SDSS imaging data and was designed to let volunteers from the general public identify clumpy galaxies and the clumps they contain. Multiple volunteers inspect images of galaxies and provide two dimensional annotations marking the locations of any clumps the galaxies contain. 

One of the most challenging aspects of collecting data using a citizen science approach is calibrating the reliability of the responses that volunteers provide. Translating astrophysical analyses into a citizen science context can be difficult because the subject matter and related concepts are often not familiar to non-experts. This unfamiliarity can result in annotations that are noisy, with large variations between the responses of different volunteers. The traditional approach for mitigating such noise is to collect a large number of independent annotations and derive an average result representing the overall consensus between volunteers. This has two obvious disadvantages: firstly, volunteer effort may be wasted if more responses are accumulated than are actually required to mitigate the variation between responses and secondly, even after a large number of responses have been collected, there is no formal guarantee that the consensus is accurate or sufficiently precise.

To address these issues, more quantitative approaches have been developed that attempt to infer statistical estimates for the reliability of consensus derived from citizen science annotations and classifications. For example, \citet{2016MNRAS.455.1171M} developed the Space Warps Analysis Pipeline (SWAP) which used a binomial model for a simple true-or-false response to derive a Bayesian estimate for the probability that astrophysical images included signatures of strong gravitational lensing. The SWAP algorithm was also used by \citet{2017MNRAS.472.1315W} to accelerate consensus for citizen-science classification of potential supernova flashes and assign false-alarm probabilities to candidate events. Later,   \citet{2018MNRAS.476.5516B} showed that applying SWAP to galaxy morphology labels collected via the \textit{Galaxy Zoo} platform \citep{2008MNRAS.389.1179L, Willett+2013} increased the rate of classification by 500\% and reduced the volunteer effort that was required by a factor of $\sim6.5$, relative to the \textit{Galaxy Zoo} standard requirement for 40 volunteers to inspect each galaxy.

In this paper we build on the principle of SWAP and develop an aggregation approach to derive quantitative estimates for the reliability of two dimensional labels of clump locations within galaxies based on annotations provided by \gzcs volunteers. Like SWAP, we rely on a statistical model to derive probabilistic estimates for several quantities that determine the reliability of a label that represents the consensus of multiple independent annotations. Two dimensional annotations are more complex than the simple binary classification tasks that SWAP was designed to process and our statistical model is necessarily also more complicated. We base our approach on an method that was initially presented by \citet{8100130} (Hereafter \bvp), who tested their algorithm on small and relatively noise-free annotation datasets that contained a few thousand annotations and were collected from paid workers on the Amazon \textit{Mechanical Turk} platform\footnote{\url{https://www.mturk.com}}. We have developed a new implementation of this algorithm that is computationally efficient enough to process millions of independent annotations provided for tens of thousands of images by the \gzcs volunteers. Our goal is to find out whether this algorithm can be used successfully to derive complicated two-dimensional labels with quantitative reliability estimates in a mass-participation citizen-science context using noisy annotations provided by a cohort of non-expert volunteers. We also aim to determine whether the reliability estimates we derive can be used to accelerate the labeling process and reduce the amount of volunteer effort that is required to accurately label the clumps in each galaxy. 

The remainder of this paper is organised as follows: In \autoref{sec:data} we describe how the imaging data presented to volunteers in \gzcs were selected and prepared. In \autoref{sec:collecting-annotations} we outline the annotation workflow that volunteers used to annotate the images and the training they received. In \autoref{sec:data_aggregation_model} we provide details of the statistical model that underpins our aggregation algorithm. In \autoref{sec:computing_aggregated_labels}, we explain how our algorithm actually computes the labels it derives. In \autoref{sec:results}, we present the results of applying our algorithm to the \gzcs data and analyse the quantitative reliability metrics that are generated. In \autoref{sec:discussion} we discuss the implications of these results in the context of the goals outlined above and the suitability of citizen science as a method for complex astrophysical image analysis. Finally, in \autoref{sec:conclusion}, we summarise our findings and conclude.

\section{Data}\label{sec:data}

In this section we briefly describe the the galaxy selection criteria and the image preparation pipeline used for \gzcs. A much more detailed description is provided by \citet{\nicospaper}. 

\subsection{Galaxy Image Selection}\label{subsec:data:selection}

The galaxy images used in \gzcs comprise three subsets of the sample that was visually inspected and morphologically classified by volunteers contributing to the \textit{Galaxy Zoo 2} (GZ2) citizen science project \citep{2013MNRAS.435.2835W}. The criteria that were used to select these subsets are described in detail in \citet{\nicospaper}. For convenience, this section summarises the most relevant properties of the galaxies that were inspected by the \gzcs volunteers.

A primary sample of 53,613 galaxies with $0.02\leq z \leq 0.25$ was selected based on the morphological labels provided by GZ2 volunteers. We anticipated that the presence of obvious star-forming clumps in images of smooth elliptical galaxies was very unlikely so for this primary sample, we limited our selection to galaxies for which more than 50\% of volunteers responded negatively\footnote{A negative response corresponds to selecting the answer ``Features or disk.''} to the question ``Is
the galaxy simply smooth and rounded, with no sign of a disk?''.

To estimate the number of clumpy galaxies that were observed by SDSS but which were excluded from our primary sample, we also include a smaller, secondary sample. This sample contains 4,937 galaxies for which \textit{fewer} than 50\% of GZ2 volunteers identified features or a disk and was selected within a more restricted redshift range $0.02\leq z \leq 0.075$.

Finally, \gzcs volunteers also annotated a sample of 26,736 galaxies matching the selection criteria used for the primary sample, but which had \textit{simulated} emission from clumps with known photometric and physical properties superimposed \citep[see][for details of the simulation procedure]{\nicospaper}. Annotations of these simulated clumps were used by \citet{\nicospaper} to derive an estimate of the \gzcs sample completeness for clumps with specified photometric properties.

Stellar mass estimates for galaxies in all three samples were taken from the SDSS DR7
MPA-JHU value-added catalog \citep{Kauffmann+2003_MPA-JHU, Brinchmann+2004_MPA-JHU}. All three samples include galaxies with stellar masses $10^{8.5}\mathrm{M}_{\odot}\lesssim M_{\star}\lesssim10^{12}\mathrm{M}_{\odot}$.

\subsection{Galaxy Image Preparation}\label{subsec:data:preparation}
For each of our selected galaxies we extract square cutouts from SDSS \textit{g}, \textit{r} and \textit{i} band FITS \citep{2010A&A...524A..42P} images that are normally 6 times larger than the galaxy's measured 90\% \textit{r}-band Petrosian radius, but have a minimum side length of 40 pixels\footnote{This minimum size criterion is designed to handle galaxies that have very small, incorrectly measured Petrosian radii.}. Experience from previous iterations of the \textit{Galaxy Zoo} projects, including GZ2, has shown that sizing cutouts relative to the host galaxy radius in this way provides sufficient angular resolution for volunteers to discern morphological features, while including enough of the surrounding context to help distinguish those features from instrumental noise and background objects. We then resample these single band images onto a common pixel grid with SDSS native resolution ($0.396\arcsec$/pixel) before combining them (without PSF-matching) into a three-channel colour composite. We assign the \textit{g}, \textit{r} and \textit{i} bands to the red, blue and green channels respectively and scale each band independently using the formula presented in \citet{Lupton+2004}. For an input pixel intensity $I_{x}$ in band $x$ the scaled intensity $I^{\prime}_{x}$ is computed using
\begin{equation}
    I^{\prime}_{x} = \frac{1}{Q}\mathrm{asinh}\left[Q\cdot\frac{\left(\frac{I_{x}}{\beta_{x}}-m\right)}{\alpha}\right] 
\end{equation}
We specify that $\{Q, \alpha, m\} = \{7, 0.2, 0\}$ for all bands and that $\{\beta_{g}, \beta_{r}, \beta_{i}\}=\{0.7, 1.17, 1.818\}$. Finally, we re-scale each colour image so that its height and width are both 400 pixels. Note that this means the angular size of the cutout image pixels varies between $0.1\arcsec\,\mathrm{pix}^{-1}$ and $\sim18\arcsec\,\mathrm{pix}^{-1}$ for different subjects depending on the angular size of the central galaxy. The number of SDSS native image pixels spanned by the SDSS imaging PSF FWHM varies between 1.5 and 7.5, with a median value $\sim2.8$, for 99\% of the unscaled cutout images. The remaining of subjects 1\% populate a tail out to $\sim18$ pixels. In the final scaled cutouts the number of pixels spanned by the PSF FWHM varies between $\sim1$ and $\sim70$ with a median value of $\sim11$. Examples of the images generated using this procedure are shown in figures \ref{fig:agg_examples}, \ref{fig:empirical_fp_example_images}, \ref{fig:agg_v_ms_more_examples_0-5} and \ref{fig:agg_v_ms_less_examples_0-5} 

\section{Collecting Annotations}\label{sec:collecting-annotations}

To identify the locations of clumps within their host galaxies, we designed a web-based citizen science project using the \textit{Zooniverse} project builder interface\footnote{\url{www.zooniverse.org/lab}}. 

\subsection{Volunteer Training}\label{subsec:volunteer-training}
For non-expert volunteers, identifying genuine clumps among the potentially complex features of their host galaxies can be daunting. To improve volunteers' confidence and help them to provide accurate annotations we provided several pedagogical and training resources. Following the approach of other \textit{Zooniverse} projects, we designed a detailed practical tutorial explaining each step of the annotation workflow. This tutorial was automatically presented to volunteers when they joined the project and remained available for reference thereafter. Additional reference images and explanatory text were provided using the \textit{Field Guide} feature of the \textit{Zooniverse} interface. A separate \textit{About} section of the project provided pedagogical material explaining the scientific motivation of the project. Finally, to guide the progress of first-time volunteers, we provided expert labels for a small subset of our galaxy images. Ten such images were interspersed with decreasing frequency among the first $\sim20$ subjects that each volunteer inspected. We implemented a system to provide real-time feedback for volunteer annotations of expert-labeled galaxy images and inform them if they missed genuine clumps or mistakenly annotated an object that experts had disregarded. This feedback system was designed to refine volunteers expectations regarding the visual appearance of genuine clumps during the early stages of their engagement with the project.

\subsection{The annotation workflow}\label{subsec:annot_wf}

Volunteers following the \textit{Galaxy Zoo: Clump Scout} workflow inspect a sequence of single subject galaxy images (hereafter ``subjects'') that are randomly drawn from a global subject set. The subject selection ensures that no volunteer inspects the same image more than once and each subject is inspected by a group of approximately 20 volunteers. Each volunteer first annotates the two-dimensional location of the central bulge of the central galaxy in the image if it is visible, before proceeding to annotate the locations of any clumps they can discern. To mitigate against the possibility that volunteers would disregard genuine clumps with appearances that confound their expectations, we provided an opportunity to mark clumps as ``unusual''. We investigate the impact of including or discarding this unusual clump subset in \autoref{sec:results}. 

The full \textit{Galaxy Zoo: Clump Scout} dataset contains 3,561,454 click locations, which constitute 1,739,259 annotations of 85,286 distinct subjects provided by 20,999 volunteers. 

\subsection{Initial annotation processing}\label{subsec:initial_processing}
We expect that even the largest individual clumps will be at best marginally resolved for the lowest redshift galaxies in our data sample. This implies that almost all clumps will appear as point-sources with a light profile equal to the instrumental point-spread function (PSF). Our data preparation procedure (\autoref{subsec:data:preparation}) results in subject images that have different pixel sampling of the PSF depending on the angular size of the central host galaxy. To account for this fact, we transform the two-dimensional point estimates for clump locations that volunteers provide into square boxes with side-length equal to twice the full width at half maximum (FWHM) of the pertinent subject's PSF. Assigning a finite, instrumentally motivated clump extension allows us to identify groups of volunteer clicks with separations that are smaller than the PSF. A prior assumption of our data aggregation approach that it is impossible for a single volunteer to mark separate clumps within the same subject that are closer than twice the PSF FWHM\footnote{Even if volunteers are able to submit such nearby marks, our algorithm is designed to only recognise one of them. The choice of which nearby clicks to discard depends on the clicks provided by other volunteers.}. It is likely that any such multiplets that volunteers do provide represent noise peaks in contrast-enhanced subject images or are simply accidents. In \autoref{sec:data_aggregation_model}, we describe how our aggregation algorithm effectively deduplicates multiple nearby annotations by individual volunteers.

\subsection{A scale-free distance metric}\label{subsec:scale_free_distance}

\begin{figure}
    \centering
    \includegraphics[width=0.2\textwidth]{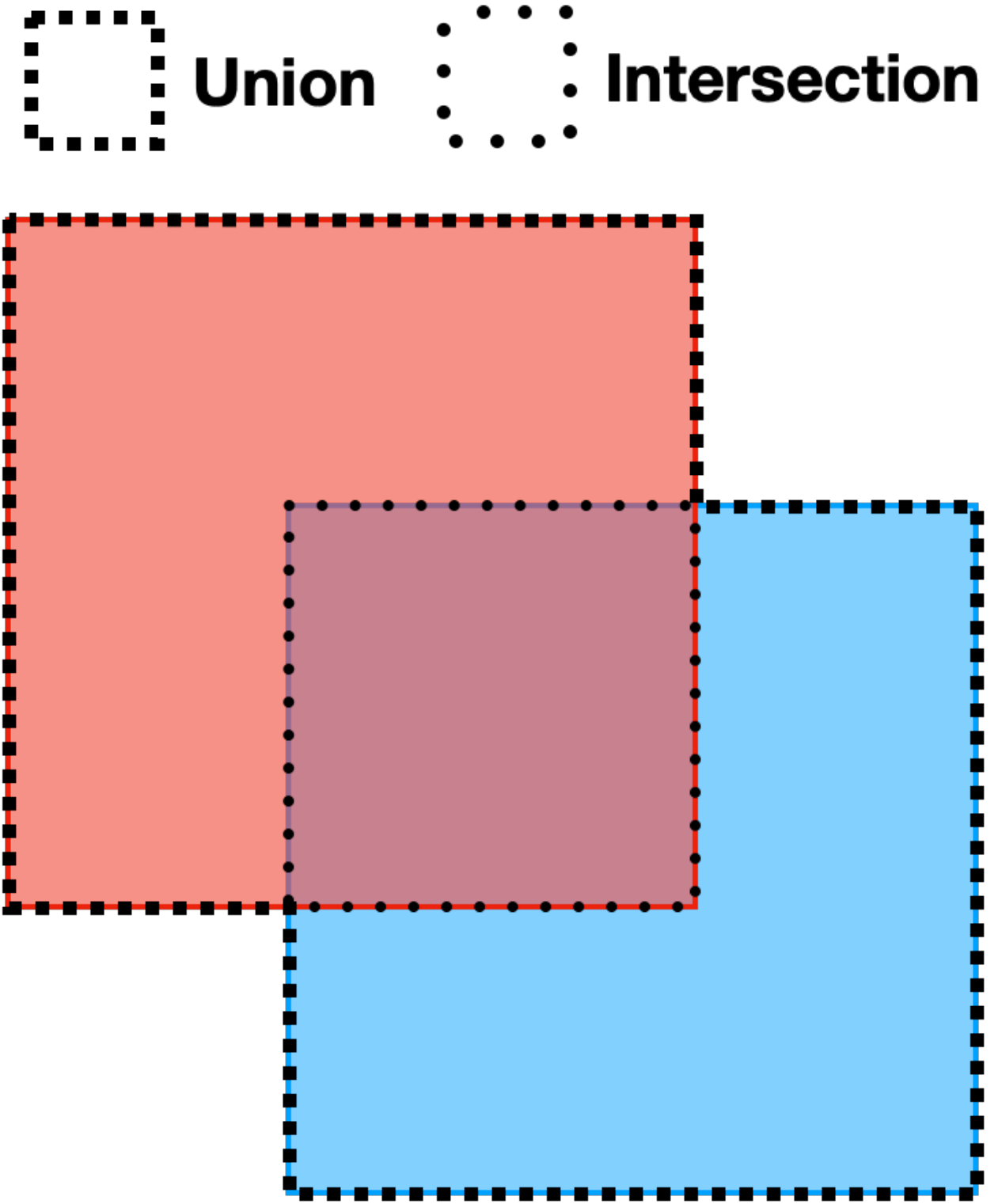}
    \caption{Geometric illustration of the ratio between the area of the intersection between two boxes (\textit{dotted region}) and the area of their union (\textit{dashed region}). We use the complement of this ratio as a scale-free distance metric bounded between zero and unity.}
    \label{fig:mark_iou}
\end{figure}

Using square boxes to define the marked clump locations allows us to inexpensively compute the ratio of the area of the intersection between pairs of boxes and the area of their union (see \autoref{fig:mark_iou}). We use the complement of this ratio, which is commonly referred to as the \textit{Jaccard distance} \citep{jaccard_distance}, as a scale-free distance metric between any volunteer-marked locations. 
\begin{equation}
    d=1-\frac{A_{\mathrm{intersection}}}{A_{\mathrm{union}}}
\end{equation}
The Jaccard distance is maximally unity if the boxes are disjoint and minimally zero if they coincide perfectly.

\section{Data Aggregation Model}\label{sec:data_aggregation_model}

The core of our data aggregation approach is based on a custom implementation of the probabilistic model and algorithm proposed by \bvp. In this section, we present a detailed description of the model and explain how it is used to  optimise the efficiency of clump detection using the volunteers' annotations. We recognise that this paper contains a lot of somewhat complicated notation, so to aid the reader we have included a reference table of the most commonly recurring symbols in Appendix \ref{sec:terms}.

\subsection{Overview}\label{subsec:overview}

We construct a global model that simultaneously considers $N_{\mathrm{S}}$ individual elements of the full subject set $S\equiv\{s_{i}\}_{i=1}^{N_{\mathrm{S}}}$ and individual members of the entire volunteer cohort $V$. Each subject $s_{i}\in S$, is inspected by a randomly selected group of volunteers $V_{i}\in V$, who each provide a set of independent two dimensional annotations of visible clump locations $Z_{i}\equiv\{z_{ij}\}_{j=1}^{|V_{i}|}$. Throughout this paper we will use the notation $|X|$ to denote the number of elements in the set $X$, so here $|V_{i}|$ denotes the number of volunteers who annotate the subject $s_{i}$. For convenience, we define $S_{j}\in S$ to denote the subset of subjects that are inspected by the $j$th volunteer. For every subject $s_{i}$, we define a \textit{true label} $y_{i}$ to encode the unknown locations of all \textit{real} clumps in the image. Using only the information provided by the global set of volunteer annotations $Z\equiv \bigcup_{i}Z_{i}$, we wish to derive a separate \textit{estimated} label $\hat{y}_{i}$ for each subject that closely approximates $y_{i}$. Our goal is to minimise the mismatch between $\hat{y}_{i}$ and $y_{i}$ while keeping the number of volunteers who annotate the subject $s_{i}$ as small as possible and thereby to optimise our use of volunteers' effort. We facilitate this aim by computing a ``risk'' metric $\mathcal{R}_{i}$ for each subject that represents a weighted combination of quantitative magnitude estimates for several sources of approximation error in the estimated label (see \autoref{subsec:computing_risk} for more details). We expect that the risk for a particular subject will decrease as the number of volunteer annotations for that subject increases. Accordingly, by choosing an appropriate global risk threshold $\mathcal{R}_{i} < \tau$, we aim to be able to confidently retire individual subjects from the classification pool as soon as the expected error is acceptably small. This approach differs from many traditional crowd-sourcing techniques, which require a fixed number of volunteers to inspect each subject. Such approaches are generally less efficient because stable consensus between volunteers is often achieved before the prescribed number of annotations have been gathered. An additional benefit of our approach is that particularly difficult subjects can be segregated for expert inspection if their risk remains high after many volunteers have inspected the subject.

\subsection{Associating subject annotations with subject labels}\label{subsec:agg_annots}
Each of the volunteer annotations $z_{ij}\in Z_{i}$ forms a set of $|B_{ij}|\geq0$ square boxes $z_{ij}=\{b_{ij}^{k}\}_{k=1}^{|B_{ij}|}$ that encodes the locations of any clumps that the volunteer perceived in the subject $s_{i}$. Analogously, we model the true clump locations for $s_{i}$ as an abstract set of $|B_{i}| \geq 0$ \textit{rectangular} boxes such that $y_{i}\equiv\{b_{i}^{l}\}_{l=1}^{|B_{i}|}$. The concrete sizes and shapes of these boxes are ultimately determined by our aggregation algorithm, but for subject $s_{i}$ they are guaranteed to be at least as large as the boxes comprising the volunteer annotations for that subject. Our goal is to associate each of the click locations corresponding to volunteer annotations for a particular subject with a single true clump location. Formally, we aim to associate each of the concrete elements of $Z_{i}$ with a single abstract element of $y_{i}$. This task is complicated for several reasons. Different volunteers may annotate different subsets of clumps and the order in which they do so is not defined nor even constrained. Volunteers may miss some real clumps, so there may be elements of $y_{i}$ that have no counterpart annotations in a particular $z_{ij}$. Conversely, the set of annotations provided by a particular volunteer, for a particular subject may contain false positives, so some elements of a particular $z_{ij}$ may not correspond with any elements of $y_{i}$. 

\begin{figure*}
    \centering
    \includegraphics[width=0.95\textwidth]{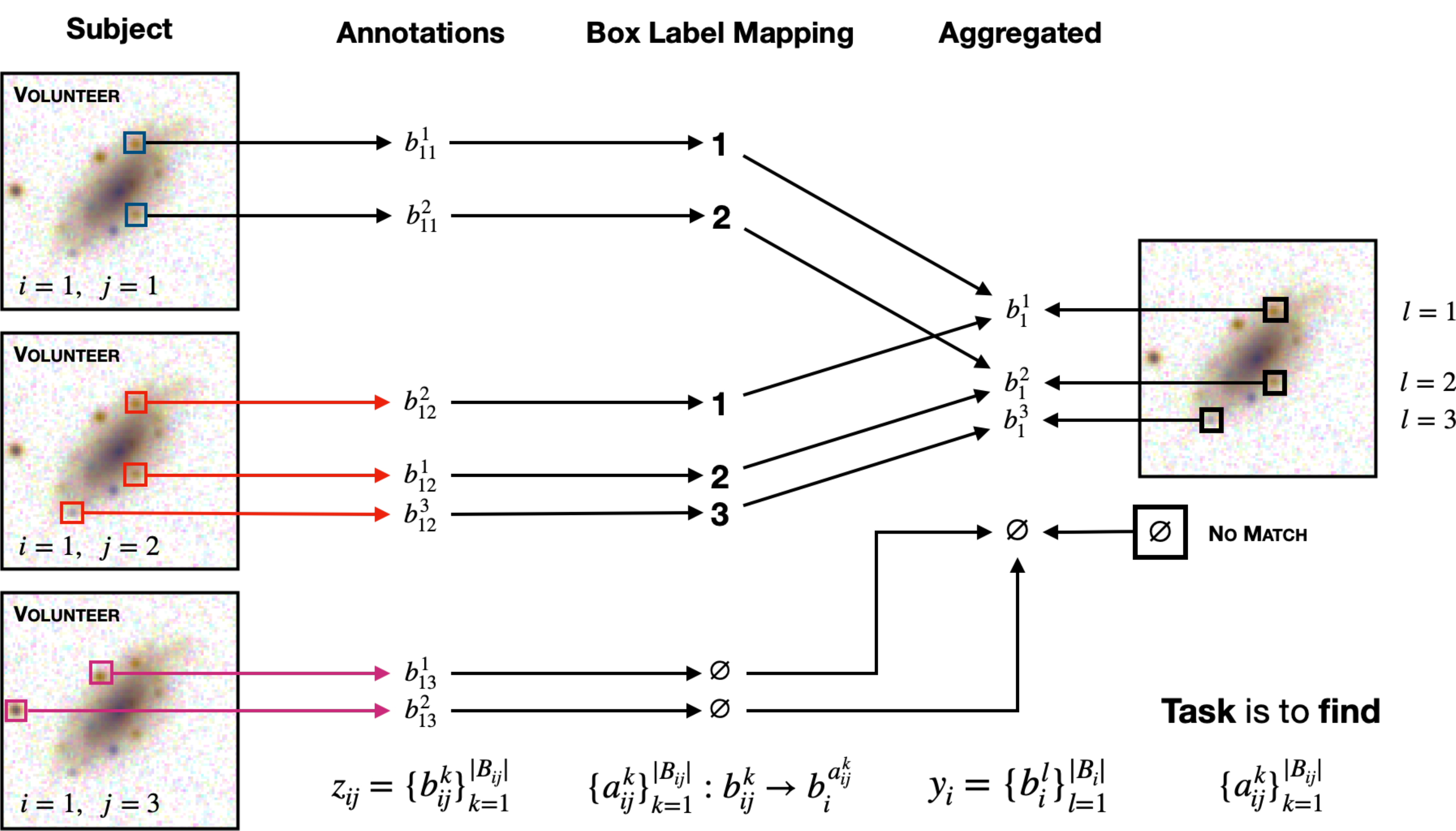}
    \caption{Schematic illustration of how elements of volunteers' annotations are associated with elements of the subject label $y_{i}$. We illustrate a case in which three volunteers have provided three independent annotations of the same subject. Volunteers 1 and 2 both annotate subsets of the real clumps in the image. Volunteer 3 mistakenly marks two foreground stars as clumps. The central column lists the value of $\{a_{ij}^{k}\}$ computed for each of the boxes forming the volunteers' annotations. For volunteers 1 and 2, these values define the index of the corresponding box in $y_{i}$. Both annotations provided by volunteer 3 probably mark foreground stars and neither is marked by another volunteer. In this toy example, the algorithm maps both to the ``$\varnothing$'' element, thereby defining them as false positives.}
    \label{fig:box_associations}
\end{figure*}

\autoref{fig:box_associations} provides a schematic illustration of the process by which we associate volunteer annotations with probable clump locations and \autoref{subsec:comp_box_assoc} explains the notation and the computational details. Formally, our aggregation algorithm computes an optimal set of mapping indices $\{a_{ij}^{k}\}_{k=1}^{|B_{ij}|}$ such that each volunteer-provided box $b_{ij}^{k}\in z_{ij}$ is associated with real clump location $b_{i}^{a_{ij}^{k}}\in y_{i}$. The possibility of false positive boxes in $z_{ij}$ is accounted for by defining a singleton ``$\varnothing$'' element to which they can be associated.

\subsection{Modelling volunteer skill}\label{subsec:volunteer_skill}
For a given subject, the visibility of clumps to a particular volunteer, and the positional accuracy with which they are able annotate the clumps they \textit{do} perceive is likely to be influenced by several factors. These may include: domain expertise, experience gained from time spent contributing to \textit{Galaxy Zoo: Clump Scout}, confusion regarding the detailed task instructions and even the screen size and resolution of device they typically use to provide annotations.

To model the impact of these factors we consider three scenarios, which relate a particular volunteer's annotations to the locations of real clumps in the subject image. Consider the annotations provided by the $j$th volunteer in our cohort.

In the first scenario, volunteer $j$ provides a \textit{true positive} by marking a location that lies close to a real clump. It is unlikely that any volunteer's mark precisely annotates the true clump location and indeed, different volunteers may have different perceptions of where the clump actually is. We model any positional offset between the volunteer $j$'s annotation and the true clump location as a random Jaccard distance $d_{j}$, drawn from a Gaussian distribution with zero mean and a volunteer-specific variance $\sigma^{2}_{j}$.
\begin{equation}
    d_{j} \sim \mathrm{Gaussian}\left(0, \sigma^{2}_{j}\right)
\end{equation}

In the second scenario, the volunteer provides a \textit{false positive} by marking a location which does \textit{not} correspond to the location of a real clump. We model the rate of false positive annotations for volunteer $j$ by considering each mark they provide as a Bernoulli trial with ``success'' probability $p_{j}^{\mathrm{fp}}$.

Finally, volunteer $j$ may provide an implicit \textit{false negative} by failing to mark the location of a real clump. We model the false negative rate for volunteer $j$ by considering each opportunity to mark a real clump location as a Bernoulli trial with ``success'' probability $p_{j}^{\mathrm{fn}}$.

Hereafter, we refer collectively to the three model parameters $\mathcal{S}_{j} \equiv \{\sigma_{j}, p_{j}^{\mathrm{fp}}, p_{j}^{\mathrm{fn}}\}$ as volunteer $j$'s ``skill'' parameters. 

\subsection{Modelling subject difficulty}\label{subsec:image_difficulty}
Notwithstanding the skill of individual volunteers, there are numerous \textit{image} characteristics that may result in varying degrees of clump visibility at different locations for different subjects in $S$. An obvious example is clump contrast; bright clumps that appear superimposed on a smooth, faint background galaxy will be easier to discern than faint clumps on a bright, noisy background. For simplicity, we assume that the impact of all such confounding factors manifests as a positional offset between the true location of a clump and any volunteer annotations that identify it. For a particular true clump location $b_{i}^{l}\in y_{i}$, we model the size of this offset as a random Jaccard distance $d_{i, l}$ drawn from a Gaussian distribution with zero mean and variance ${\sigma^{l}_{i}}^{2}$. 
\begin{equation}
    d_{i, l} \sim \mathrm{Gaussian}\left(0, {\sigma^{l}_{i}}^{2}\right)
\end{equation}
Hereafter, we refer to the set $\mathcal{D}_{i}\equiv\{\sigma_{i}^{l}\}$ as the subject ``difficulty''.

\subsection{Modelling volunteer annotations}\label{subsec:clump_annotations}

We combine our volunteer skill and image difficulty models to define a compound model for the annotation $z_{ij}$ that each volunteer provides for a each subject $s_{i}$. 
\begin{equation}\label{eq:anno_model}
\begin{aligned}
p(z_{ij}|y_{i}, \mathcal{D}_{i},\mathcal{S}_{j}) =
&
\,(p_{j}^{\mathrm{fn}})^{n_{\mathrm{fn}}}(1-p_{j}^{\mathrm{fn}})^{n_{\mathrm{tp}}}\\
&
\cdot(p_{j}^{\mathrm{fp}})^{n_{\mathrm{fp}}}(1-p_{j}^{\mathrm{fp}})^{n_{\mathrm{tp}}}\\
&\cdot\displaystyle\prod\limits_{\substack{k=1,\\ l=a_{ij}^{k}\neq\varnothing}}^{|B_{ij}|}
\mathrm{Gaussian}\left(\vert d_{kl}\vert^{2};{\sigma_{ij}^{k}}^{2}\right)
\end{aligned}
\end{equation}
The first and second terms represent binomial models, which compute the probability that $z_{ij}$ contains $n_{\mathrm{fn}}$ false negatives, $n_{\mathrm{fp}}$ false positives and $n_{\mathrm{tp}} = |B_{ij}|-n_{\mathrm{fp}}$ true positives, given $p_{j}^{\mathrm{fp}}$ and $p_{j}^{\mathrm{fn}}$.

The third term considers the Jaccard distances $d_{kl}$ between any \textit{true positive} (i.e. $a_{ij}^{k}\neq\varnothing$) box $b_{ij}^{k}\in z_{ij}$ and their counterparts $b_{i}^{l}\in y_{i}$ as well as the subject's difficulty $\mathcal{D}_{i}$ and the volunteer's skill $\mathcal{S}_{j}$.

We combine the Gaussian components of the volunteer skill and image difficulty models by computing a combined variance parameter
\begin{equation}\label{eq:combined_variance}
{\sigma_{ij}^{k}}^{2} = (1-\eta)\cdot {\sigma_{i}^{a_{ij}^{k}}}^{2} + \eta\cdot {\sigma_{j}}^{2}
\end{equation}
where $\eta$ weights the relative impact of volunteer skill and image difficulty according to the $p$-values computed by their respective probability models. Formally, we model $\eta$ as the expected value of a binary indicator variable, $e$
\begin{equation}
    e = 
    \begin{cases}
    \textrm{1 if volunteer skill dominates } d_{kl}\\
    \textrm{0 if image difficulty dominates } d_{kl}
    \end{cases}
\end{equation}
We assume that both sources of variance are equally likely to dominate for any particular volunteer annotation (i.e. $P(e=1) = P(e=0)$), which implies \citep[e.g.][]{2019sdmm.book.....I}
\begin{equation}
\eta = \mathbb{E}(e) = \frac{\displaystyle\sum_{e=0}^{1}eP(d_{kl}|e)}{\displaystyle\sum_{e=0}^{1}P(d_{kl}|e)} = \frac{p_{j}}{p_{i}+p_{j}}\label{eq:eta_def}
\end{equation}
where
\begin{align}
p_{j}&=P(d_{kl}|e=1)=\mathrm{Gaussian}\left(\left|d_{kl}\right|^{2};{\sigma_{j}}^{2}\right)\notag\\
p_{i}&=P(d_{kl}|e=0)=\mathrm{Gaussian}\left(\left|d_{kl}\right|^{2};{\sigma_{i}^{l}}^{2}\right)\notag
\end{align}

\subsection{Global model and parameter priors}\label{subsec:global_mod_and_priors}

Our combined model for a single volunteer annotation of a single subject (i.e. $p(z_{ij}|y_{i}, \mathcal{D}_{i},\mathcal{S}_{j})$, \autoref{eq:anno_model}) forms the kernel of a joint model for the set of all subject \textit{true} labels $Y\equiv\{y_{i}\}_{i=1}^{|S|}$, the set of all subject difficulties $\mathcal{D}\equiv\{\mathcal{D}_{i}\}_{i=1}^{|S|}$ and the set of all volunteer skills $\mathcal{S}\equiv\{\mathcal{S}_{j}\}_{j=1}^{|V|}$ given the union of all volunteer annotations, which we denote $Z$.
\begin{equation}\label{eq:joint_model}
\begin{aligned}
    P(Y, \mathcal{D},\mathcal{S}|Z) =& \prod_{i}\pi(y_{i})\pi(\mathcal{D}_{i})\\
    &\cdot\prod_{j}\pi(\mathcal{S}_{j})p(z_{ij}|y_{i}, \mathcal{D}_{i},\mathcal{S}_{j})
\end{aligned}
\end{equation}
The additional terms in \autoref{eq:joint_model} represent prior distributions for the parameters of our model.
\begin{enumerate}
    \item $\pi(\mathcal{D}_{i})$ models the prior probabilities of observing the difficulty parameters associated with the $i$th subject.
    \item $\pi(\mathcal{S}_{j})$ models the prior probability of observing the volunteer skill parameters associated with the $j$th volunteer.
    \item $\pi(y_{j})$ models the prior probability that the unknown true label for $s_{i}$ is $y_{i}$. For simplicity, we assume that all possible labels are equally likely.
\end{enumerate}
For practical reasons, we choose prior distributions for each parameter that are the \textit{conjugate priors}\footnote{Specifying a conjugate prior $\pi(\theta)$ for parameter $\theta$ in Bayes's rule yields a posterior distribution $p(\theta|z) \propto \pi(\theta)\cdot p(z|\theta)$ that has the same functional form as the prior itself. Note that in general the conjugate prior depends on both the likelihood model and the parameter of interest. For example, the variance and mean of a Gaussian likelihood function have different conjugate priors.} of that parameter for the corresponding likelihood model distribution. This choice facilitates straightforward computation of model parameter updates when new annotations are collected. 

Specifically, we use Beta distribution priors for the binomial-distributed parameters $\{p_{j}^{\mathrm{fp}}, p_{j}^{\mathrm{fp}}\}$ 
\begin{equation}
    \pi(p_{j}^{k}) \sim \mathrm{Beta}(p_{j}^{k};n_{\beta}^{k}p_{0}^{k}, n_{\beta}^{k}(1-p_{0}^{k})):\;k\in[\mathrm{fp},\mathrm{fn}].
\end{equation}
Intuitively, this prior simulates the information gained by performing $n_{\beta}$ Bernoulli trials with success probability $p_{0}^{k}$. 

For the parameters that are modeled as variances of Gaussian likelihood models $\{\sigma_{j}^{2}, {\sigma_{i}^{l}}^{2}\}$, we specify scaled inverse chi-squared priors 
\begin{equation}
\begin{aligned}
      \pi({\sigma_{i}^{l}}^{2}; n_{\chi,S}, \sigma_{0,S}^{2})= \mathrm{Scale-inv-}\chi^{2}(\sigma^{2};  n_{\chi, S}, \sigma_{0,S}^{2})\\
      \pi(\sigma_{j}^{2}; n_{\chi, V}, \sigma_{0,V}^{2})= \mathrm{Scale-inv-}\chi^{2}(\sigma^{2};  n_{\chi, V}, \sigma_{0,V}^{2})
\end{aligned}
\end{equation}
which simulate the information gained from a sample of $n_{\chi}$ previous observations drawn from a Gaussian distribution with zero mean and variance $\sigma_{0}^{2}$.

The \textit{initial} values for parameters of our prior models $\{p_{0}^{\mathrm{fp}}, p_{0}^{\mathrm{fp}}, n_{\beta}^{\mathrm{fp}}, n_{\beta}^{\mathrm{fn}}, \sigma_{0,S}^{2}, n_{\chi, S}, \sigma_{0,V}^{2}, n_{\chi, V}\}$ are hyper-parameters of our algorithm which must be chosen \textit{a-priori}. \autoref{tab:prior_param_values} lists the values that we assign to each of these hyper-parameters when processing the \gzcs dataset. 
\begin{table}
    \centering
	\caption{Framework hyper-parameter values used to process the \gzcs dataset.}
	\label{tab:prior_param_values}
	\begin{tabular}{lc}
		\hline
		Parameter & Value \\
		\hline
		$p_{0}^{\mathrm{fp}}$&0.1\\ $p_{0}^{\mathrm{fp}}$&0.1\\ $n_{\beta}^{\mathrm{fp}}$&500\\ $n_{\beta}^{\mathrm{fn}}$&50\\ $\sigma_{0,S}^{2}$&0.1\\
		$n_{\chi, S}$&10\\ $\sigma_{0,V}^{2}$&0.1\\ 
		$n_{\chi, V}$&10\\
		$f_{\mathrm{V}}$&0.1\\
		$d_{\max}$&0.9\\
		\hline
	\end{tabular}
\end{table}

In Appendix \ref{sec:appendix_priors} we provide detailed rationale for our choice of prior distribution models and show how they yield estimates for our likelihood model parameters that become increasingly data-dominated as more annotations are collected.

\section{Computing Aggregated Labels}\label{sec:computing_aggregated_labels}
\begin{figure*}
    \centering
    \includegraphics[width=\textwidth]{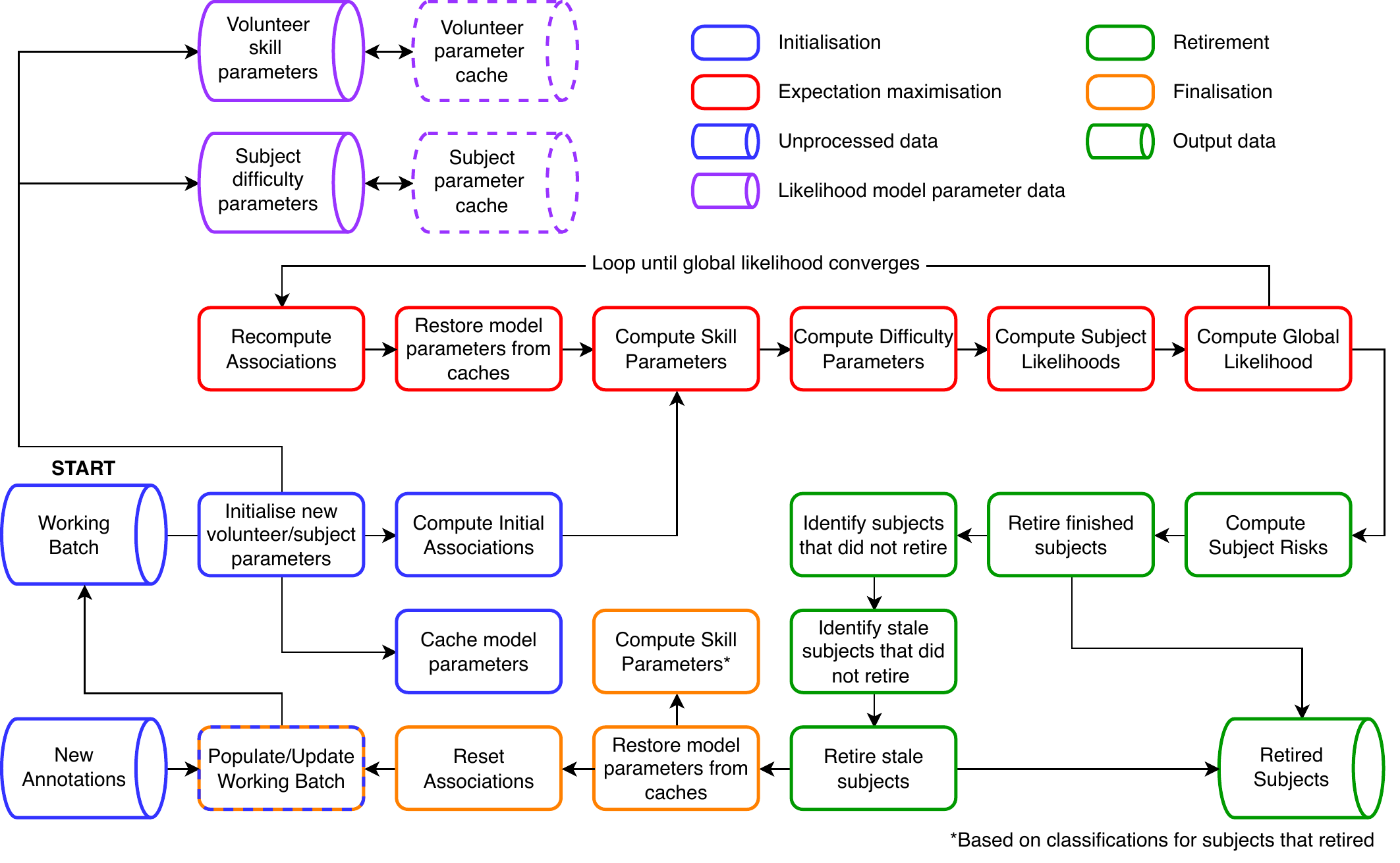}
    \caption{Schematic overview of the aggregation algorithm.}
    \label{fig:agg_algo}
\end{figure*}

\autoref{fig:agg_algo} provides a schematic overview of how our implementation computes aggregated labels for subjects. In subsequent subsections we describe the illustrated operations in detail.

\subsection{The Working Batch}

To minimise the dependence of aggregated clump locations on our choice of model prior hyper-parameters we design our aggregation framework to process elements from a dynamically maintained \textit{working batch} containing data and metadata for $\lesssim25$ thousand classifications.\footnote{Although our implementation does not explicitly limit batch sizes, in practice we found that model data storage requirements for batches containing $\gtrsim25$ thousand classifications exhausted the 32 GB memory capacity of our available hardware.} Each element in the working batch represents a single click location marking a clump as part of the annotation provided by a single volunteer.

To populate the working batch, we select subjects that have been inspected by at least three volunteers and have at least one annotated clump. For each selected subject, we assemble all its available annotation data and append them to the working batch in a single block of elements. This ensures that any subject retirement decision is made on the basis of all available information. We specify a minimum target batch size and new blocks are added until the size of working batch exceeds this target. If five or more volunteers inspect a subject and none annotate a clump, we assume that no clumps are present and preemptively retire the subject instead of adding its data to the working batch. Whenever a volunteer inspects a subject that has at least one clump annotation, but does not annotate any clumps themselves, we append a single \textit{empty classification} element to the working batch. We require records of these empty classifications in order to compute the probability that a particular volunteer fails to annotate a real clump, i.e. $p_{j}^{\mathrm{fn}}$.

After processing a single batch of classification data, the most likely outcome is that only a subset of the corresponding subjects will have $\mathcal{R}_{i}<\tau$ (see \autoref{subsec:overview} and \autoref{subsec:computing_risk}) and be deemed sufficiently low-risk for retirement. We update the working batch by removing the classification data for retired subjects and replenishing them with new blocks of classification data for active subjects. Once a subject is retired, the aggregated estimated label $\hat{y}_{i}$ is considered final and any subsequently submitted classifications for that subject will not be included in subsequent batches.

We impose a maximum lifetime for any data element by specifying the maximum number of batch replenishment cycles that they can persist within the working batch. Subjects whose data remain after this lifetime has expired are retired and flagged for inspection by experts. This forced retirement strategy prevents the working batch becoming stale and dominated by inherently difficult or high-risk subjects that never retire normally.

\subsection{Initialisation}

Processing of each working batch begins with an initialisation phase. Adding new blocks of data to the working batch implies introducing new subjects to our likelihood model. We initialise the subject difficulty parameters of all new subjects to the same value, which we specify as a hyper-parameter of our aggregation framework.
\begin{equation}
    \sigma^{2}_{i, \mathrm{init}}=\sigma_{\mathrm{S},0}^{2}\;\forall i
\end{equation}
The newly added data blocks \textit{may} include annotations that were provided by previously unknown volunteers. If so, we initialise the skill parameters for all new volunteers identically using three of the hyper-parameters that were introduced in \autoref{subsec:global_mod_and_priors}.
\begin{align}
    p_{j, \mathrm{init}}^{\mathrm{fp}}&=p_{0}^{\mathrm{fp}}\;\forall j\\
    p_{j, \mathrm{init}}^{\mathrm{fn}}&=p_{0}^{\mathrm{fn}}\;\forall j\\
    \sigma^{2}_{j, \mathrm{init}}&=\sigma_{\mathrm{V},0}^{2}\;\forall j
\end{align}
A subset of elements in the working batch correspond with subject blocks from earlier batches that did not retire. We re-initialise the parameters for these subjects, and re-compute the skill parameters of returning volunteers to reflect only their annotations for subjects that \textit{have} retired. This parameter propagation strategy allows us to use information that we have learned about volunteers' skills, while ensuring that the subjects that persist between batches are processed identically to new subjects that happen to have received annotations from returning volunteers. After initialising or propagating the model parameters for all elements of the working batch, we cache their values. 

To complete the initialisation phase for each new working batch we use the algorithm described in \autoref{subsec:comp_box_assoc} to perform preliminary clustering of overlapping volunteer annotations for each subject. The subsequent subsections explain how we apply iterative expectation maximisation to refine the initial clusters, while simultaneously computing the maximum likelihood solution of \autoref{eq:joint_model}.

\subsection{Computing box associations}\label{subsec:comp_box_assoc}

For each subject $s_{i}\in S$, we follow the approach of \bvp and implement a \textit{Facility Location} algorithm \citep{Mahdian01greedyfacility} to approximately\footnote{The chosen algorithm implements approximate computation of the maximum log-likelihood solution and is guaranteed to find a solution for which the log-likelihood is at most 1.61 times the optimal one.} derive the maximum likelihood mapping $A=\{a_{ij}^{k}\}_{k=1}^{|B_{ij}|}$ between the click locations comprising individual volunteers' annotations $z_{ij} = \{b_{ij}^{k}\}_{k=1}^{|B_{ij}|}$ and the set $y_{i} = \{b_{i}^{l}\}_{l=1}^{|B_{i}|}$ (see \autoref{subsec:agg_annots} and \autoref{fig:box_associations}).

\begin{figure}
    \centering
    \includegraphics[width=0.35\textwidth]{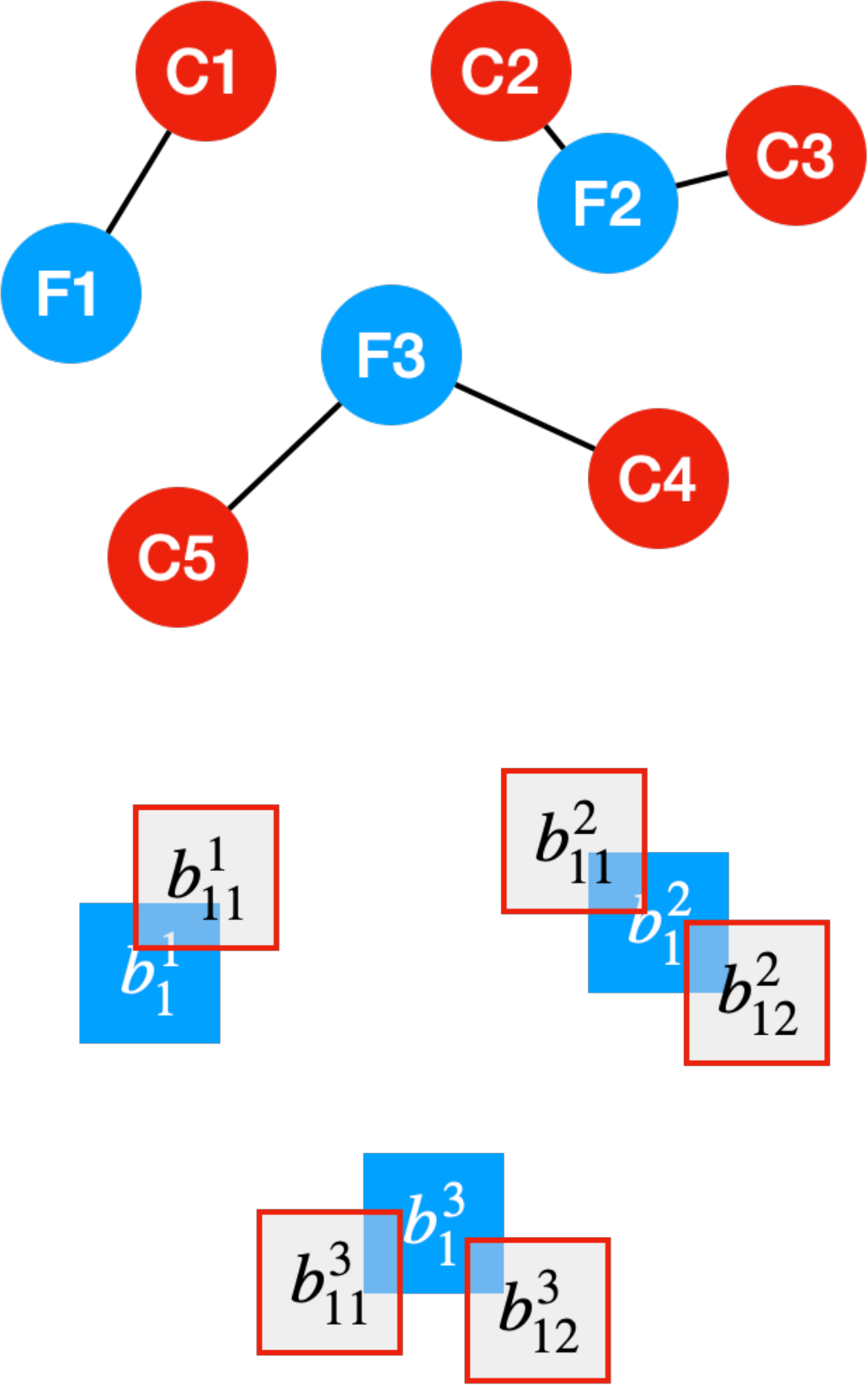}
    \caption{\textit{Top:} The topology of the clusters that are assembled by the Facility Location algorithm. In this case the set of boxes has been partitioned into three clusters. Within each cluster, the central \textit{facility} (F 1-3) is connected to one or more \textit{cities} (C 1-5). Each city is connected to exactly one facility. \textit{Bottom:} Possible arrangement of aggregated box clusters corresponding to the illustrated topology for an image after inspection by three volunteers. Blue boxes $b_{i}^{l}$ correspond to facilities (F 1-3) and red boxes $b_{ij}^{k}$ correspond with the cities (C 1-5). Note that each volunteer may contribute at most one box to each cluster and in this case the same volunteer contributed the boxes that were assigned facility status.}
    \label{fig:facloc_topology}
\end{figure}

Facility location algorithms form clusters with a specific topology comprising one or more \textit{cities}, uniquely connected to a single, central \textit{facility}.\footnote{This nomenclature reflects a common application of facility location algorithms to optimise distribution of some essential commodity from facilities located at a small number of locations within a larger network of cities.} This topology is illustrated in \autoref{fig:facloc_topology}.

Our implementation identifies disjoint, spatially concentrated subsets of the boxes in $Z_{i}$ which we then identify with true clump locations $b_{i}^{l}\in y_{i}$. We label each of these aggregated clusters with the index $l$ and denote them as $Z_{i}^{l}$. Establishing a new cluster entails labelling a particular box $b_{ij}^{k}\in Z_{i}$ as a facility and connecting at least one other box $b_{ij^{\prime}}^{k^{\prime}}$ that was provided by a different volunteer. Note that by associating box $b_{ij}^{k}$ with cluster $Z_{i}^{l}$ as either a city or a facility, we establish the mapping $a_{ij}^{k} = l$. Each box in the set of volunteer annotations is associated with \textit{at most} one true clump and each subset may contain \textit{at most} one box per volunteer. These constraints reflect our assumption that separate marks provided by the same volunteer are intended to indicate separate clumps.

We specify that assigning facility status to a particular box incurs a real-valued cost 
\begin{equation}\label{eq:fac_cost} 
    C^{\mathrm{f}}(b_{ij}^{k}) = -\displaystyle\sum_{j=1}^{ |V_{i}|}\ln(p_{j}^{\mathrm{fn}})
\end{equation}
and connecting another box $b_{ij^{\prime}}^{k^{\prime}}$ to an established facility $b_{ij}^{k}$ incurs a cost
\begin{align}
     C^{\mathrm{fc}}(b_{ij}^{k}, b_{ij^{\prime}}^{k^{\prime}}) = &\ln(p_{j^{\prime}}^{\mathrm{fn}})-\ln(1-p_{j^{\prime}}^{\mathrm{fn}})~-\notag\\ 
     &\ln(1-p_{j^{\prime}}^{\mathrm{fp}})-
     \ln\left[\mathrm{Gaussian}(\vert d\vert^{2}, \sigma_{j^{\prime}}^{2})\right]\notag\\
     = &\ln(p_{j^{\prime}}^{\mathrm{fn}})-\ln(p_{j^{\prime}}^{\mathrm{tn}})~-\notag\\ 
     &\ln(p_{j^{\prime}}^{\mathrm{tp}})-
     \ln\left[\mathrm{Gaussian}(\vert d\vert^{2}, \sigma_{j^{\prime}}^{2})\right]\label{eq:city_fac_cost}
\end{align}
where $d$ represents the Jaccard distance between $b_{ij}^{k}$ and $b_{ij^{\prime}}^{k^{\prime}}$. 

Combining these cost definitions yields the assembly cost for an individual cluster 
\begin{equation}\label{eq:cluster_cost}
C(Z_{i}^{l}) = C^{\mathrm{f}}(b_{ij}^{k}) + \displaystyle\sum\limits_{
\substack{b_{ij^{\prime}}^{k^{\prime}}\in Z_{i}^{l}\\ 
j\neq j^{\prime}}} 
C^{\mathrm{fc}}(b_{ij}^{k}, b_{ij^{\prime}}^{k^{\prime}})
\end{equation}

Some boxes may represent false positive annotations. To handle these cases we follow the approach of \bvp and establish a \textit{dummy} facility at zero cost. Connections to the dummy facility identify boxes as false positives and incur box-specific costs 
\begin{equation}\label{eq:city_dummy_cost}
    C^{\varnothing \mathrm{c}}(b_{ij^{\prime}}^{k}) = -\ln(p_{j^{\prime}}^{\mathrm{fp}})
\end{equation}

Let $Z^{\star}_{i}$ be the set of all established clusters for subject $s_{i}$. The definitions of $C^{\mathrm{f}}(b_{ij}^{k})$,  $C^{\mathrm{fc}}(b_{ij}^{k}, b_{ij^{\prime}}^{k^{\prime}})$ and $C^{\varnothing \mathrm{c}}(b_{ij^{\prime}}^{k})$ imply an expression for the total cost $C_{i}$ of all established clusters and all connections to the dummy facility that closely approximates the negative natural logarithm of the product over volunteers $\prod_{j}\pi(\mathcal{S}_{j})p(z_{ij}|y_{i}, \mathcal{D}_{i},\mathcal{S}_{j}))$ defined in \autoref{eq:anno_model}.\footnote{The correspondence is approximate because $d_{kl}$ in equation \autoref{eq:anno_model} represents the Jaccard distance between a volunteer box and the \textit{true} clump location, whereas $d$ in equation \autoref{eq:city_fac_cost} is the Jaccard distance between two volunteer boxes, one of which is labeled as a facility.}
\begin{align}\label{eq:total_cost}
   C_{i} &= \displaystyle\sum_{Z_{i}^{l}\in Z^{\star}_{i}} C(Z_{i}^{l}) +\displaystyle\sum_{b_{ij^{\prime}}^{k}\in Z_{i}\setminus Z^{\star}_{i}} C^{\varnothing \mathrm{c}}(b_{ij^{\prime}}^{k}) \notag \\
   &\approx -\ln\left(\prod_{j}p(z_{ij}|y_{i}, \mathcal{D}_{i},\mathcal{S}_{j}))\right)
\end{align}

The facility location algorithm is designed to compute the box-to-cluster mapping that minimises $C_{i}$, which simultaneously yields the approximate maximum likelihood solution of \autoref{eq:anno_model} for given volunteer skill and image difficulty parameters.

To derive the aggregated estimate for the subject label $\hat{y}_{i}$, we merge the individual boxes comprising each cluster by computing the mean coordinates of their corresponding vertex indices.\footnote{Concretely, let $\mathbf{r}$ be a generalised two dimensional coordinate and the index $m$ enumerate the corner vertices of a box, beginning in the upper-left and proceeding along the box edges in a clockwise direction, then \begin{equation}\mathbf{r}_{i,m}^{l} = \frac{1}{|Z_{i}^{l}|}\displaystyle\sum_{b_{ij}^{k}\in Z_{i}^{l}}\mathbf{r}_{ij,m}^{k}\notag\end{equation}} This yields a rectangular representation for each true clump location that is at least as large as each of the boxes comprising the set of annotations for the $i$th subject, $Z_{i}$.

During the initialisation phase, we use a simplified set of facility location costs that do not depend on the volunteer skill parameters or the image difficulties. We specify that establishing a new facility during initialisation incurs the same cost for any volunteer annotation
\begin{equation}\label{eq:init_open_cost} 
    C^{\mathrm{f, init}}_{i} = f_{\mathrm{V}}|V_{i}|
\end{equation}
where $f_{\mathrm{V}}\in [0,1]$ is a hyper-parameter that represents the fraction of volunteers who inspected a subject that must contribute a box to an assembled cluster and we remind readers that $|V_{i}|$ denotes the number of volunteers who inspected the $i$th subject. The initialisation-phase cost of connecting box $b_{ij^{\prime}}^{k^{\prime}}$ to an established facility $b_{ij}^{k}$ still depends on the Jaccard distance $d$ between them.
\begin{equation}\label{eq:init_conn_cost}
    C^{\mathrm{fc, init}}(b_{ij}^{k}, b_{ij^{\prime}}^{k^{\prime}}) = {\begin{cases}0~&{\text{ if }}~d \leq d_{\max}\\\infty~&{\text{ if }}~d > d_{\max}\end{cases}}
\end{equation}
where $d_{\max}\in [0,1]$ is a hyper-parameter that represents the maximum Jaccard distance between any city in a cluster and its central facility. Finally, connecting any box to the dummy facility during initialisation incurs unit cost
\begin{equation}\label{eq:init_city_dummy_cost}
    C^{\varnothing \mathrm{c, init}}(b_{ij^{\prime}}^{k}) = 1
\end{equation}

\autoref{tab:prior_param_values} lists the values we adopt for $f_{\mathrm{V}}$ and $d_{\max}$.

\subsection{Computing Image Difficulty}\label{subsec:compute_image_diff}

For each rectangular box $\hat{b}_{i}^{l}\in\hat{y}_{i}$ comprising the estimated label for the $i$th subject, we use the global hyper-parameter $\sigma^{2}_{S,0}$ to define a subject-specific minimum difficulty
\begin{equation}
    \sigma^{2}_{\min, i} = \frac{\sigma^{2}_{S,0}}{\vert\hat{y}_{i}\vert}
\end{equation}
Intuitively, if a subject's label includes more identified clump locations then we assume that clumps are easier to precisely locate and the minimum difficulty is reduced. We then update the minimum value to reflect the scatter between the subset of volunteer boxes ${b}_{ij}^{k}\in Z_{i}^{l}$ that were associated with the corresponding ground truth cluster. For each of these \textit{true positive} boxes we compute the Jaccard distance ${d_{ij}^{k}}^{l}$ between it and its corresponding rectangular box in the estimated subject label, $\hat{b}_{i}^{l}$. Using these distances in conjunction with \autoref{eq:variance_param_estimate} we estimate
\begin{equation}\label{eq:difficulty_estimate}
    {\sigma^{l}_{i}}^{2}\approx\frac{n_{\chi, S}\sigma_{S, 0}^{2}}{\vert Z_{i
    }^{l}\vert+ n_{\chi, S}+2}\displaystyle\sum_{\hat{b}_{ij}^{k}\in Z_{i
    }^{l}}\vert\Delta\vert^{2}
\end{equation}
where $\Delta^{2} = \sigma^{2}_{\min, i} + \vert{d_{ij}^{k}}^{l}\vert^{2}$, and $n_{\chi, S}$ is another hyper-parameter or our algorithm (see \autoref{subsec:global_mod_and_priors} and Appendix \ref{sec:appendix_priors}).

\subsection{Computing Volunteer Skill}\label{subsec:compute_volunteer_skill}

We compute each volunteer's skill parameters $p_{j}^{\mathrm{fp}}$, $p_{j}^{\mathrm{fp}}$ and $\sigma_{j}^{2}$ (see \autoref{subsec:volunteer_skill}) by comparing their individual clump annotations $z_{ij}\in Z$ for each subject in $s_{i}\in S_{j}$ with the corresponding label estimate $\hat{y}_{i}$. For each volunteer, we compute the number of \textit{false positives} by counting the subset of their annotation boxes that were associated with the dummy cluster.
\begin{equation}\label{eq:num_fp}
    n_{\mathrm{fp}, j}=\displaystyle\sum_{s_{i}\in S_{j}}
    \displaystyle\sum_{z_{ij}\in Z_{i}}\mathbf{1}[a_{ij}^{k}=\varnothing]
\end{equation}
We compute the number of \textit{false negatives} for a volunteer by summing the number of established clusters for each image they inspected that do not contain one of their boxes.
\begin{equation}\label{eq:num_fn}
    n_{\mathrm{fn}, j}=
    \displaystyle\sum_{s_{i}\in S_{j}}
    \displaystyle\sum_{Z_{i}^{l}\in Z_{i}}\mathbf{1}[z_{ij}\cap Z_{i}^{l}=\emptyset]
\end{equation}
Note that $\emptyset$ in \autoref{eq:num_fn} represents the empty set and not our notation for the dummy facility $\varnothing$. Analogously, we compute the number of \textit{true positives} by counting the total number of clusters to which the volunteer contributed.
\begin{equation}\label{eq:num_tp}
    n_{\mathrm{tp}, j}=
    \displaystyle\sum_{s_{i}\in S_{j}}
    \displaystyle\sum_{Z_{i}^{l}\in Z_{i}}\mathbf{1}[z_{ij}\cap Z_{i}^{l}\neq\emptyset]
\end{equation}
We use the expressions in \autoref{eq:num_fp} and \autoref{eq:num_fn} in conjunction with \autoref{eq:skill_prob_estimate} to compute estimates for $p^{\mathrm{fp}}_{j}$ and $p^{\mathrm{fn}}_{j}$.
\begin{align}
p_{j}^{\mathrm{fp}}&\approx \frac{n_{\beta}p_{0}^{\mathrm{fp}}+n_{\mathrm{fp}, j}}{n_{\beta} + \vert Z_{j}\vert }\label{eq:prob_fp}\\
p_{j}^{\mathrm{fn}}&\approx \frac{n_{\beta}p_{0}^{\mathrm{fn}}+n_{\mathrm{fn}, j}}{n_{\beta} + \vert Z_{j}\vert}\label{eq:prob_fn}
\end{align}
To compute $\sigma_{j}^{2}$ for each volunteer we follow a similar approach to that used when computing image difficulties. We compute the Jaccard distances $\{{d_{ij}^{k}}^{l}\}_{l=1}^{n_{\mathrm{tp},j}}$ between all \textit{true positive} boxes and the merged rectangular box $\hat{b}_{i}^{l}$ that was derived from the cluster to which they are associated. We then use these distances in conjunction with \autoref{eq:num_tp} and \autoref{eq:variance_param_estimate} to estimate
\begin{equation}\label{eq:volunteer_variance}
    \sigma^{2}_{j}\approx\frac{n_{\chi,V}\sigma_{0,V}^{2}}{n_{\mathrm{tp}, j}+ n_{\chi,V}+2}\displaystyle\sum_{l=1}^{n_{\mathrm{tp}, j}}\vert{d_{ij}^{k}}^{l}\vert^{2}.
\end{equation}

As a consequence of our prior specifications the formulations of \autoref{eq:prob_fp}, \autoref{eq:prob_fp} and \autoref{eq:volunteer_variance} can all be factored into terms that depend only on the current working batch and terms that depend only on prior information. This allows us to straightforwardly update the skill parameters of returning volunteers without having to reconsider the annotations they contributed to previous working batches. 

\subsection{Computing Maximum Likelihood Labels}

Once the associated clusters have been defined and the subject difficulties and volunteer skills have been computed we are able to compute the likelihood of each subject's estimated label using \autoref{eq:eta_def}, \autoref{eq:combined_variance} and \autoref{eq:anno_model}. Practically, we compute the log-likelihood for each subject and sum these to derive a global likelihood for all annotation data that comprise the current working batch. 

Recall (\autoref{subsec:comp_box_assoc}) that we use a simplified set of facility location costs to derive an initial clustering solution for each new working batch. These costs are used for initialisation because they can be computed without having estimated volunteer skills or subject difficulties, but they will generally not yield a set of clusters that correspond with the maximum likelihood solution of \autoref{eq:anno_model} for any subject. Similarly, the likelihood model parameters that we compute based on the initial clustering solution are unlikely to be good estimates of the subject difficulties or volunteer skills. As illustrated by the red boxes in \autoref{fig:agg_algo}, we use an iterative approach to derive the maximum likelihood solution for \autoref{eq:anno_model} and the corresponding best estimates of the likelihood model parameters.

After the initial set of volunteer skills have been computed, we recompute the box associations for all subjects using the nominal facility location costs specified in \autoref{eq:fac_cost}, \autoref{eq:city_fac_cost}, \autoref{eq:city_dummy_cost}. Using these clusters we recompute the likelihood model parameters and the corresponding subject label likelihoods. We repeat this procedure until the sum of log-likelihoods for all subjects converges to its maximum value.  

\subsection{Computing Subject Risks}\label{subsec:computing_risk}

In \autoref{subsec:overview} we introduced the concept of a ``risk'' metric $\mathcal{R}_{i}$ that can be computed for any subject $s_{i}$ and used to quantitatively determine whether the estimated label $\hat{y}_{i}$ is sufficiently representative of the \textit{unknown} true label $y_{i}$ to be scientifically useful. Specifying a risk that decreases monotonically as the reliability of $\hat{y}_{i}$ increases enables a principled decision to retire the subject $s_{i}$ when its risk falls below a predefined threshold value which we denote $\tau$.

To compute the risk for the $i$th subject, we follow the approach of \bvp and define $\mathcal{R}_{i}$ for each subject as the weighted sum of three separate terms.
\begin{equation}\label{eq:risk_expression}
    \mathcal{R}_{i} = \alpha_{\mathrm{fp}}N_{i}^{\mathrm{fp}} + \alpha_{\mathrm{fn}}N_{i}^{\mathrm{fn}} + \alpha_{\mathrm{\sigma}}N_{i}^{\mathrm{\sigma}}(\delta)
\end{equation}
The first term, $N_{i}^{\mathrm{fp}}$, represents an estimate of the number of detected clumps that are spurious, while $N_{i}^{\mathrm{fn}}$ estimates the number of genuine clumps that have not been detected.  Finally, $N_{i}^{\mathrm{\sigma}}(\delta)$ estimates the number detected clump locations that are genuine but insufficiently accurate in the sense that their Jaccard distance from the true clump location is likely to exceed a threshold value $\delta$, which we specify as a hyper-parameter. 

The weight terms $\alpha_{\mathrm{fp}}$, $\alpha_{\mathrm{fn}}$ and  $\alpha_{\mathrm{\sigma}}$ are hyper-parameters that allow the properties of the clump sample for retired subjects to be tuned for particular scientific investigations. For a specific value of $\tau$, increasing the value of $\alpha_{\mathrm{fp}}$ relative to the other weights will result in a purer clump sample, while a relative increase in $\alpha_{\mathrm{fn}}$ increases the sample completeness. Specifying a larger value for $\alpha_{\mathrm{\sigma}}$ will result in more accurate clump locations, which may be useful for studies considering the radial distribution of clumps within their host galaxies.

To estimate the expected number of genuine clumps in the estimated label for the $i$th subject, we consider each established cluster $Z_{i}^{l}\in Z_{i}$ and identify two subsets $V_{i}^{\mathrm{mark},l}, V_{i}^{\mathrm{miss},l}\in V_{i}$ of the volunteers who inspected the $i$th subject $s_{i}$. The volunteers in $V_{i}^{\mathrm{mark},l}$ are those who inspected $s_{i}$ and contributed a box to the $l$th cluster $Z_{i}^{l}$. Conversely, $V_{i}^{\mathrm{miss},l}$ contains the volunteers who inspected $s_{i}$ but missed the clump associated with $Z_{i}^{l}$. To estimate the overall probability that $Z_{i}^{l}$ represents a false positive detection, we combine the probability that all volunteers in $V_{i}^{\mathrm{miss},l}$ correctly omitted the detected clump from their annotation with the probability that all volunteers in $V_{i}^{\mathrm{match},l}$ provided a spurious annotation. 
\begin{equation}\label{eq:cluster_false_pos_prob}
p_{l}^{\mathrm{fp}} = \displaystyle\prod_{j=1}^{\left|V_{i}^{\mathrm{miss},l}\right|}(1-p_{j}^{\mathrm{fn}})\cdot\displaystyle\prod_{j=1}^{\left|V_{i}^{\mathrm{mark},l}\right|}p_{j}^{\mathrm{fp}}
\end{equation}
Similarly, to estimate the overall probability that the cluster $Z_{i}^{l}$ represents a true positive clump detection, we combine the probability that all volunteers in $V_{i}^{\mathrm{miss},l}$ missed a genuine clump with the probability that the associated boxes provided by all volunteers in $V_{i}^{\mathrm{mark},l}$ were correct.  
\begin{equation}\label{eq:cluster_true_pos_prob}
p_{l}^{\mathrm{tp}} = \displaystyle\prod_{j=1}^{\left|V_{i}^{\mathrm{miss},l}\right|}p_{j}^{\mathrm{fn}}\cdot\displaystyle\prod_{j=1}^{\left|V_{i}^{\mathrm{mark},l}\right|}(1-p_{j}^{\mathrm{fp}})
\end{equation}
Finally, we use \autoref{eq:cluster_false_pos_prob} and \autoref{eq:cluster_true_pos_prob} to estimate the number of clusters in $Z_{i}$ that are false positives by summing the expected value of an indicator variable that equals 1 when the cluster is a false positive and 0 otherwise for all clusters $Z_{i}^{l}\in Z_{i}$ (Recall that we used an analogous approach to compute the parameter $\eta$ in \autoref{subsec:clump_annotations}).
\begin{equation}
    N_{i}^{\mathrm{fp}} = \displaystyle\sum_{l=1}^{|Z_{i}|}{\frac{p_{l}^{\mathrm{fp}}}
    {p_{l}^{\mathrm{fp}}+p_{l}^{\mathrm{tp}}}}
\end{equation}

To estimate the expected number of clumps in the estimated label for the $i$th subject that are genuine, but have insufficiently accurate locations\footnote{Recall that insufficient accuracy implies  that the Jaccard distance between the estimated and true clump locations is likely to exceed the value of the hyper-parameter $\delta$}, we consider the sets of true positive boxes, supplied by the volunteers in $V_{i}^{\mathrm{mark},l}$, that were associated with each cluster $Z_{i}^{l}\in Z_{i}$. We model the Jaccard distance $d_{l}$ between estimated clump location $\hat{b}_{i}^{l}\in\hat{y}_{i}$ and the true clump location $b_{i}^{l}\in y_{i}$ as a random sample from a Gaussian distribution with zero mean and variance derived by summing the constituent box variances defined in \autoref{eq:combined_variance}.
\begin{equation}
    {\sigma_{i}^{l}}^{2} = \displaystyle\sum_{b_{ij}^{k}\in Z_{i}^{l}} {\sigma_{ij}^{k}}^{2}
\end{equation}
Using this Gaussian model, we estimate the expected number of estimated clump locations that are inaccurate by more than $\delta$ by summing the probabilities $\{p_{l}^{\mathrm{\sigma}}\}_{l=1}^{|\hat{y}_{i}|}$ that the errors in the individual clump locations exceed this threshold. 
\begin{align}
    N_{i}^{\mathrm{\sigma}} &= \displaystyle \sum_{\hat{b}_{i}^{l}\in\hat{y}_{i}} p_{l}^{\mathrm{\sigma}}\notag\\
    &=\sum_{\hat{b}_{i}^{l}\in\hat{y}_{i}}1-\mathrm{erf}\left(\frac{\delta}{\sqrt{2{\sigma_{i}^{l}}^{2}}}\right)
\end{align}

Our approach for estimating the expected number of genuine clumps that are not represented in estimated label for the $i$th subject (i.e. the number of false negatives) emulates the one used by \bvp. We begin by using the facility location algorithm to re-cluster the annotations for each subject, subject to three additional constraints that are based on the original maximum likelihood solution.
\begin{enumerate}
    \item Volunteer boxes that were originally associated with true positive clusters are \textit{not} considered as potential cities. This means that the only way that true positive annotations can contribute to clusters is by becoming facilities.
    \item \textit{Only} annotations that were \textit{not} defined as facilities originally are considered as potential facilities. This prevents rediscovery of the clumps that were indicated by the maximum likelihood solution for the subject.
    \item There is no dummy facility available, so all annotations must either become a facility or connect to an existing facility, regardless of how high the connection or establishment costs are.
\end{enumerate}
We assume that each of the assembled clusters $\{{Z^{\prime}}_{i}^{l^{\prime}}\}_{l^{\prime}=1}^{\vert {Z^{\prime}}_{i}\vert}$ comprising the constrained facility location solution ${Z^{\prime}}_{i}$ represents a potentially missed clump detection. For each new cluster we compute its assembly cost $C_{l^{\prime}}$ using \autoref{eq:cluster_cost} and compare this with the cost of connecting all the cities it contains to the dummy facility.
\begin{equation}
    C_{l^{\prime}}^{\varnothing} = \displaystyle\sum_{b_{ij}^{k}\in {Z^{\prime}}_{i}^{l^{\prime}}} C^{\varnothing \mathrm{c}}(b_{ij}^{k})
\end{equation}
To compute an initial estimate for $N_{i}^{\mathrm{fn}}$ we sum the expected value of an indicator variable that equals 1 when the cluster is a false negative and 0 otherwise for all clusters in ${Z^{\prime}}_{i}$. 
\begin{equation}\label{eq:est_num_fn}
    N_{i}^{\mathrm{fn, init}} = \displaystyle\sum_{{Z^{\prime}}_{i}^{l^{\prime}}\in {Z^{\prime}}_{i}}\frac{p_{l^{\prime}}^{\mathrm{fn}}}{p_{l^{\prime}}^{\mathrm{fn}}+p_{l^{\prime}}^{\mathrm{tn}}}
\end{equation}
where $p_{l^{\prime}}^{\mathrm{fn}}$ estimates the probability that cluster ${Z^{\prime}}_{i}^{l^{\prime}}$ identifies a real clump $b_{i}^{l^{\prime}}$ that was originally missed by the maximum likelihood solution. By analogy with \autoref{eq:total_cost}
\begin{equation}
    p_{l^{\prime}}^{\mathrm{fn}} = e^{-C_{l^{\prime}}}
    \approx \prod_{z_{ij}\in{Z^{\prime}}_{i}^{l^{\prime}}} p(z_{ij}|y_{i}\ni b_{i}^{l^{\prime}},\mathcal{D}_{i},\mathcal{S}_{j})
\end{equation} 
Furthermore, $p_{l^{\prime}}^{\mathrm{tn}}$ is the probability that the boxes in ${Z^{\prime}}_{i}^{l^{\prime}}$ all correspond with false positive clicks.
\begin{equation}
    p_{l^{\prime}}^{\mathrm{tn}} = e^{-C_{l^{\prime}}^{\varnothing}}
    = \prod_{z_{ij}\in{Z^{\prime}}_{i}^{l^{\prime}}} p_{j}^{\mathrm{fp}}
\end{equation} 

\begin{figure*}
    \centering
    \includegraphics[width=\textwidth]{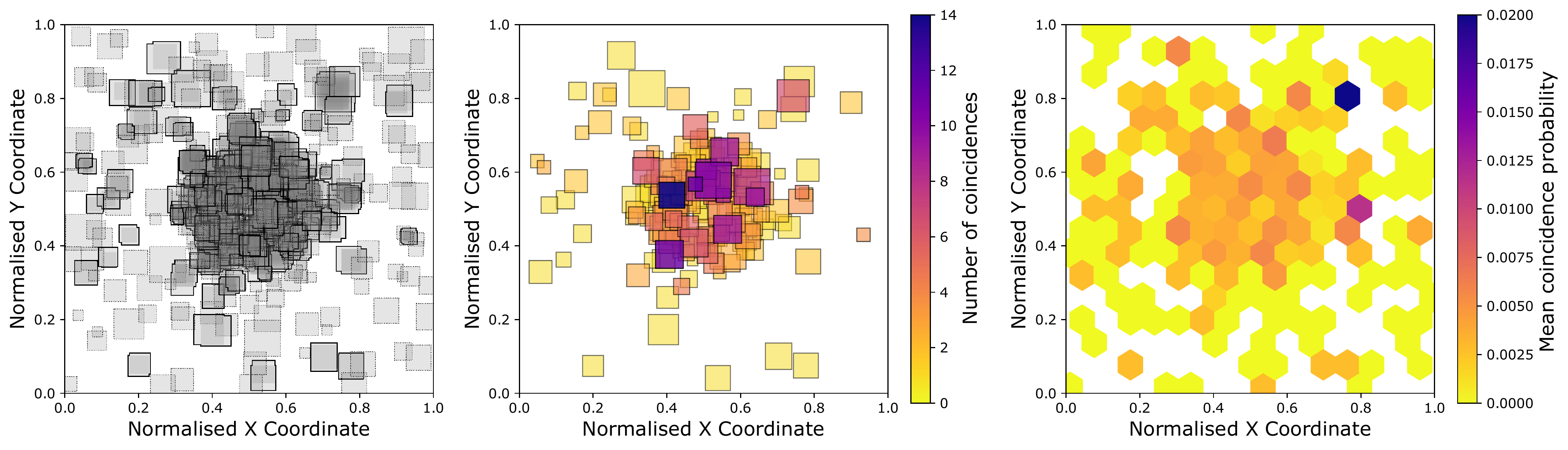}
    \caption{Computing the random coincidence probability using all boxes in the working batch. \textit{Left panel:} Shaded boxes represent all elements in the first working batch. Solid boundaries indicate groups of boxes that coincided using the $d_{\max}=0.9$ criterion. Note that large boxes may validly encompass all or most of smaller ones without coinciding if the ratio of the box areas areas in normalised coordinates less than $0.9 d_{\max}$. Boxes that did not coincide with any others are shown using dashed lines. \textit{Middle panel:} The elements of $B^{\cap}$ coloured according to the number of boxes they were found to coincide with. \textit{Right panel}: Two dimensional map showing the mean probability that one or more boxes in the working batch will accidentally coincide at a given two-dimensional location.}
    \label{fig:fn_map}
\end{figure*}

This initial estimate cannot be computed for a subject if no volunteer boxes were originally connected to the dummy facility. However the absence of nominally false positive boxes does not imply that no clumps have been missed. To estimate how many clumps might have been missed when no false positives are present we consider intersections between the global set of all annotations provided by all volunteers for all subjects in the working batch. 
We use this global set to estimate a subject-agnostic probability that two boxes coincide at any particular location within a subject. The higher this probability is for a particular location, the more likely it is that a clump will be located there. We define a coincidence when two boxes are separated by a Jaccard distance less than $d_{\max}$.\footnote{As described in \autoref{subsec:initial_processing}, volunteer boxes have a side-length equal to twice the FWHM of subject's PSF, and may have different absolute pixel dimensions. When computing  $N_{i}^{\mathrm{fn}}$ we account for this by using normalised image coordinates $\{x^{\prime}, y^{\prime}\} \equiv \{x/x_{\max}, y/y_{\max}\}$ to define box boundaries when we compute the Jaccard distance between boxes in the global set.} We begin by randomly shuffling the elements of the working batch. We then process the randomised elements sequentially to find any mutually coinciding subsets of volunteer boxes. For each box, we check for coincidence with any of the previously processed elements. If the boxes coincide we increment a coincidence count, which we denote ${n^{\cap}}_{ij}^{k}$, for the previously processed element and remove the current element from the shuffled batch. If no coincidences are found we retain the current element, which allows coincidences between it and subsequent elements to be identified and counted. After the shuffled working batch has been processed, we estimate the probability of a coincidence with each of the remaining elements by computing the ratio between its accumulated coincidence count and the total number of annotations $n_{z}$ comprising to the working batch.\footnote{Recall (\autoref{subsec:agg_annots}) that we define an annotation $z_{ij}=\{b_{ij}^{k}\}_{k=1}^{|B_{ij}|}$ to be the set of box markings provided by a particular volunteer when they inspect a particular subject, so the number of annotations is generally less than the size of the working batch.}
\begin{equation}
    {p^{\cap}}_{ij}^{k}=\frac{{n^{\cap}}_{ij}^{k}}{n_{z}}
\end{equation}
\autoref{fig:fn_map} illustrates the different stages in our computation of the ${p^{\cap}}_{ij}^{k}$ using all boxes in the working batch.  

Let $B^{\cap}$ represent the remaining elements of the shuffled working batch, for which a value of ${p^{\cap}}_{ij}^{k}$ has been computed. For each subject in the original working batch, we find the subset $B^{\star}$ of elements in $B^{\cap}$ that constitute that do not coincide with any of the boxes $\hat{b}_{i}^{l}\in \hat{y}_{i}$ that comprise the estimated subject label. For each box in $B^{\star}$, we increment $N_{i}^{\mathrm{fn, init}}$ by the product of the probability that a coincidence occurs at that location and the probability that all volunteers who inspected the image would have missed the clump. 
\begin{equation}\label{eq:exp_num_fn}
    N_{i}^{\mathrm{fn}} = N_{i}^{\mathrm{fn, init}} + \displaystyle\sum_{b_{ij}^{k}\in B^{\star}}{p^{\cap}}_{ij}^{k}\cdot e^{-C^{\mathrm{f}}(b_{ij}^{k})}
\end{equation}
Note that for subjects that have no clumps identified in their estimated labels, $B^{\star}\rightarrow B^{\cap}$. In practice, we find that $N_{i}^{\mathrm{fn, init}}$ always dominates the estimate of $N_{i}^{\mathrm{fn}}$ and that the second term in \autoref{eq:exp_num_fn} is always $\ll 1$.

\subsection{Subject retirement and batch finalisation}\label{subsec:retirement_and_finalisation}

Computing the expected false positive, false negative and inaccurate true positive counts (i.e. $N_{i}^{\mathrm{fp}}$, $N_{i}^{\mathrm{fn}}$ and $N_{i}^{\sigma}(\delta)$) independently for each subject allows us to define a compound retirement criterion that specifies maximum permissible values , $N_{i,\max}^{\mathrm{fp}}$, $N_{i,\max}^{\mathrm{fn}}$ and $N_{i,\max}^{\sigma}$, for each of these quantities as well as a threshold $\tau$ on the overall subject risk. \autoref{tab:subject_retirement_params} lists the thresholds we use in practise as well as the values we adopt for the coefficients specified in \autoref{eq:risk_expression}. 

\begin{table}
	\centering
	\caption{Parameters used to determine subject retirement and compute overall subject risk}
	\label{tab:subject_retirement_params}
	\begin{tabular}{lc}
		\hline
		Parameter & Value \\
		\hline
		$\alpha_{\mathrm{fp}}$ & 1\\
		$\alpha_{\mathrm{fn}}$ & 1\\
		$\alpha_{\sigma}$ & 2\\
		$\delta$ & 0.5\\
		$N_{i,\max}^{\mathrm{fp}}$ & 1\\ $N_{i,\max}^{\mathrm{fn}}$ & 0.3\\ $N_{i, \max}^{\sigma}$ & 3\\
		$\tau$ & 5\\
		\hline
	\end{tabular}
\end{table}

Once the subject risks have been computed we retire those subjects for which the overall risk $\mathcal{R}_{i}<\tau$ and $N_{i}^{\mathrm{fp}}$, $N_{i}^{\mathrm{fn}}$ and $N_{i}^{\sigma}$ are \textit{all} less than their specified maximum permissible values, before removing their elements from the working batch. We also identify and remove any stale subject data that have persisted for the maximum allowed number of batch replenishment cycles without retiring. Such subjects are likely very difficult or complicated so we mark them for expert inspection, assessment and labelling. For the remaining subjects that were not retired, we re-initialise their difficulty parameters and discard any associated clusters that were established when the working batch was processed. 

Annotation data that were provided by a single volunteer for different subjects can appear in separate working batches, especially if volunteers return to the project regularly over an extended period of time. It is also possible that only a subset of the subjects annotated by a volunteer in a single working batch are retired when batch processing completes. If a volunteer's annotation data persist between batches, those persistent data should not be used to update volunteer skills multiple times during multiple batch processing cycles. This could lead to pathological subjects unfairly inflating or reducing the skill parameter values ($p_{j}^{\mathrm{fp}}$, $p_{j}^{\mathrm{fn}}$, $\sigma^{2}_{j}$) for a particular volunteer. To avoid this scenario, we restore the volunteer skills that were cached at the start of the latest cycle and update them using only annotation for subjects that did retire.

The batch processing cycle then restarts by acquiring new annotation data and repopulating the working batch. 

\section{Results}\label{sec:results}

\begin{figure*}
    \centering
    \includegraphics[width=\textwidth]{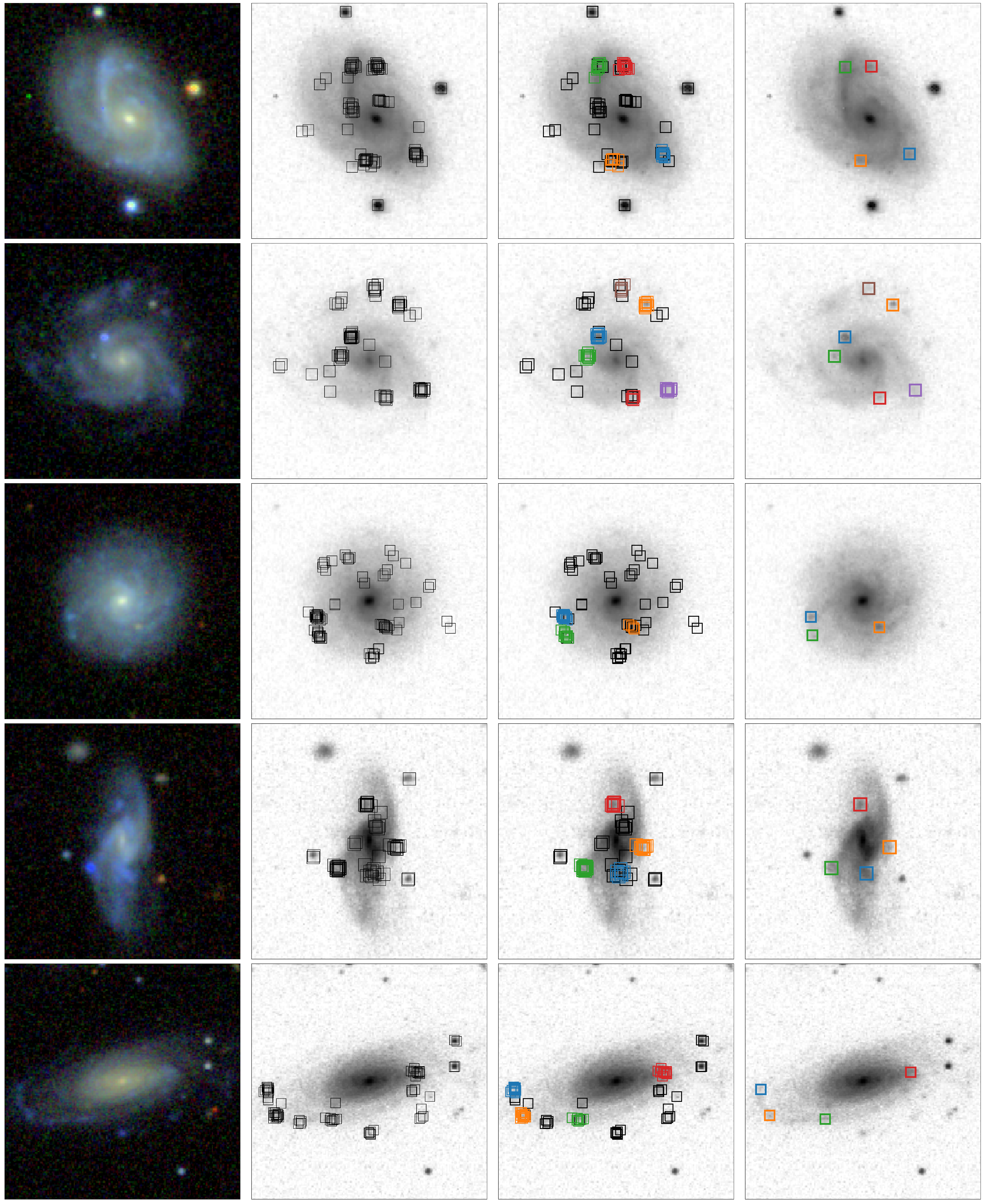}
    \caption{Examples of clump-hosting galaxies, illustrating the ability of our framework to exclude false-positive annotations. The \textit{left hand column} shows galaxy images as they were seen by volunteers. The \textit{second column} overlays all volunteer annotations on a grey-scale image of the same galaxy. In the \textit{third column} volunteer annotations that were assigned to a facility and identified as clumps are shown in colour. Annotations that were assigned to the dummy facility are shown in black. The \textit{fourth column} shows the clump locations that we ultimately identify.}
    \label{fig:agg_examples}
\end{figure*}

Recall that the full \gzcs dataset ($Z$) contains 3,561,454 click locations, which constitute 1,739,259 annotations of 85,286 distinct subjects provided by 20,999 volunteers and that approximately 20 volunteers inspected each subject. Using this dataset, we identify 128,100 potential clumps distributed among 44,126 galaxies. \autoref{fig:agg_examples} shows five examples of galaxies in which clumps were detected.

\subsection{Testing the effect of volunteer multiplicity}\label{subsec:vol_multiplicity_effect}
We expect that the performance of our aggregation framework will vary depending upon the number of volunteers who inspect each subject. To investigate this dependence we assemble 17 subsamples of annotations $\{\tilde{Z}_{n}\}_{n=3}^{20}\in Z$, that contain between 3 and 20 annotations per galaxy. Each $\tilde{Z}_{n}$ is constructed by randomly sampling $n$ annotations for each subject $s_{i}\in S$. For example, $\tilde{Z}_{5}$ includes 5 randomly sampled annotations for each galaxy in the \gzcs subject set. We then use our aggregation framework to derive the set of corresponding estimated subject labels $\hat{Y}(\tilde{Z}_{n})\equiv\{\hat{y}_{i,n} \}_{i=1}^{\vert S\vert}$ where $\hat{y}_{i,n} = \hat{y}_{i}(Z=\tilde{Z}_{n})$ is the label for $s_{i}$ based only on the $n$ annotations for that subject within $\tilde{Z}_{n}$ \footnote{Note that the labels for each subject may in principle depend on \textit{all} annotations in $\tilde{Z}_{n}$ via those annotations' influence on the volunteers' skill parameters.}. In subsequent sections, we will examine the differences between results derived using these different restricted datasets. Note that the dataset containing 20 annotations per subject, denoted $\tilde{Z}_{20}$, is not quite the full \gzcs dataset $Z$ because the Zooniverse interface occasionally collects more than 20 annotations per subject.

\subsection{Aggregated clump properties}

\begin{figure*}
    \centering
    \includegraphics[width=\textwidth]{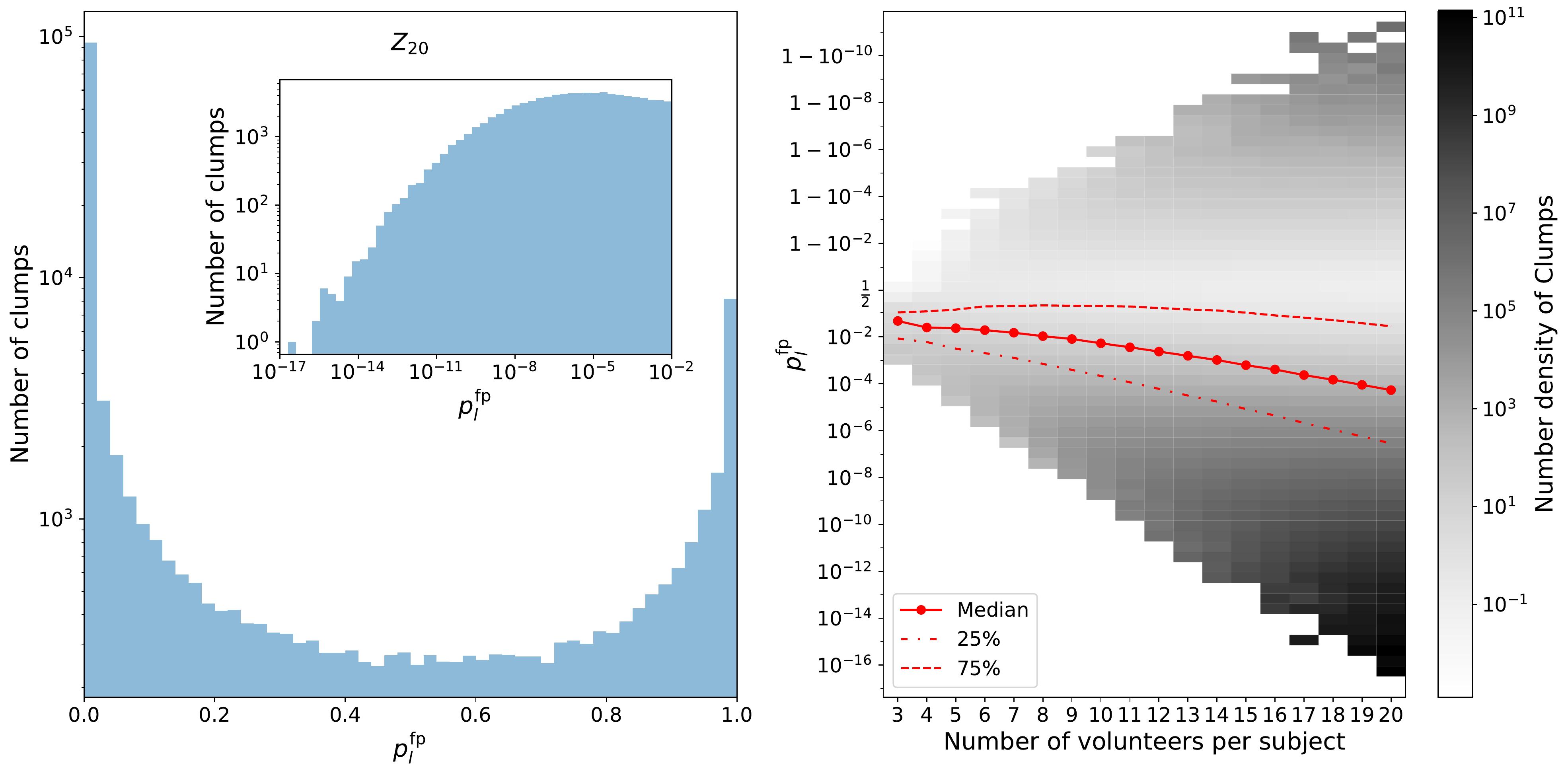}
    \caption{\textit{Left panel:} Distribution of estimated false positive probability $p^{\mathrm{fp}}_{l}$ for clumps identified using 20 annotations per subject (i.e. using $\tilde{Z}_{20}$). The distribution is strongly bimodal with $\approx90\%$ of clumps having $0.2<p^{\mathrm{fp}}_{l}>0.8$. The inset shows the distribution in for $p^{\mathrm{fp}}_{l}<0.01$. \textit{Right panel:} Distributions of $p^{\mathrm{fp}}_{l}$ corresponding to $n$ between 3 and 20 volunteer annotations per subject. The distribution medians decrease monotonically from $\approx0.04$ for $n=3$ to $\approx5\times10^{-5}$ for $n=20$, while the distribution interquartile ranges become wider as more volunteers annotate each subject. We use a ``logistic'' scaling for the \textit{y} axis to highlight the development of the bimodal structure for large $n$. Note that the colour scale shows the number \textit{density} of clumps to account for the fact that the two-dimensional histogram bins cover different areas.}
    \label{fig:clump_fpp_dist}
\end{figure*}

Our aggregation algorithm assigns a separate false positive probability $p^{\mathrm{fp}}_{l}$ to each clump it identifies (see \autoref{subsec:computing_risk}). The  left-hand panel of \autoref{fig:clump_fpp_dist} shows the distribution of this false positive probability for clumps detected using 20 annotations per subject, which is strongly bimodal with $\approx90\%$ of clumps having $0.2<p^{\mathrm{fp}}_{l}>0.8$. The right hand panel shows how the distribution of the false positive probabilities for all identified clumps evolves as more volunteers annotate each subject. For fewer than 5 annotations per subject (i.e. $n\lesssim5$) the estimates for the clumps' false positive probabilities remain somewhat prior-dominated and the distributions are unimodal with medians close to the hyper-parameter value $p_{0}^{\mathrm{fp}}=0.1$. For more than 5 annotations per subject (i.e. $n>5$), the distributions become progressively more bimodal which increases their interquartile ranges. The distribution medians decrease monotonically as the number of annotations per subject $n\rightarrow20$, which indicates that providing more volunteer annotations per subject allows our framework to more confidently predict the \textit{presence} of clumps.

\begin{figure*}
    \centering
    \includegraphics[width=\textwidth]{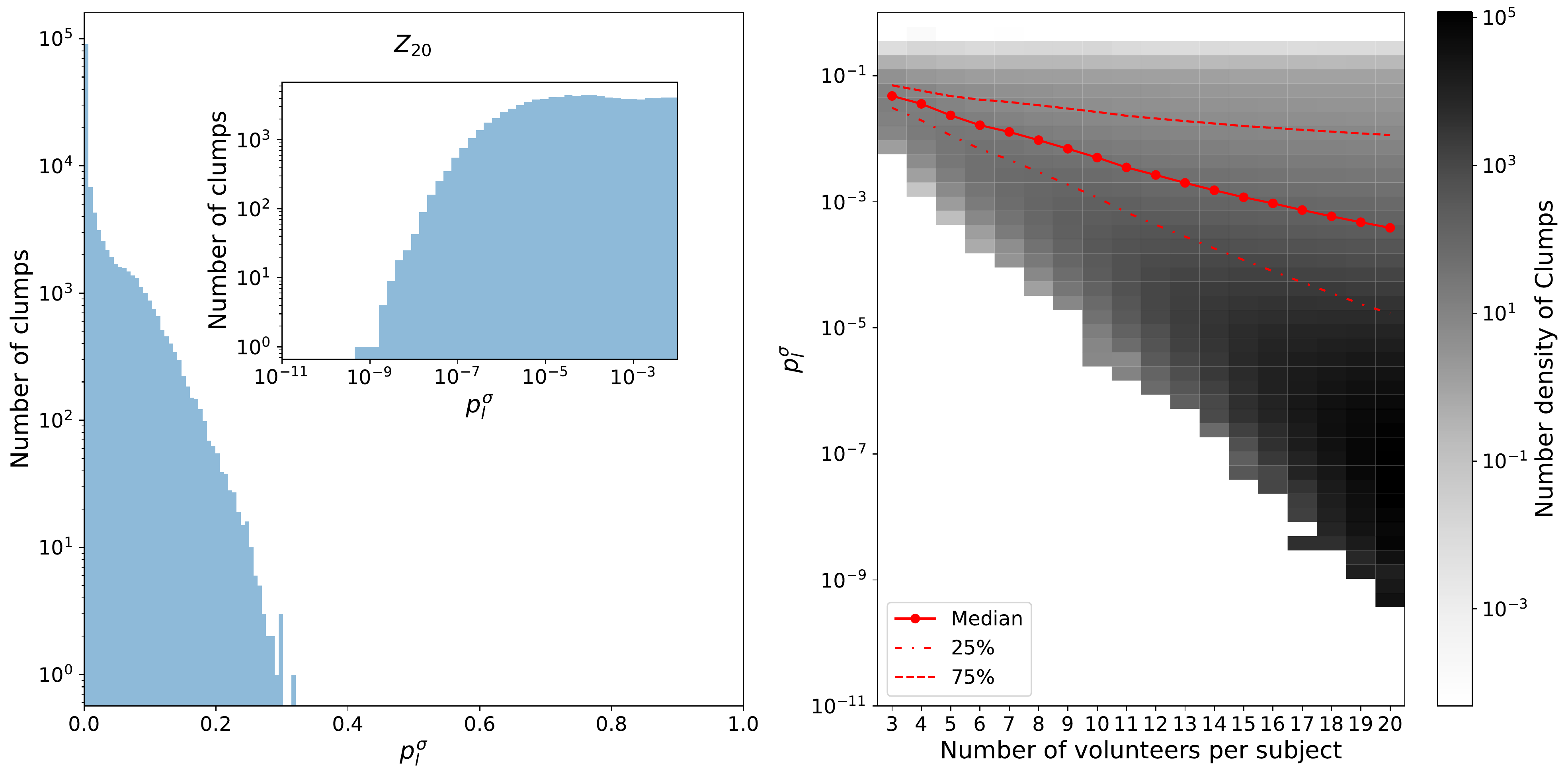}
    \caption{\textit{Left panel:} Distribution of the estimated probability that an individual clump location is inaccurate ($p^{\sigma}_{l}$)  for clumps identified using 20 annotations per subject (i.e. using $\tilde{Z}_{20}$). The distribution is concentrated close to zero with all clumps having $p^{\sigma}_{l}\lesssim 0.3$. The inset shows the distribution in for $p^{\sigma}_{l}<0.01$. \textit{Right panel:} Distributions of $p^{\sigma}_{l}$ corresponding to $n$ between 3 and 20 volunteer annotations per subject. The distribution medians decrease monotonically from $\approx0.05$ for $n=3$ to $\approx4\times10^{-4}$ for $n=20$, while the distribution interquartile ranges become wider as more volunteers annotate each subject. Note that the colour scale shows the number \textit{density} of clumps to account for the fact that the two-dimensional histogram bins cover different areas.}
    \label{fig:clump_variance_dist}
\end{figure*}

For every bounding box in each subject's maximum likelihood label, we also compute the probability $p^{\sigma}_{l}$ that the Jaccard distance between it and the unknown true location of the clump exceeds $\delta = 0.5$. The left hand panel of \autoref{fig:clump_variance_dist} shows the distribution of $p^{\sigma}_{l}$ for clumps detected using 20 annotations per subject, while the right hand panel shows how the distribution $p^{\sigma}_{l}$ of  evolves as more volunteers annotate each subject. Again, our model priors appear to dominate for fewer than 5 annotations per subject  and the distribution medians decrease monotonically as the number of annotations per subject $n\rightarrow20$. This pattern indicates that providing more volunteer annotations per subject allows our framework to more precisely determine the \textit{locations} of clumps. 

\begin{figure*}
    \centering
    \includegraphics[width=\textwidth]{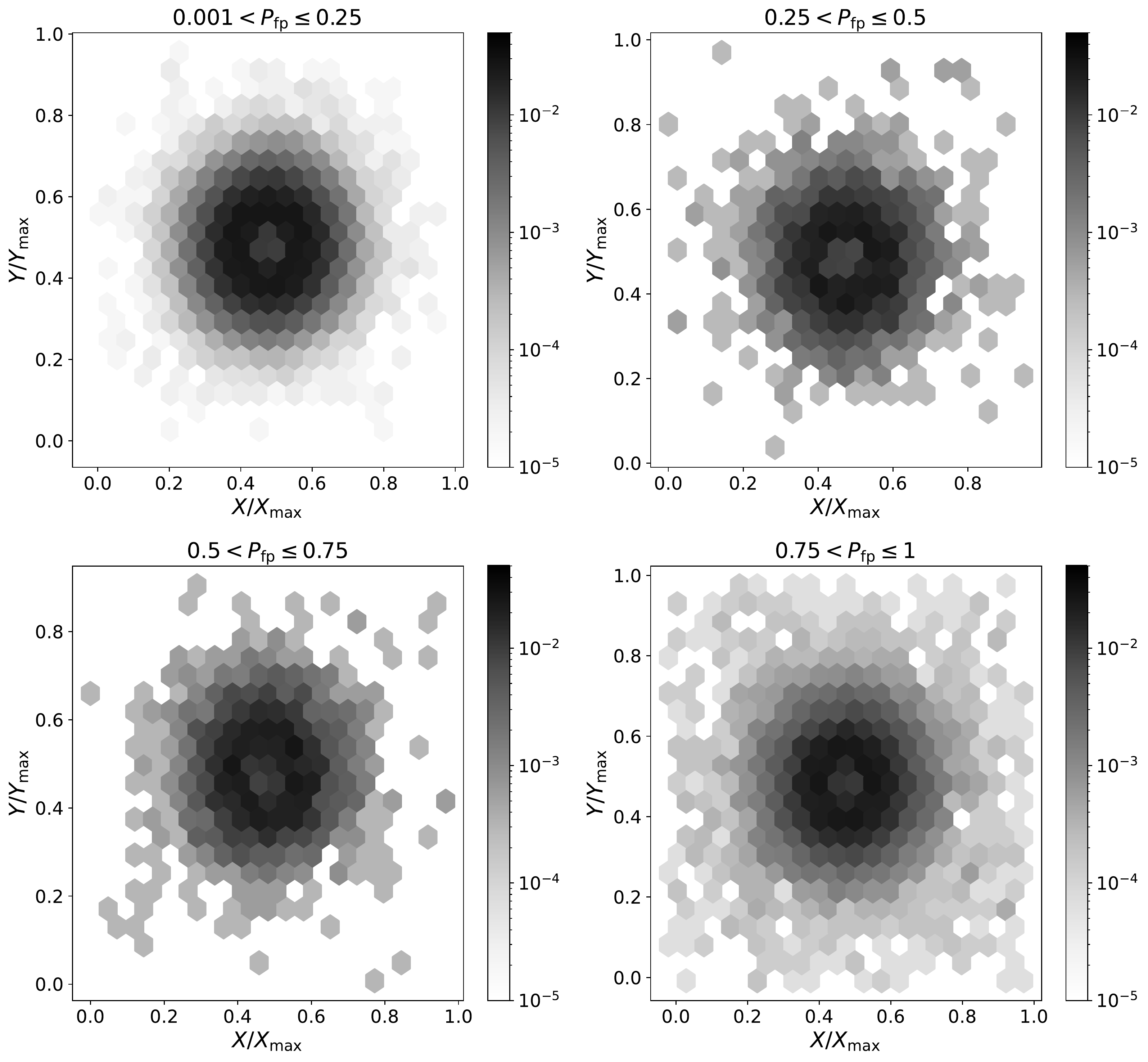}
    \caption{Detected clump locations in normalised image coordinates, in bins of estimated clump false positive probability, $p_{\mathrm{fp}}$. For $p^{\mathrm{fp}}\lesssim 0.5$, 99.9\% of detected clumps have $R_{\mathrm{clump}}=\sqrt{(X_{\mathrm{clump}}/X_{\max})^{2}+(Y_{\mathrm{clump}}/Y_{\max})^{2}}<0.25$. In contrast 10 times more clumps ($\sim1$\%) with $p^{\mathrm{fp}}_{l}\gtrsim 0.5$ have $R_{\mathrm{clump}} > 0.25$.}
    \label{fig:clump_location_vs_pfp}
\end{figure*}

\autoref{fig:clump_location_vs_pfp} illustrates the spatial distribution of the detected clump locations, in bins of estimated clump false positive probability $p^{\mathrm{fp}}_{l}$. We observe that 99.9\% of clumps with $p^{\mathrm{fp}}_{l}\lesssim 0.5$ (i.e. likely true positives) are located within a central circular region occupying 20\% of the area of their corresponding images. In contrast, clumps with $p^{\mathrm{fp}}_{l}\gtrsim 0.5$ (i.e. likely false positives) are 10 times more likely to fall outside this region. This central concentration of confidently identified clumps is reassuring because it reflects the typical footprints of the target galaxies in each subject image, which is where we would reasonably expect to find genuine clumps. For all clumps, regardless of their estimated false positive probability, we observe a clear under-density at the centre of the distribution, which likely reflects the fact that most volunteers correctly distinguish the target galaxies' central bulges from clumps.

\subsection{Comparison with expert annotations}\label{subsec:expert_comparison}

To quantify the degree of correspondence between the clumps identified by volunteers and those identified by professional astronomers, we used the \gzcs interface to collect annotations from three expert astronomers for 1000 randomly selected subjects and compared the recovered clump locations with those derived from volunteer clicks by our aggregation framework. 

For each subject in this expert-annotated image set, we consider the 17 different estimated labels $\hat{y}_{i,n}$ that were computed using $3\leq n\leq20$ volunteer annotations per subject (see \autoref{subsec:vol_multiplicity_effect}). We then filter each of these 17 labels by selecting a subsample of its bounding boxes that have associated false positive probabilities $p^{\mathrm{fp}}_{l}$ that are less than a selectable threshold value, which we denote $p^{\star,\mathrm{fp}}$. By setting $p^{\star,\mathrm{fp}}$ close to one, we expect to select only the bounding boxes that mark real clumps. Conversely, we expect that setting $p^{\star,\mathrm{fp}}$ close to zero results in a subsample that is likely to contain more false positive bounding boxes. We use the symbol $\hat{Y}^{\star}_{n}(p^{\star,\mathrm{fp}})$ to denote the set of estimated labels for all expert-annotated subjects that were computed using $n$ volunteer annotations per subject and filtered to include only those bounding boxes with false positive probabilities less than $p^{\star,\mathrm{fp}}$. 

For a \textit{particular} false positive filtering threshold $p^{\star,\mathrm{fp}}$ and number of annotations per subject $n$, we consider the filtered labels for all 1000 expert-annotated subjects and define $N_{n}^{\mathrm{FP}}$ to be the total number of \textit{empirically} false positive aggregated clump bounding boxes in $\hat{Y}^{\star}_{n}(p^{\star,\mathrm{fp}})$ that contain zero expert click locations. Conversely, $N_{n}^{\mathrm{FN}}$ denotes the total number of expert clicks located outside of any aggregated box, which we designate as false negatives. We identify the remaining $N_{n}^{\mathrm{TP}}$ aggregated boxes that coincided with an expert click location as true positives. 

Using the set of aggregated clump designations we compute the aggregated $p^{\mathrm{fp}}$-threshold-dependent clump sample completeness  
\begin{equation}
    \mathcal{C}_{n}(p^{\star,\mathrm{fp}})=\frac{N_{n}^{\mathrm{TP}}}{N_{n}^{\mathrm{TP}} + N_{n}^{\mathrm{FN}}}
\end{equation}
and purity
\begin{equation}
    \mathcal{P}_{n}(p^{\star,\mathrm{fp}})=\frac{N_{n}^{\mathrm{TP}}}{N_{n}^{\mathrm{TP}} + N_{n}^{\mathrm{FP}}}
\end{equation}

\begin{figure*}
    \centering
    \includegraphics[width=0.48\textwidth]{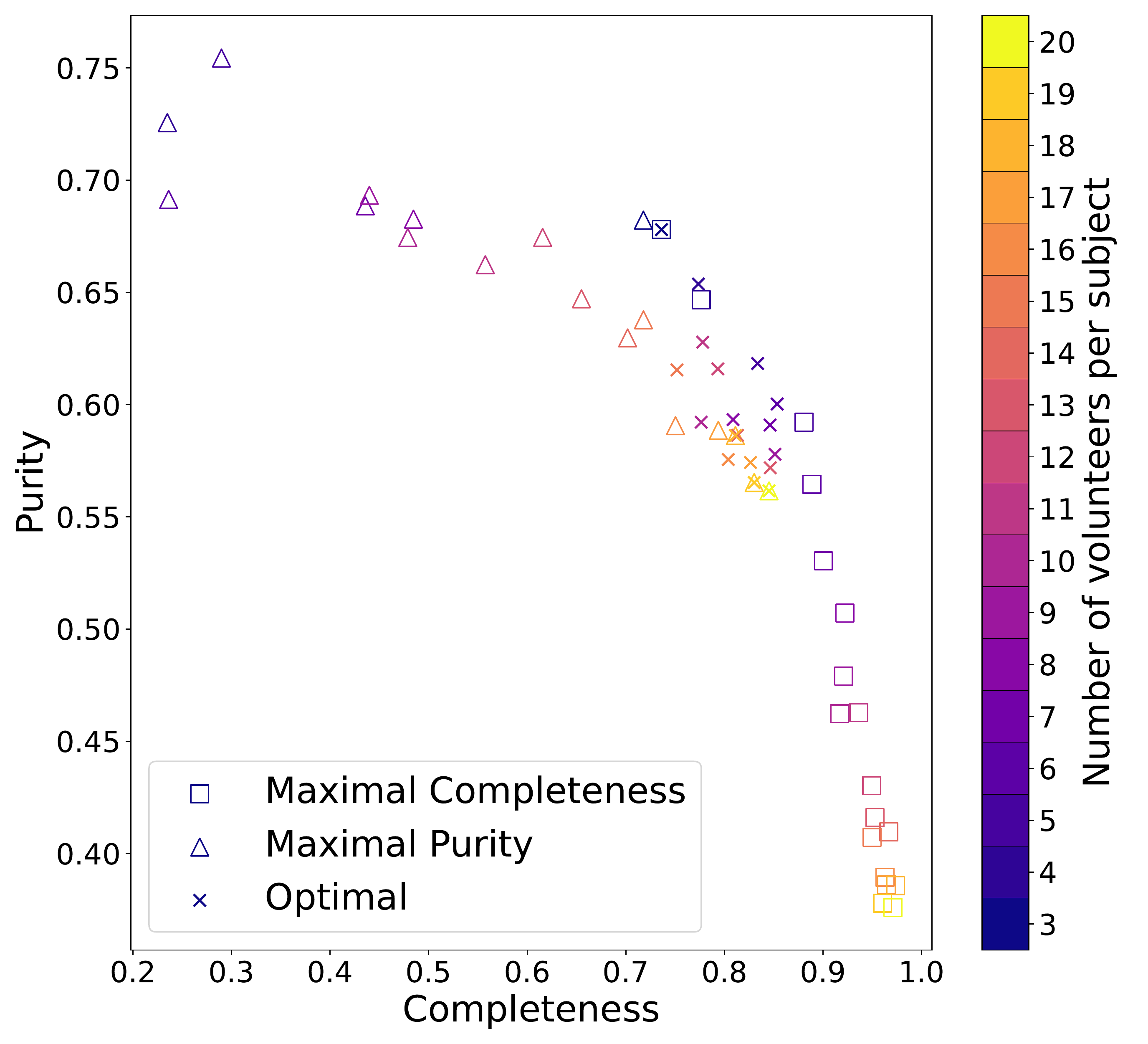}
    \includegraphics[width=0.48\textwidth]{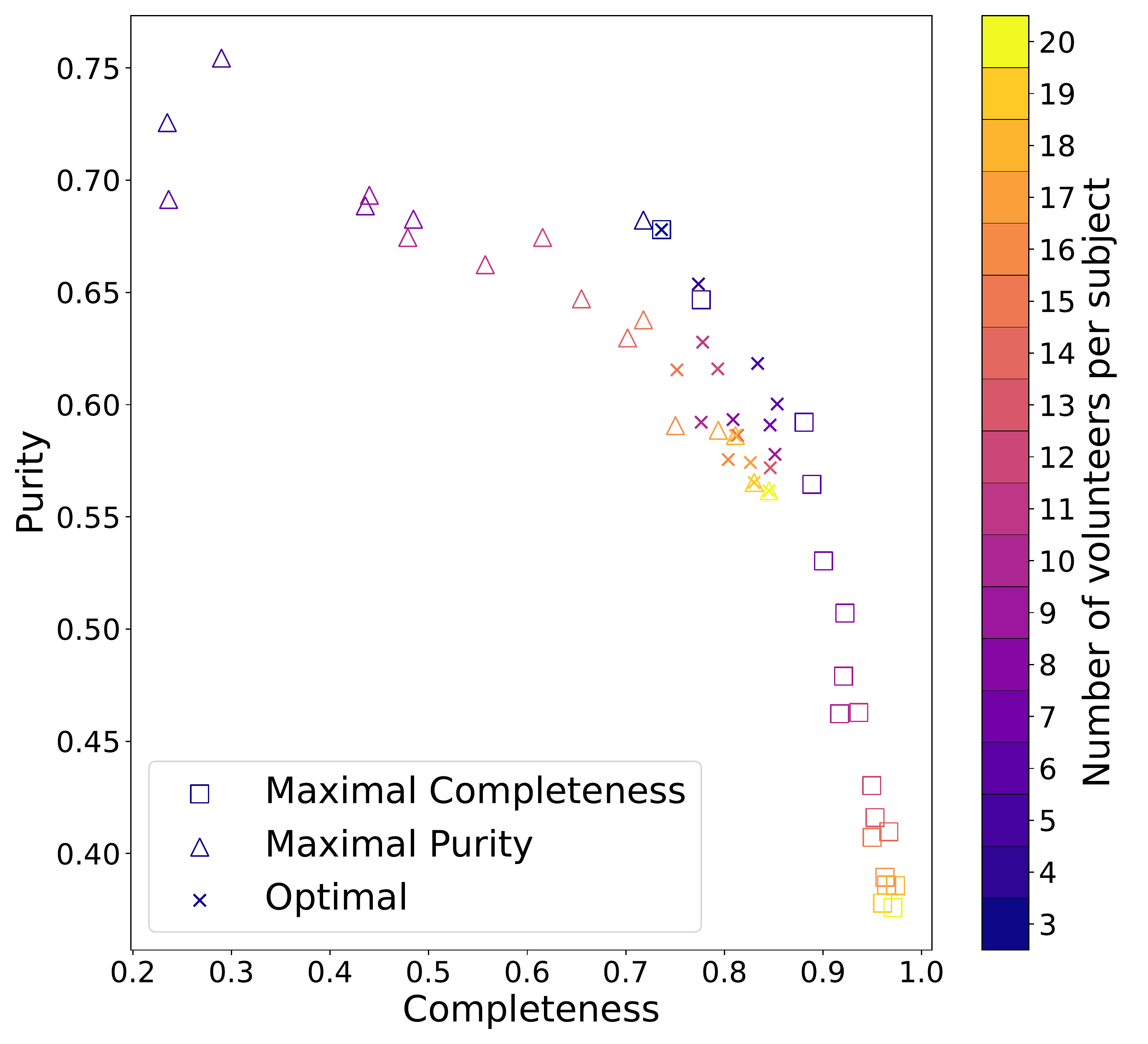}
    \caption{Purity versus completeness for different numbers of volunteers per subject. The \textit{left-hand} panel shows values derived using the full volunteer label sets  and \textit{all} expert-identified clumps as a benchmark sample, while the values shown in the \textit{right-hand} panel are derived by comparing the sets of clumps which experts and volunteers identified as ``normal''. Squares indicate the values of the maximum possible completeness and purity for each number of volunteers, which can generally not be realised simultaneously. Crosses indicate the optimal completeness and purity values  that can be simultaneously realised for each volunteer count.}
    \label{fig:pseudo_roc_volunteer_counts}
\end{figure*}

\autoref{fig:pseudo_roc_volunteer_counts} illustrates how the completeness and purity of our aggregated clump sample depend on $n$. In the \textit{left-hand} panel we plot $\mathcal{C}_{n}$ and $\mathcal{P}_{n}$ values derived using the whole expert-identified clump sample as a ground-truth set. The values plotted in the \textit{right-hand} panel are derived by comparing a restricted set of nominally normal ground-truth clumps which experts did not identify as ``unusual'' (see \autoref{subsec:annot_wf}) with aggregated clumps that the majority of volunteers who identified the clump classified it as being normal in appearance. In both panels, the crosses show the ``optimal'' completeness and purity values that maximise the hypotenuse $\sqrt{\mathcal{C}(p^{\star,\mathrm{fp}})^{2} + \mathcal{P}(p^{\star,\mathrm{fp}})^{2}}$ over all possible $p^{\star,\mathrm{fp}}$ thresholds. For comparison, the square and triangular points in \autoref{fig:pseudo_roc_volunteer_counts} respectively illustrate the maximum values of completeness and purity that can be achieved independently. 

For both the full and the restricted ground truth sets, we observe a general trend that increasing the number of volunteers who inspect each subject increases the optimal sample completeness at the expense of reducing purity. Using the expert classifications as a benchmark it is clear that our most complete aggregated clump samples suffer substantial contamination. In the most extreme case, using the ``normal'' clump comparison sets for $n=20$ and letting  $p^{\star,\mathrm{fp}}=1$ yields $\sim97\%$ completeness, but only $\sim35\%$ purity. The high level of contamination indicates that volunteers are much more optimistic than experts when annotating clumps i.e. volunteers will mark features that experts will ignore. Moreover, while completeness values generally improve when comparing the restricted ``normal'' clumps, the corresponding purity values are substantially worse than those derived from the full clump samples. This degradation in purity for the ``normal'' clump subset likely indicates that volunteers and experts disagree about the definition of a ``normal'' clump with volunteers being less likely to label a clump as unusual.

\begin{figure*}
    \centering
    \includegraphics[width=\textwidth]{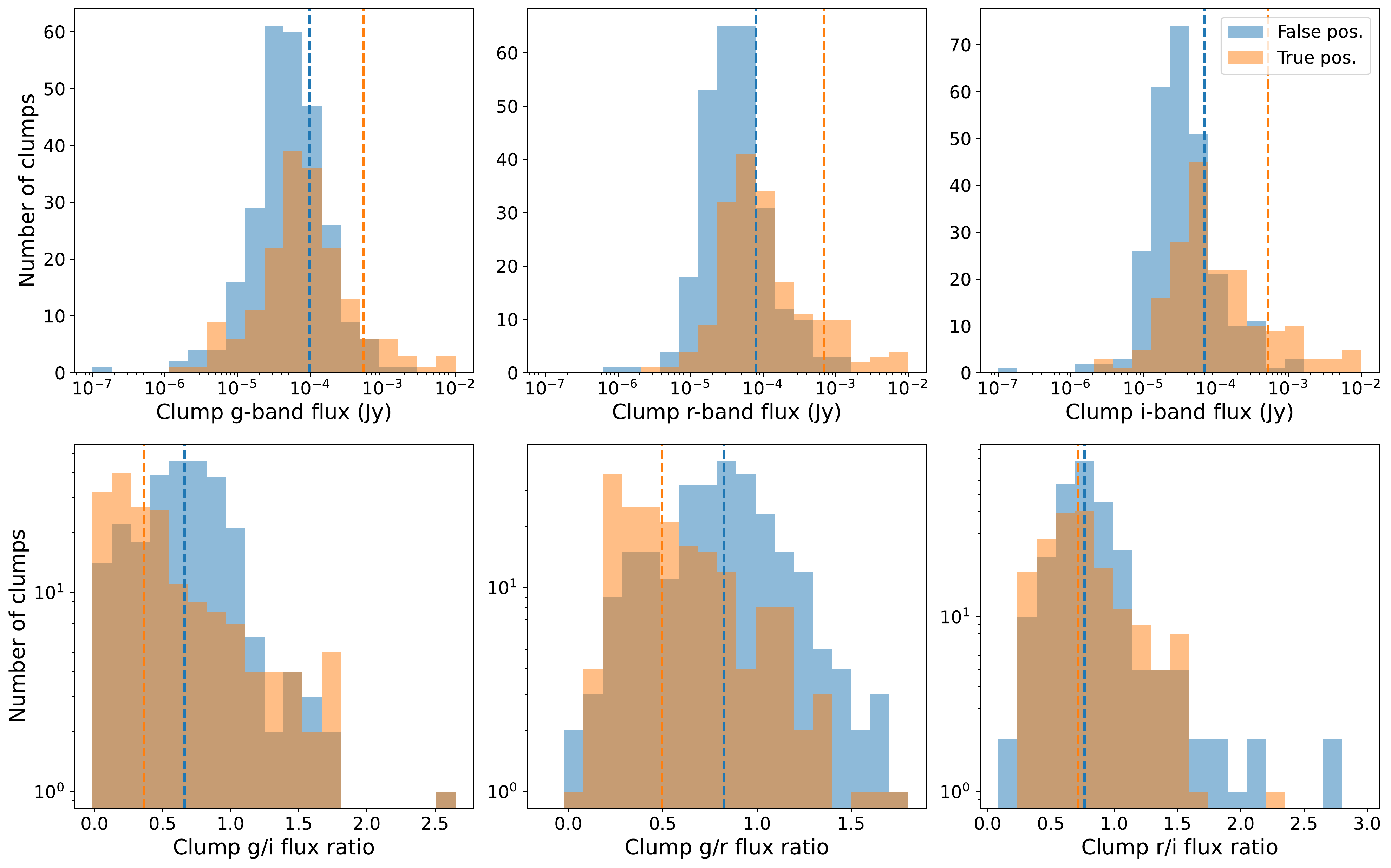}
    \caption{\textit{Top row:} Flux distributions in \textit{g}, \textit{r} and \textit{i} bands for clumps that are empirically determined to be false positive or true positive by comparing with expert clump annotations. Dashed vertical lines indicate the distribution means. \textit{Bottom row:} Flux ratio distributions for clumps that are empirically determined to be false positive or true positive by comparing with expert clump annotations. Dashed vertical lines indicate the distribution medians. In both rows, the fluxes in each band are scaled in the same way as the corresponding bands of the subject images (see \autoref{subsec:data:preparation}) to better reflect the data that volunteers actually see.}
    \label{fig:expert_all_set_hitmiss_band_flux_dists_lscale}
\end{figure*}

The top row of \autoref{fig:expert_all_set_hitmiss_band_flux_dists_lscale} shows the \textit{g}, \textit{r} and \textit{i} band flux distributions\footnote{See $\S$3 in \cite{\nicospaper} for a detailed explanation of how clump fluxes are computed.} for aggregated clumps that are \textit{empirically} determined to be false positive and true positive when comparing them with expert clump annotations. To better represent the appearance of the clumps that volunteers and experts actually see, the band-limited fluxes shown in \autoref{fig:expert_all_set_hitmiss_band_flux_dists_lscale} are independently scaled in the same way as the corresponding bands of the \gzcs subject images (see \autoref{subsec:data:preparation}). The distributions reveal that empirically false-positive clumps are $\sim5-10$ times fainter on average than empirically true positive clumps. The bottom row of \autoref{fig:expert_all_set_hitmiss_band_flux_dists_lscale} shows all non-redundant flux ratios for the \textit{g}, \textit{r} and \textit{i} bands. In general the empirically false positive clumps are brighter in the \textit{g} band and would appear bluer in the subject images. Overall, the distributions in \autoref{fig:expert_all_set_hitmiss_band_flux_dists_lscale} suggest that volunteers are more likely to mark faint features than experts, particularly when those features appear blue. \autoref{fig:empirical_fp_example_images} shows typical examples of the faint blue features that volunteers annotate but experts ignore.

\begin{figure}
    \centering
    \includegraphics[width=0.45\textwidth]{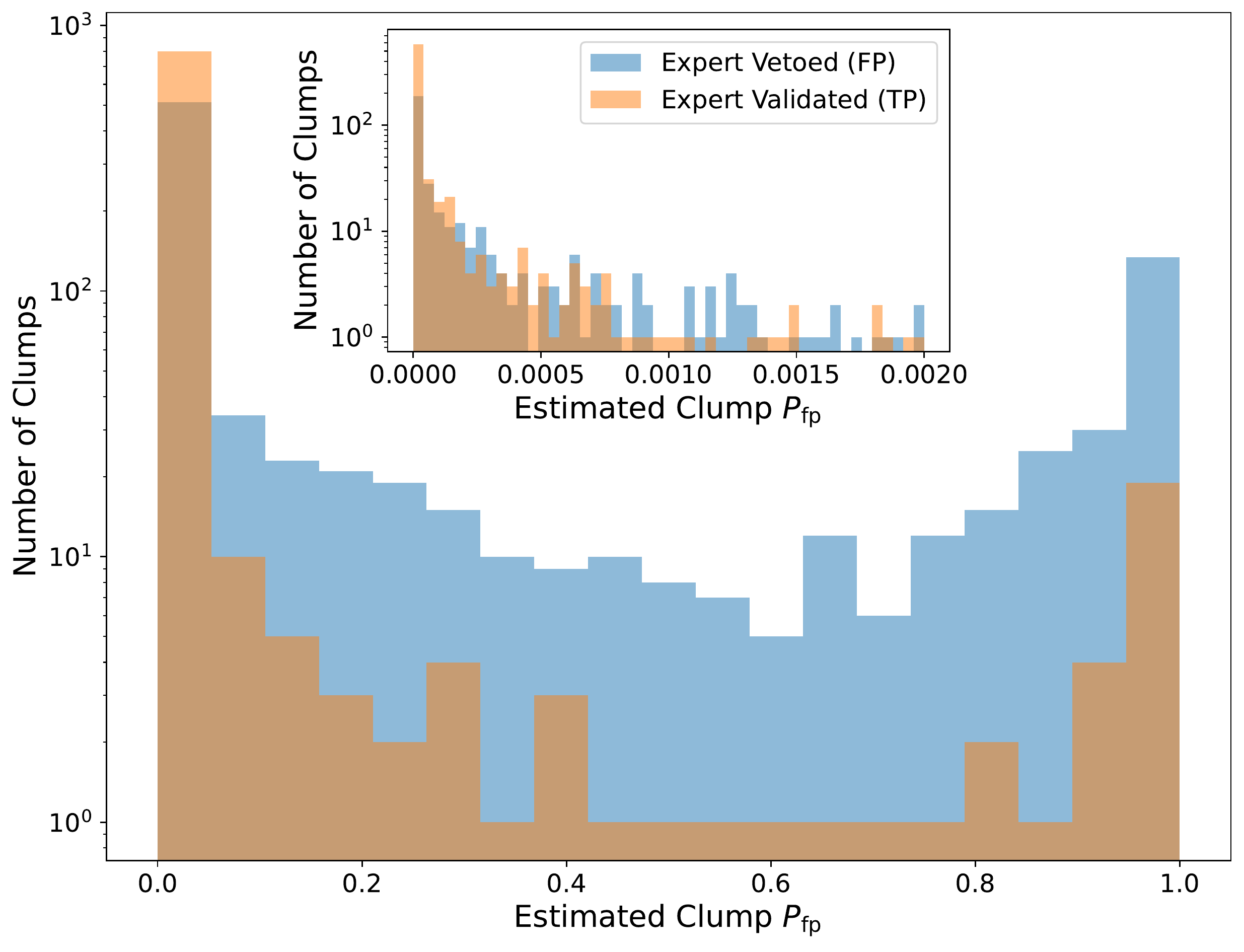}
    \caption{Distribution of estimated clump false positive probability ($p^{\mathrm{fp}}_{l}$) values for aggregated clump locations that coincide with expert annotations (orange) and those that did not (blue). We use coincidence with any expert clump to establish the true-positive or false-positive categories, but only aggregated clumps that the majority of volunteers labeled as ``normal'' are considered. The inset shows a zoomed view of the distributions for $p^{\mathrm{fp}}_{l}<0.01$.}
    \label{fig:volunteer_normal_set_fpp_dist}
\end{figure}

\autoref{fig:volunteer_normal_set_fpp_dist} illustrates the degree of correspondence between the value of $p^{\mathrm{fp}}_{l}$ assigned to each clump by our aggregation framework and their empirical categorisation as true or false positives. The figure compares the distributions of $p^{\mathrm{fp}}_{l}$ for empirically true positive and false clumps identified using all available annotations in for the expert-annotated subject set. The distributions represent the restricted subset of clumps in $\hat{Y}_{20}$ that the majority of volunteers labeled as ``normal''. However, we recognise that volunteers and experts may disagree about what criteria define a ``normal'' clump. Therefore, to avoid conflating this categorical disagreement with genuine cases when experts and volunteers mark different features (regardless of the annotation tool used) we consider any expert identified clump when assigning true-positive or false-positive labels. The majority of aggregated clumps in both categories have very low estimated false positive probabilities ($p^{\mathrm{fp}}_{l}\ll 1$), indicating a high degree of consensus between volunteers, albeit that this consensus disagrees with the expert annotations. Although clumps in both empirical categories have estimated $p^{\mathrm{fp}}_{l}$ values spanning the full range $[0, 1]$, we note that 95\% of empirically true-positive clumps have $p^{\mathrm{fp}}_{l}<0.3$ compared with only 68\% of empirical false positives. This reinforces the evidence implicit in \autoref{fig:pseudo_roc_volunteer_counts} that the aggregated clump sample can be made purer with respect to the expert sample by applying a threshold on $p^{\mathrm{fp}}_{l}$.

\subsection{Volunteer Skill Parameters}\label{subsec:volunteer_skills}

\begin{figure}
    \centering
    \includegraphics[width=0.45\textwidth]{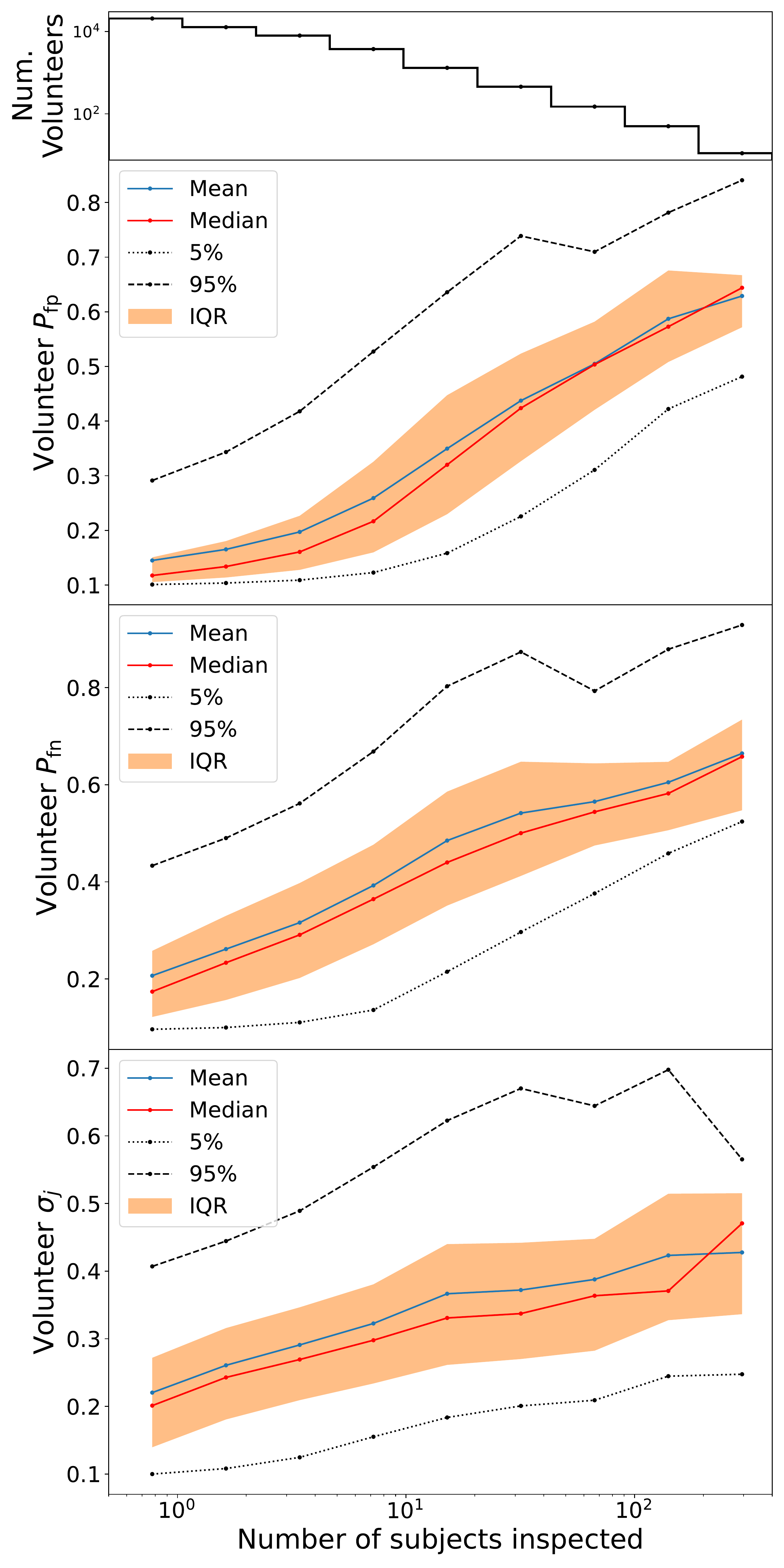}
    \caption{Evolution of volunteer skill parameter statistics versus number of subjects inspected. The \textit{top} panel show the distribution of the number of volunteers who have inspected at least as many subjects as indicated by the upper boundary of each bin. This means volunteers who annotate many subjects will contribute to several bins. However, their skill parameters are sampled at the point that they had inspected the maximum number of subjects represented by a particular bin. Statistics for the different volunteer skill parameters $p_{j}^{\mathrm{fp}}$, $p_{j}^{\mathrm{fn}}$ and $\sigma_{j}$ are shown in the \textit{upper-middle}, \textit{lower-middle} and \textit{bottom} panels respectively. \textit{Red} and \textit{blue} markers plot the median and mean skill parameter of all volunteers contributing to a particular bin, respectively. The \textit{orange} band illustrates the inter-quartile ranges of the bin-wise distributions. \textit{Dotted} and \textit{dashed} lines indicate the 5th and 95th percentiles respectively.}  
    \label{fig:worker_skill_evolution}
\end{figure}

Our aggregation framework allows us to monitor the evolution of volunteers' skill parameters as they spend time in the project. The top panel of \autoref{fig:worker_skill_evolution} shows the distribution of the \textit{Galaxy Zoo: Clump Scout} volunteers' subject classification counts. The distribution is  bottom-heavy with a median of 3 subjects per volunteer and 19,859 volunteers ($\sim95\%$) annotating fewer than 10 images, and only 176 volunteers ($\sim0.08\%)$ annotating more than 200\footnote{This skewed non-uniform distribution for the nuber of annotations per volunteer is also seen in many other Zooniverse projects \citep[e.g.][]{spiers_gini}.}. The remaining panels of \autoref{fig:worker_skill_evolution} illustrate how our estimates of the volunteers' skill parameters evolve as volunteers inspect and annotate increasing numbers of subjects. For all three skill parameters, the mean and median of the maximum likelihood estimates increase monotonically from their prior values as volunteers annotate more subjects. The relatively slow evolution of $p^{\mathrm{fp}}_{j}$ for subject inspection counts below $\sim10$ reflects the strong regularisation that results from setting the hyper-parameter $n_{\beta}^{\mathrm{fp}}=500$ (see \autoref{tab:prior_param_values}).

\subsection{Subject risk and its components}\label{subsec:results_risk_and_components}

\begin{figure*}
    \centering
    \includegraphics[width=\textwidth]{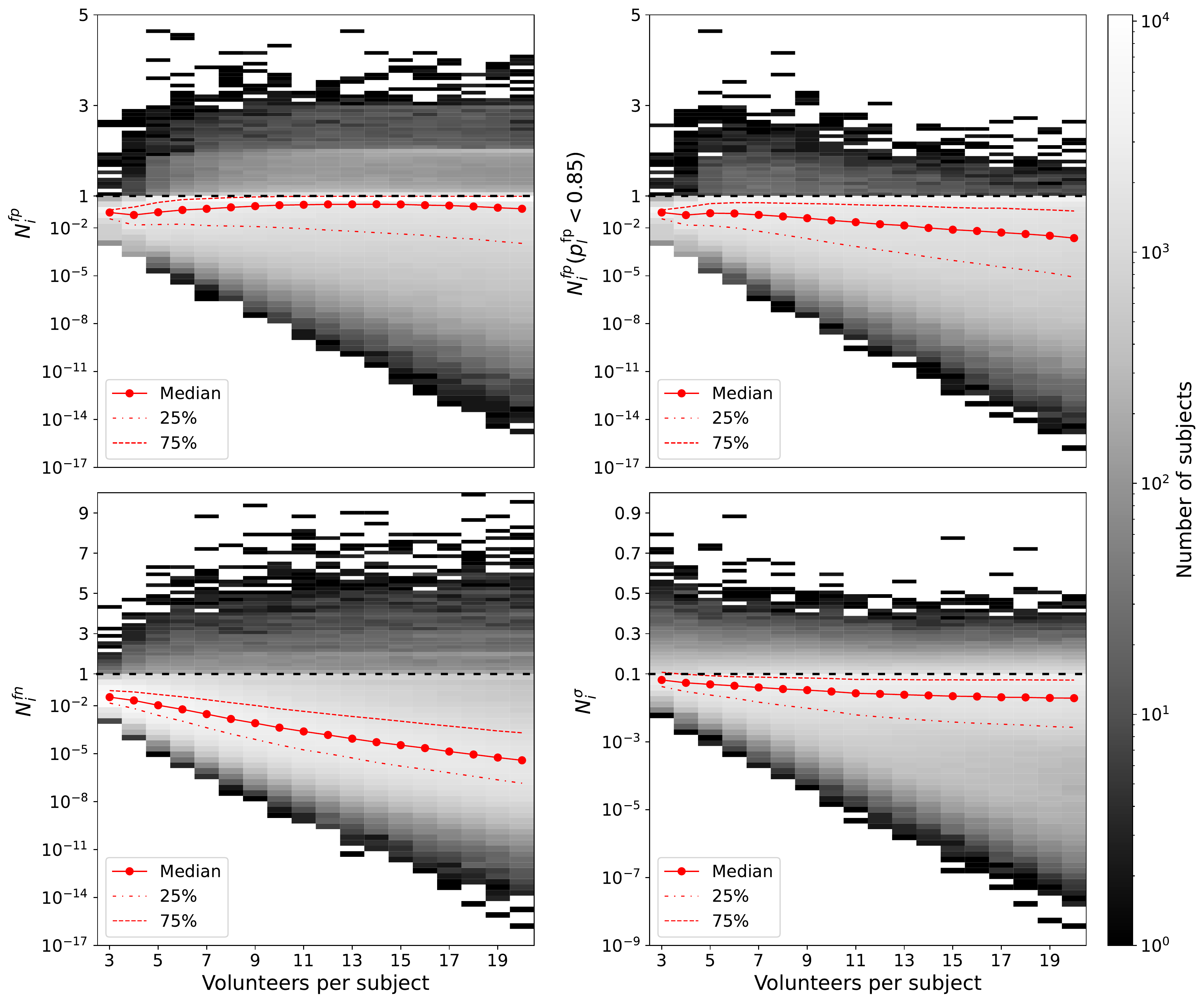}
    \caption{Evolution of the distributions for components of subject risk as the number of volunteer annotations per subject increases. Distributions for the \textit{expected} numbers of false positive bounding boxes $N_{i}^{\mathrm{fp}}$, missed clumps (or false negatives) $N_{i}^{\mathrm{fn}}$ and inaccurate clump locations $N_{i}^{\mathrm{\sigma}}$ are shown in the \textit{upper left}, \textit{lower left} and \textit{lower right} panels respectively. The \textit{upper right} panel shows the distributions of $N_{i}^{\mathrm{fp}}$ after discarding individual clumps with false positive probabilities $p^{\mathrm{fp}}_{l} > 0.85$. Note that the $y$ axis changes from logarithmic to linear scaling at the values indicated by the black horizontal dashed lines to better illustrate the evolution of structures in each distribution.}
    \label{fig:subject_risk_param_evolution}
\end{figure*}
The distributions shown in \autoref{fig:subject_risk_param_evolution} reveal how the \textit{expected} numbers of false positive bounding boxes $N_{i}^{\mathrm{fp}}$, missed clumps (or false negatives) $N_{i}^{\mathrm{fn}}$ and inaccurate clump locations $N_{i}^{\mathrm{\sigma}}$ (see \autoref{subsec:computing_risk}) evolve for the subjects in the the \gzcs subject set as more volunteers annotate them. For the majority of subjects, our framework estimates values less than one for all risk components, regardless of how many volunteers annotated them. The distributions of $N_{i}^{\mathrm{fp}}$, $N_{i}^{\mathrm{fn}}$ and $N_{i}^{\mathrm{\sigma}}$ become broader and their median values decrease monotonically as $n\rightarrow20$. This pattern indicates that for the majority of subjects, increasing the number of volunteers who annotate each subject improves the reliability of their consensus labels. 

A minority of subjects have estimated values for one or more of $N_{i}^{\mathrm{fp}}$, $N_{i}^{\mathrm{fn}}$ or $N_{i}^{\mathrm{\sigma}}$ that are greater than one. For this subset of subjects, their associated risk component distributions appear to stabilise after five or more volunteers have annotated each subject. We suggest that estimates for subjects that are annotated by fewer than 5 volunteers (i.e. for $n\lesssim5$) are noise-dominated or prior-dominated and somewhat unreliable. The structure that is visible in the distributions of $N_{i}^{\mathrm{fp}}$ in the \textit{upper-left} panel is produced by a strong bimodality in the distribution of false positive probabilities ($p^{\mathrm{fp}}_{l}$) for the clumps in the corresponding sets of estimated labels (i.e. the clumps in the corresponding $\hat{Y}_{n}$ -- see \autoref{fig:clump_fpp_dist}). For each clump in the estimated label for a particular subject, its false positive probability is very likely to be close to zero or one. The expected number of false positive clumps in a subject's estimated label is derived by summing a term that includes these probabilities in its denominator, so the distributions of  will naturally be concentrated into peaks around integer values of $N_{i}^{\mathrm{fp}}$. Similar structures that are visible in the distributions of $N_{i}^{\mathrm{fn}}$ are produced by a strong bimodality in the summand in \autoref{eq:est_num_fn}. The fraction of subjects for which $N_{i}^{\mathrm{fp}}>1$ peaks at $\sim10\%$ for $n=15$ and decreases to $\sim8\%$ for $n\sim20$.  In contrast, the fraction of subjects for which $N_{i}^{\mathrm{fn}}>1$ does not peak, but increases quasi-monotonically to reach $\sim2\%$ as $n\rightarrow20$. The fraction of subjects for which $N_{i}^{\sigma}>1$ is negligible and $<0.05\%$ for all $n$.

The overall median values for the estimated numbers of missed clumps and inaccurate clump locations per subject both decrease monotonically as the number of volunteers who inspect each subject increases. However the overall median for the expected number of false positive clumps per subject increases slowly until $n=13$ before beginning to decrease. We assess the feasibility of reducing $N_{i}^{\mathrm{fp}}$ by discarding aggregated clumps with high individual false positive probabilities. The upper right panel of the \autoref{fig:subject_risk_param_evolution} shows the effect of filtering clumps with $p^{\mathrm{fp}}_{l}>0.85$ on the distribution of $N_{i}^{\mathrm{fp}}$. Applying this filter substantially reduces the estimated number of false positive clumps after 5 or more volunteers annotate each subject and moreover, the fraction of subjects for which the expected number of false positive clumps per subject exceeds one now peaks at $\sim0.1\%$ for $n=5$ and decreases rapidly thereafter.

\begin{figure}
    \centering
    \includegraphics[width=0.45\textwidth]{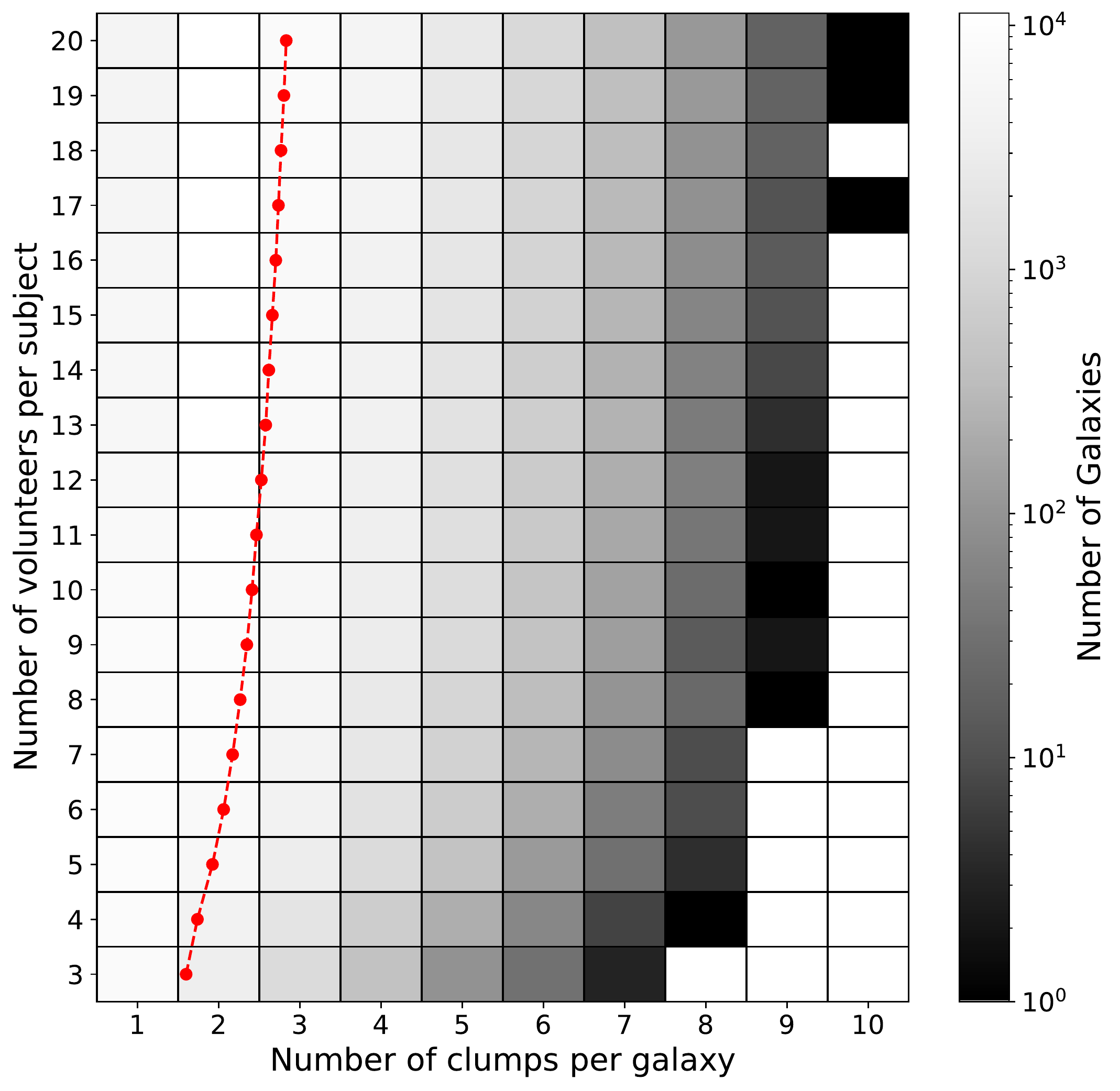}
    \includegraphics[width=0.45\textwidth]{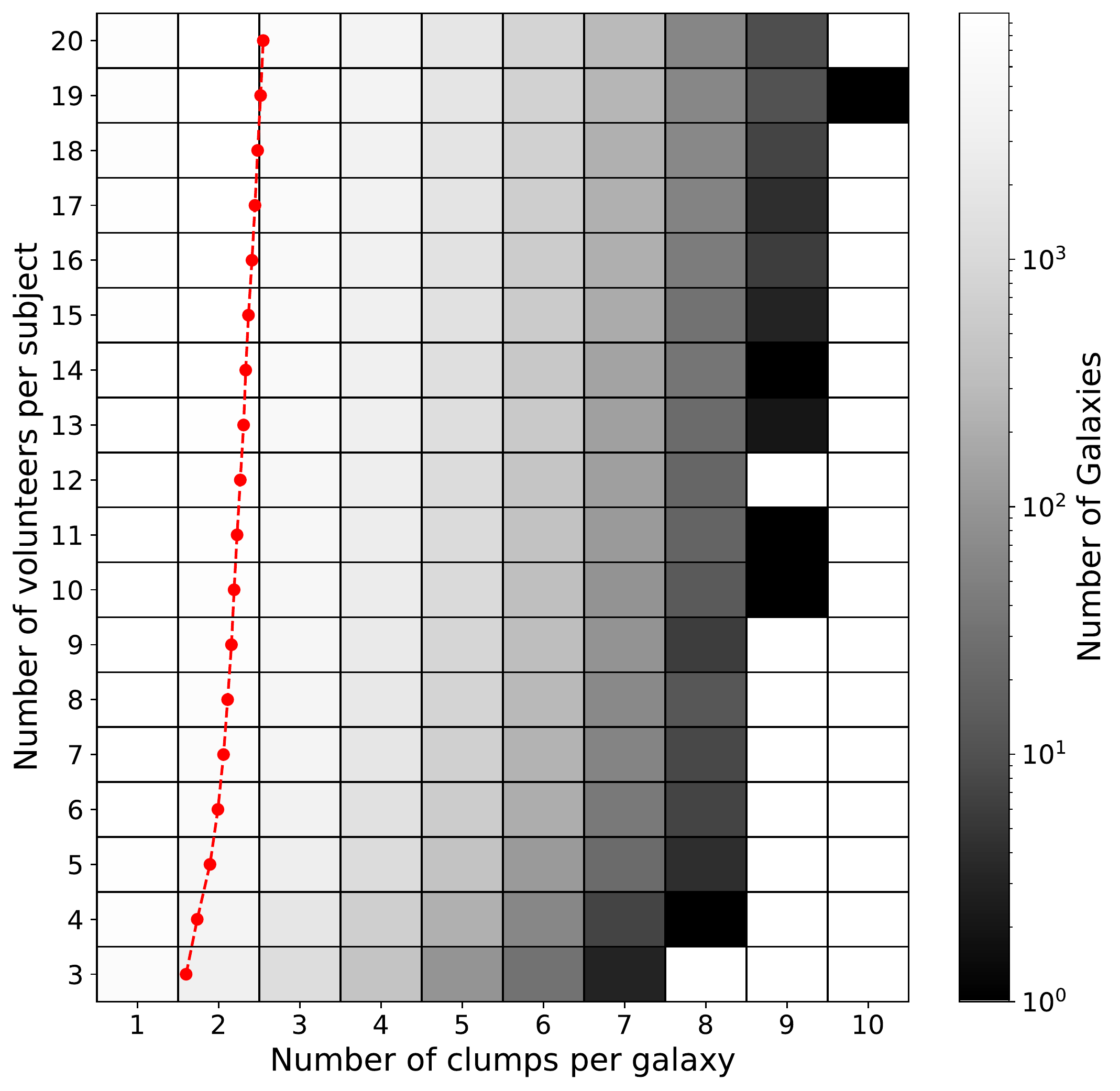}
    \caption{Evolution of the distribution of the number of clumps per galaxy as more volunteers inspect and annotate each subject. The red markers and lines plot the distribution means for the different numbers of volunteers per subject. \textit{Top panel:} Number of clumps per galaxy with any value for their estimated false positive probability $p^{\mathrm{fp}}_{l}$. \textit{Bottom panel:} Number of clumps per galaxy with $p^{\mathrm{fp}}_{l}<0.85$}
    \label{fig:clumps_per_galaxy_filtered}
\end{figure}

\begin{figure*}
    \centering
    \includegraphics[width=0.48\textwidth]{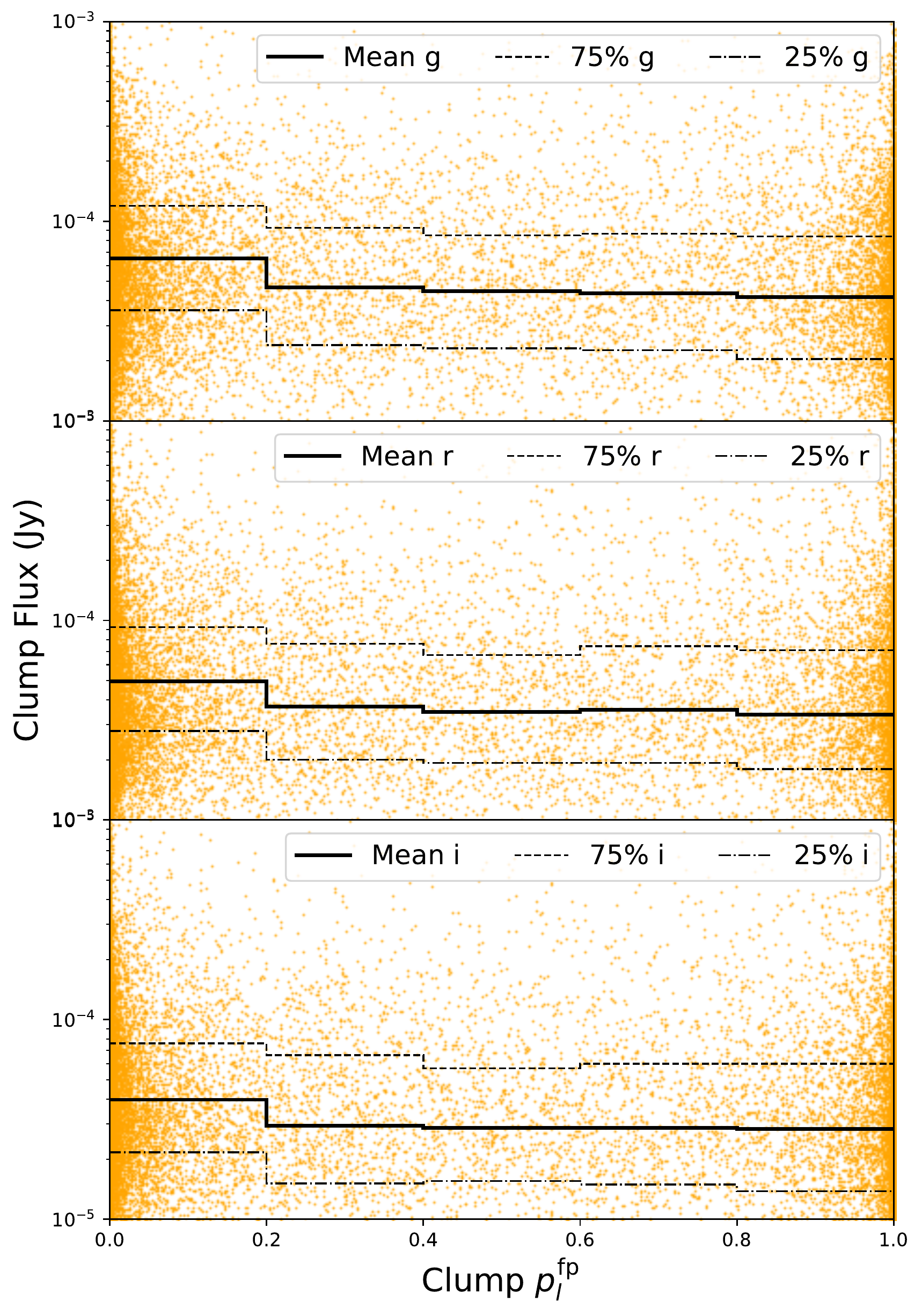}
    \includegraphics[width=0.48\textwidth]{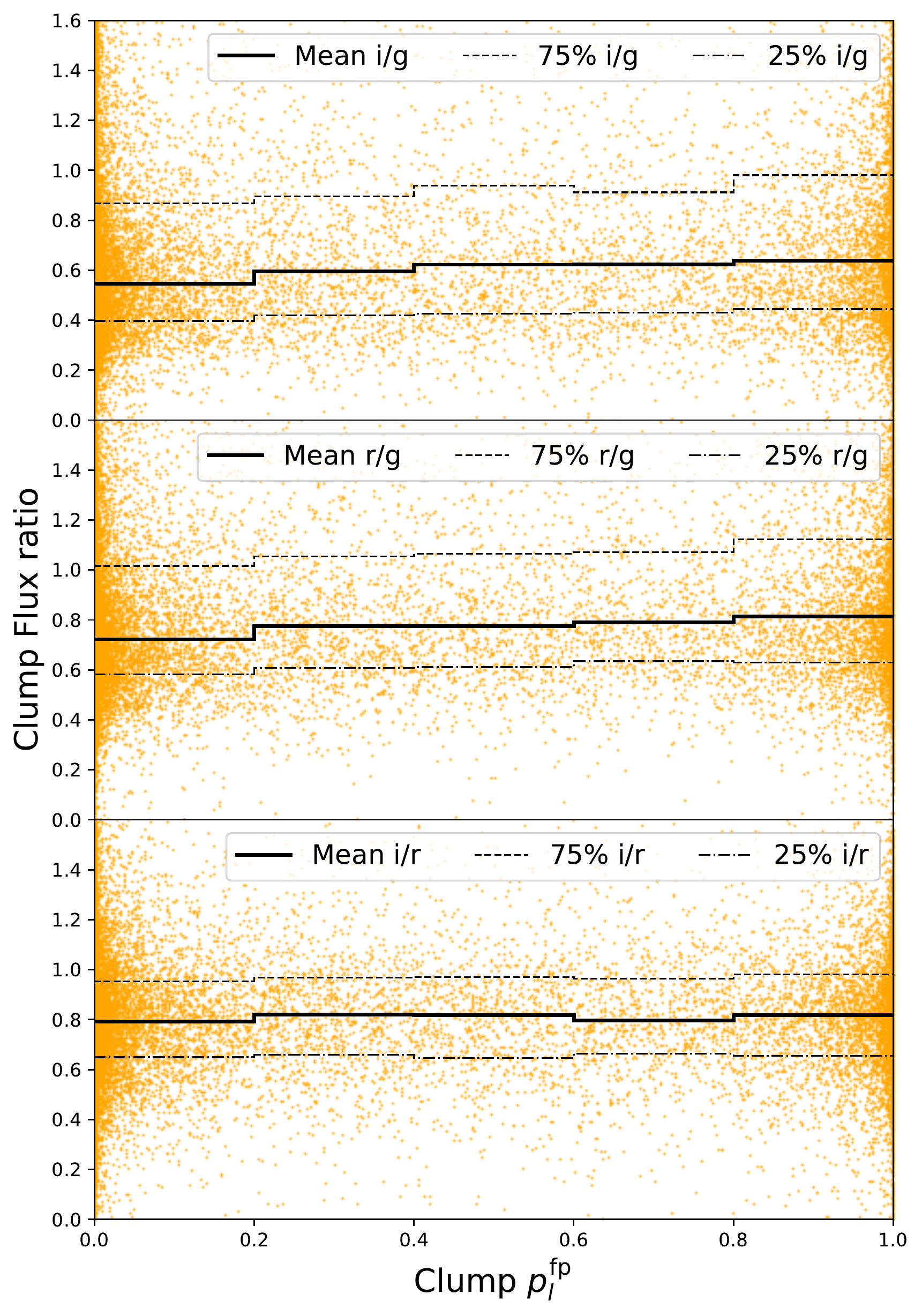}
    \caption{\textit{Left panel:} Clump flux in \textit{g}, \textit{r} and \textit{i} bands versus estimated clump false positive probability $p^{\mathrm{fp}}_{l}$. \textit{Right panel:} Clump flux ratios \textit{g}, \textit{r} and \textit{i} bands versus estimated clump $p^{\mathrm{fp}}_{l}$. In both panels, the fluxes in each band are scaled in the same way as the corresponding bands of the subject images (see \autoref{subsec:data:preparation}) to better reflect the data that volunteers actually see. On average, clumps with low $p^{\mathrm{fp}}_{l}$ appear brighter and bluer in the subject images.}
    \label{fig:flux_versus_fpp}
\end{figure*}

We note that filtering clumps based solely on their estimated false-positive probabilities may inadvertently discard real clumps if $p^{\mathrm{fp}}_{l}$ does not correlate appropriately with observable quantities like brightness and colour that can indicate whether a particular feature is a genuine clump or spurious\footnote{Indeed, for this reason \citet{\nicospaper} apply a very permissive $p^{\mathrm{fp}}_{l}$ threshold before filtering further based on observable clump parameters.}. \autoref{fig:clumps_per_galaxy_filtered} illustrates the overall effect of discarding clumps with individual false positive probabilities larger than 0.85 on the number of clumps per galaxy that our framework identifies using different numbers of volunteer annotations per subject. The impact is strongest for $n\gtrsim7$ but the overall effect is small with $\lesssim0.5$ fewer clumps identified per galaxy.
The \textit{left hand} panel of \autoref{fig:flux_versus_fpp} plots fluxes in the \textit{g}, \textit{r} and \textit{i} bands versus the estimated individual false positive probability ($p^{\mathrm{fp}}_{l}$) for all clumps that our framework identifies using 20 annotations per subject. In all three bands, the mean flux of clumps with $p^{\mathrm{fp}}_{l} < 0.2$ is $\sim1.5$ times larger than the mean flux for clumps with $p^{\mathrm{fp}}_{l} > 0.2$. The \textit{right hand} panel of \autoref{fig:flux_versus_fpp} plots the non-redundant flux ratios $i/g$, $r/g$ and $i/r$ versus $p^{\mathrm{fp}}_{l}$. On average, clumps with low estimated false positive probability appear brighter in bluer bands. Overall, we observe a pattern whereby clumps that appear brighter and bluer in the subject images tend to have lower $p^{\mathrm{fp}}_{l}$. We verified that this pattern does not change significantly when clumps are filtered according to the fraction of volunteers that labeled them as ``unusual''. This is reassuring because real clumps are expected to be bright and blue in colour and suggests that filtering clumps based on $p^{\mathrm{fp}}_{l}$ is well motivated physically. The correlations with flux and colour also resemble the \textit{empirical} patterns described in  \autoref{subsec:expert_comparison} where we observed that the sample of clumps that coincided with expert clump annotations were brighter and bluer than the sample of clumps that did not.

\begin{figure}
    \centering
    \includegraphics[width=0.48\textwidth]{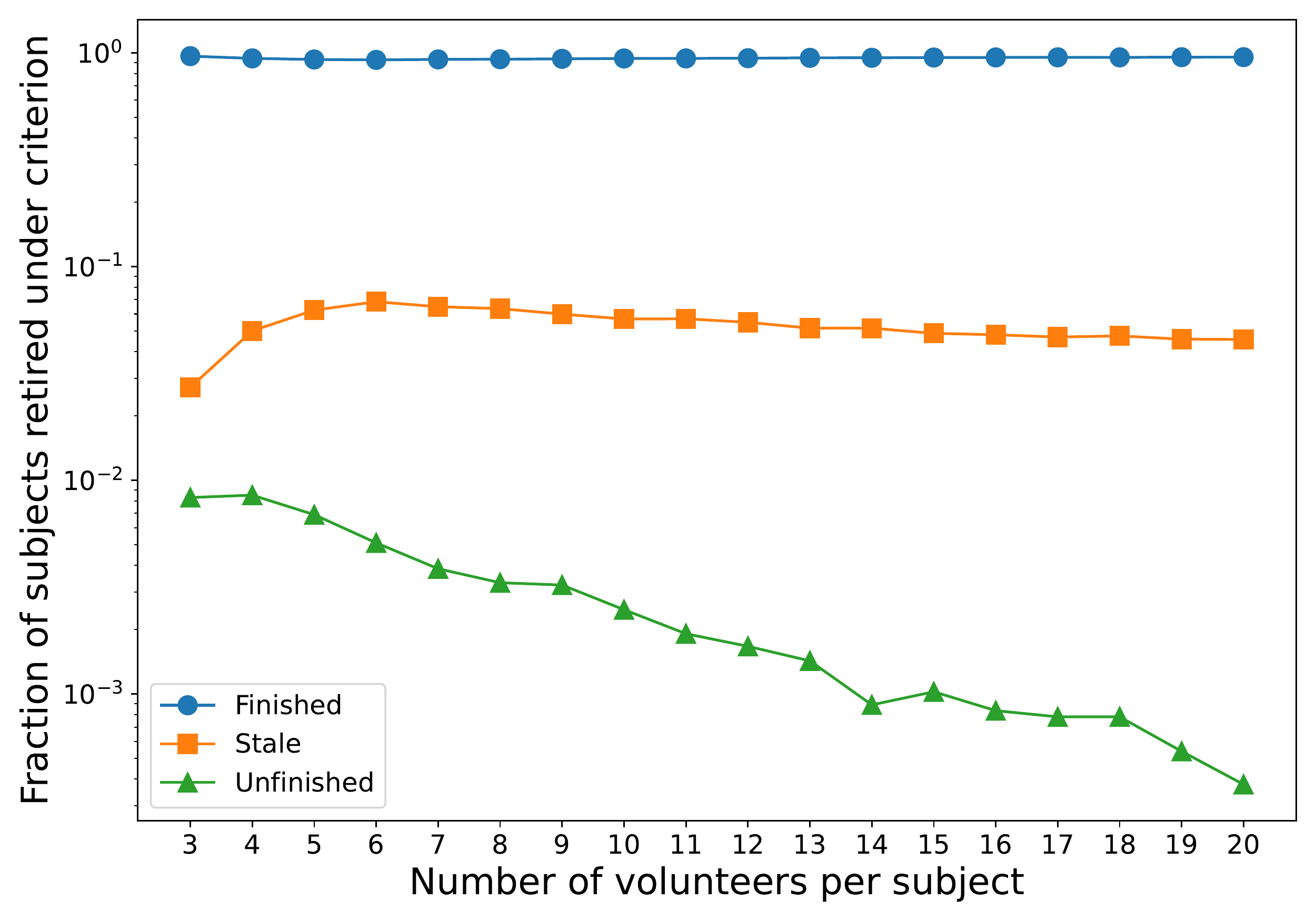}
    \caption{Fraction of subjects retired for different reasons versus number of volunteers per subject}
    \label{fig:retirement_reason_fraction}
\end{figure}

\autoref{fig:retirement_reason_fraction} shows how the fractions of subjects that are retired for different reasons vary as more volunteers annotate each subject. More than 90\% of subjects meet the subject retirement criterion specified in \autoref{subsec:retirement_and_finalisation}  regardless of how many volunteers annotate each subject. Of the remaining subjects, $\sim7-9\%$ become stale after persisting in the working batch for more than 10 replenishment cycles and are removed. The fraction of stale subjects peaks for $n=6$ annotations per subject and decreases monotonically thereafter as more annotations per subject are used. Fewer than 1\% of subjects failed to retire for any $n$. The fraction of unretired subjects is maximally  0.9\% for $n=3$ and falls to $<0.1\%$ for $n=20$. We comment that for $n<6$ the computation of $\mathcal{R}$ and its components is likely to be dominated by our model priors and therefore the apparent decrease in the number of stale subjects should probably not be interpreted as improved performance within this domain.

\section{Discussion}\label{sec:discussion}

Using the annotations provided by the \gzcs volunteers our framework has identified a large catalogue of potential clumps. In addition, our aggregation framework provides quantitative metrics for the reliability of the estimated subject labels it computes. These diagnostics allow us to better understand how volunteers interpreted the definition for a clump that they were provided with and how they execute the annotation task.

The observable properties of the clumps we detect appear plausible, both in terms of their spatial distribution within the subject images and their fluxes in the SDSS \textit{g}, \textit{r} and \textit{i} bands. The central concentration of confidently identified clumps in \autoref{fig:clump_location_vs_pfp} is reassuring because it reflects the typical footprints of the target galaxies in each subject image, which is where we would reasonably expect to find genuine clumps. For clumps with any estimated false positive probability $p^{\mathrm{fp}}_{l}$, we observe a clear under-density at the centre of the distribution, which likely reflects the fact that most volunteers correctly distinguish the target galaxies' central bulges from clumps.

The clump flux and colour distributions in \autoref{fig:flux_versus_fpp} reveal that brighter, bluer clumps tend to have lower false positive probabilities ($p^{\mathrm{fp}}_{l}$). This trend is also reassuring because real clumps are expected to be bright and blue in colour and suggests that filtering clumps based on $p^{\mathrm{fp}}_{l}$ is well motivated physically. The correlations with flux and colour also resemble the \textit{empirical} patterns described in  \autoref{subsec:expert_comparison} where we observed that the sample of clumps that coincided with expert clump annotations were brighter and bluer than the sample of clumps that did not.

By comparing expert labels for 1000 subjects with those estimated by our framework using volunteer annotations, we showed that volunteers are much more optimistic that experts when annotating clumps. Overall, the distributions in \autoref{fig:expert_all_set_hitmiss_band_flux_dists_lscale} suggest that volunteers are more likely to mark faint features than experts, particularly when those features appear blue. This results in aggregated clump samples for the 1000 test subjects that appear quite heavily contaminated with respect to the expert labels. Moreover, this apparent contamination worsens if clumps that experts or the majority of volunteers labeled as ``unusual'' are discarded. This degradation in purity for the ``normal'' clump subset likely indicates that volunteers and experts disagree about the definition of a ``normal'' clump with volunteers being less likely to label a clump as unusual. 

Using \autoref{fig:volunteer_normal_set_fpp_dist} we illustrated that our framework tends to estimate lower false positive probabilities for clumps that were marked by both volunteers and experts. The formulation of our likelihood model means that smaller estimated false positive probabilities correlate broadly with a greater degree of consensus between skilled volunteers that a clump exists at a particular location. Therefore, it seems that while many volunteers mark features that experts would not identify as clumps, features that experts \textit{do} mark tend to have also been marked by a majority of more skilled volunteers who inspected the corresponding subject. The correlation between clumps' false positive probabilities and their expert classifications also reinforces the evidence implicit in \autoref{fig:pseudo_roc_volunteer_counts} that the aggregated clump sample can be made purer with respect to the expert sample by applying a threshold on $p^{\mathrm{fp}}_{l}$.

\begin{figure*}
    \centering
    \includegraphics[width=\textwidth]{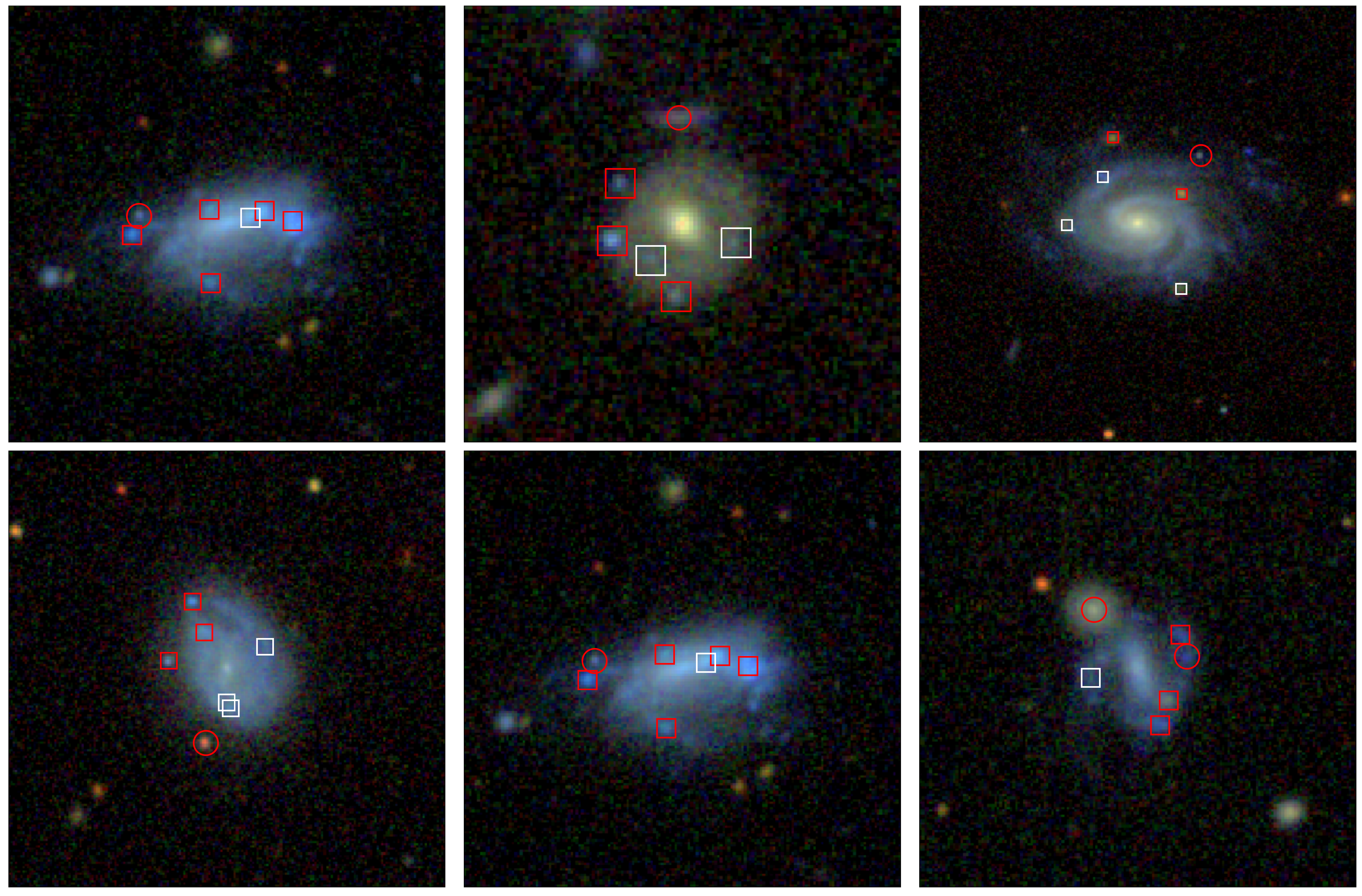}
    \caption{Six curated but representative examples of subject images that show agreements and disagreements between volunteers and experts. Features labeled as clumps by volunteers but ignored by experts are highlighted by \textit{white} boxes. \textit{Red} boxes highlight features that were annotated by both experts \textit{and} volunteers. \textit{Red} circles highlight features that were annotated by experts but \textit{not} by volunteers. Volunteers tend to mark fainter features than experts, particularly if those features appear blue in colour. None of the features highlighted in this figure were labeled as ``odd'' by a majority of volunteers or the experts who marked them.}
    \label{fig:empirical_fp_example_images}
\end{figure*}

We note that while using a visual labelling approach to identify clumps provides more flexibility than relying on a fixed set of brightness or colour thresholds, it is also unavoidably subjective. To illustrate how this subjectivity may be impacting the empirically determined purity and completeness of our clump sample, \autoref{fig:empirical_fp_example_images} shows typical examples of the faint blue features that volunteers annotate but experts ignore. Many of these any do appear clump-like and it is not always obvious why experts have not marked them. Based on these observations, we suggest that the sample of clumps identified by our framework using volunteer annotations may not be as severely contaminated  as \autoref{fig:pseudo_roc_volunteer_counts} implies. We also note that the clump samples our framework derives are generally very complete and include the majority of expert-labeled clumps. This means that that subsets of clumps for particular scientific analyses can be selected from a nominally impure sample using physically motivated criteria based on directly observable or derived characteristics of the individual clumps. For example, \citet{\nicospaper} derive samples of bright clumps by using criteria based on photometry extracted from clumps and their host galaxies.

In addition to providing quantitative estimates for the reliability of individual clump labels, our framework allows us to investigate the performance of individual volunteers and the entire volunteer cohort. The positive gradients of skill parameter evolution curves in \autoref{fig:worker_skill_evolution} decrease with increasing number of subjects inspected (their second derivatives are negative except in the final bin which contains relatively few volunteers). This suggests that the the volunteer skill parameters may converge to stable asymptotic values for very large numbers of inspected subjects. The fact that this convergence was not achieved for the \textit{Galaxy Zoo: Clump Scout} dataset likely indicates that the global maximum likelihood solution is dominated by the large number of volunteers who inspect very few images and may provide noisy annotations due to their relative inexperience.

The noisiness of volunteer annotations probably indicates that identifying clumps within star forming galaxies, which can have complex underlying morphologies, is relatively difficult for inexperienced non-experts. In \autoref{subsec:volunteer_skills} we noted that most volunteers only annotated a small number of galaxies and may not have had time to learn the visible characteristics of genuine clumps. While it may be the case that the task of clump identification is too difficult for typical Zooniverse volunteers, this seems unlikely and there are several plausible strategies for making complex and subtle image analysis tasks more feasible for citizen scientists. The most obvious is to improve the amount and quality of the initial training that is provided to volunteers. However, Zooniverse volunteers are accustomed to participating in projects with minimal tutorial material so imposing a more rigorous training requirement may discourage widespread participation. As  discussed in \autoref{subsec:volunteer-training}, the volunteers to who contributed to \gzcs received real-time feedback for a small number of expert-labeled subjects that they annotated during the early stages of their participation. Providing more detailed feedback for a larger sample of subjects may help volunteers to better understand the task they are being asked to perform. 

Some \textit{Zooniverse} projects also provide a dedicated tutorial workflow with an accompanying video tutorial in which experts annotate the same subjects that volunteers see and explain their reasoning\footnote{e.g. \url{https://www.zooniverse.org/projects/chrismrp/radio-galaxy-zoo-lofar}}. When using feedback as a training tool, it is important that the feedback subjects contain galaxies and clumps that are properly representative of the global populations within the full subject set, but it is difficult to ensure that this is the case unless the experts themselves inspect a large number of subjects. Moreover, the feedback messages that volunteers receive must be carefully chosen to avoid discouraging volunteers if their annotations disagree with those of experts.

An alternative to explicit training and feedback that was pioneered by the \textit{Gravity Spy} project\footnote{\url{https://www.zooniverse.org/projects/zooniverse/gravity-spy}} involves incrementally increasing the difficulty of subjects that volunteers inspect and annotate as they spend longer engaged with the project and their skill improves \citep{2017CQGra..34f4003Z}. Using this ``leveling up'' approach requires an \textit{a priori} metric for the relative difficulty of subjects for volunteers, as well as ongoing assessment of volunteers' skills. While our framework naturally fulfills the latter requirement, it does not facilitate prior segregation of subjects to populate the different difficulty levels. It might be possible to formulate a heuristic approach to estimating subject difficulty based on observable properties of the clumps' host galaxies, but that is beyond the scope of this paper.

As we discuss in \autoref{subsec:overview}, the consensus reliability metrics that our framework computes may enable quantitatively motivated early retirement of subjects if it can be established that a stable consensus solution has been reached. In \autoref{subsec:computing_risk} we described how our framework formulates a subject retirement criterion based on estimated metrics that are proxies for the completeness ($N_{i}^{\mathrm{fn}}$), purity ($N_{i}^{\mathrm{fp}}$) and accuracy ($N_{i}^{\sigma}$) of that subject's label. \autoref{fig:retirement_reason_fraction} seems to show that more than 90\% of subjects fulfil this criterion, even when only $n=3$ volunteers inspect each subject. However, the distributions shown in  \ref{fig:subject_risk_param_evolution} appear to be noise or prior dominated for $n\lesssim5$ and we suggest that estimates of the subject risk $\mathcal{R}$ and its components $\{N_{i}^{\mathrm{fn}},N_{i}^{\mathrm{fn}},N_{i}^{\sigma}\}$ for that domain should be treated with some caution. 

In \autoref{fig:subject_risk_param_evolution}, we showed that discarding clumps with estimated individual false positive probabilities $p_{l}^{\mathrm{fp}}>0.85$ substantially reduces the  number of subject labels that are expected to include one or more false positive clumps and that this number reduces rapidly once more than 7 volunteers have annotated each subject. 

We interpret the fact that the estimated number of missed clumps per subject ($N_{i}^{\mathrm{fn}}$) increases as more volunteers annotate each subject as an effect of some of those volunteers marking very faint features. Potential false-negative clumps identified by the second, constrained run of the facility location algorithm (see \autoref{subsec:computing_risk}) are typically on the threshold of identification by our framework, which normally means that several volunteers have marked them\footnote{The precise number of marks required depends on the skill parameters of the volunteers who provide them.}. If the fraction of highly optimistic volunteers within the overall cohort is small, then a relatively large number of volunteers must inspect each subject for faint features to reach the threshold where they are considered potential false negatives. The increase in $N_{i}^{\mathrm{fn}}$ as the number of annotations per subject $n\rightarrow 20$ is then an indication that more faint features are reaching, but not surpassing, our framework's detection threshold. \autoref{fig:clumps_per_galaxy_filtered} provides an empirical estimate for the number of clumps per galaxy that are missed when fewer volunteers inspect each subject. Although the mean number of identified clumps per galaxy does increase in the interval $7<n<20$, the rate of increase is very slow and increasing $n$ from 7 to 20 results in just 0.5 more clumps with individual false positive probabilities $p_{l}^{\mathrm{fp}}<0.85$ per galaxy on average. In line with our previous observations regarding volunteer optimism, we suggest that many of these additional clumps may in fact be faint, blue features within the target galaxies. As \autoref{fig:pseudo_roc_volunteer_counts} illustrates, our comparison with expert labels also suggests that $n\sim7$ provides that best compromise between the completeness and purity of our aggregated clump sample. 

Empirically, it seems like at least 5 volunteers must inspect each subject to obtain a stable solution for the subject labels and that the majority of genuine clumps could be identified by our framework for most subjects using the annotations provided by $\sim7$ volunteers. Increasing the number of volunteers beyond this threshold seems to introduce more noise into the annotation data and also results in progressively fainter features being identified. Retiring the majority of subjects after inspection by 7 volunteers, if it could have been well motivated, would have reduced the volunteer effort required for the \gzcs project by a factor $>2$. Unfortunately, we must acknowledge that the reliability metrics computed by our framework do not seem to converge in a way that is useful to facilitate an early retirement decision. For most subjects, our framework predicts expected numbers of false positives, false negatives and inaccurate true positives that are less than one for any number of annotations  (i.e. $N_{i}^{\mathrm{fp}},N_{i}^{\mathrm{fn}},N_{i}^{\sigma}\ll1\;\forall\,n$) and so these subjects would have been retired when $n<7$ based on the thresholds listed in \autoref{tab:subject_retirement_params}. As we show in \autoref{fig:pseudo_roc_volunteer_counts}, retiring subjects this early would yield a lower sample completeness, even for the brighter clumps that experts also identified. 

Moreover, while predicted numbers of subject labels containing false positive or inaccurate clump locations both decrease for $n\gtrsim7$ as $n\rightarrow20$, the predicted number of subjects labels that are missing real clumps increases. Using any retirement criterion predicated on $N_{i}^{\mathrm{fn}}\ll1$, considering the annotations from more volunteers would result in more subjects becoming stale in the working batch and therefore requiring inspection by experts. Fortunately in the case of \gzcs, the fraction of subjects for which the estimated number of false positive clumps $N_{i}^{\mathrm{fn}}>1$ for any $n$ is $<3\%$ of the overall dataset ($\sim 2500$ subjects), so visual inspection by experts would be feasible.

\section{Summary and Conclusion}\label{sec:conclusion}

In this paper we have presented a software framework that uses a probabilistic model to aggregate multiple annotations that mark  two-dimensional locations in images of distant galaxies and derive a consensus label based on those annotations. The annotations themselves were provided via the \gzcs citizen science project by non-expert volunteers who were asked to mark the locations of giant star forming clumps within the target galaxies. Among a sample of 85,286 galaxy images that were inspected by volunteers, our software framework identified 44,126 that contained at least one visible clump and detected 128,100 potential clumps overall.

To empirically evaluate the validity of the clumps we identify, we compared our aggregated labels with annotations provided by expert astronomers for a subset of 1000 galaxy images. We found that \gzcs volunteers are much more optimistic than experts, and are willing to mark much fainter features as potential clumps, particularly if those features appear blue in colour. However, volunteers also mark the vast majority of bright clumps that experts identify, so although the sample of clumps we identify is $\sim50\%$ contaminated with respect to the expert identifications, it is $\gtrsim90\%$ complete. 

In addition to our empirical evaluation, we have used the statistical model that underpins our framework to compute quantitative metrics for the reliability of the overall aggregated labels that we derive for each image. These metrics suggest that stable consensus for most images' labels is achieved after $\sim7$ volunteers have annotated it, which is $<50\%$ of the 20 annotations that were collected for each image via \gzcs and would represent a significant saving in volunteer effort. However, the annotation data are quite noisy with large variation between the numbers of locations that are marked by different volunteers and this noise makes it difficult to define a \textit{robust} ``early retirement'' criterion that could be used to safely curtail collection of annotations before 20 have been acquired.

We suggest that the noisy annotation data reflect the fact that inexperienced non-experts find the task of identifying clumps difficult, or that the task was not properly explained. In \autoref{sec:discussion}, we discuss how different approaches to volunteer training could be used to help volunteers better distinguish the visible characteristics of genuine clumps from those of the faint, blue features that many ultimately marked. On the other hand, one of the benefits of using citizen science to identify clumps is that it avoids being overly prescriptive regarding the definition of a clump. \gzcs represents the first extensive wide-field search for clumpy galaxies in the local Universe and it may be that low-redshift clumps have different properties to their more distant counterparts. Using strict thresholds on brightness or colour might result an unexpected population of fainter clumps being missed. Moreover, the sample of clumps identified by volunteers appears to be very complete and so, if a subset of bright clumps is required for science analysis, such a sample can be straightforwardly constructed using photometric measurements for each clump \citep[e.g.][]{\nicospaper}.

Although our framework was developed to aggregate annotations for a specific citizen science project, its applicability is more general. A large number of projects running on the \textit{Zooniverse} platform collect two dimensional image annotations. Many of those projects consider subjects that are more familiar to non-experts and may be less prone to noise. In such cases, our framework may be able to substantially reduce the amount of effort and time taken to reach consensus for each subject. 

\section{Data availability}
The data underlying this article were used in \citet{\nicospaper} and can be obtained as a machine-readable table by downloading the associated article data from \url{https://doi.org/10.3847/1538-4357/ac6512}.  

\section*{Acknowledgements}

HD and SS were partly supported by the ESCAPE project; ESCAPE - The European Science Cluster of Astronomy \& Particle Physics ESFRI Research Infrastructures has received funding from the European Union's Horizon 2020 research and innovation programme under Grant Agreement no. 824064. SS also thanks the Science and Technology Facilities Council for financial support under grant ST/P000584/1. MW gratefully acknowledges support from the Alan Turing Institute, grant reference EP/V030302/1. This research is partially supported by the National Science Foundation under grant AST 1716602. This material is based upon work supported by the National Aeronautics and Space Administration (NASA) under Grant No. HST-AR-15792.002-A. This publication uses data generated via the Zooniverse.org platform, development of which is funded by generous support, including a Global Impact Award from Google, and by a grant from the Alfred P. Sloan Foundation.

This research made use of the open-source Python scientific computing ecosystem, including NumPy \citep{harris2020array}, Matplotlib \citep{Hunter2007}, and Pandas \citep{McKinney2010}.
This research made use of Astropy, a community-developed core Python package for Astronomy \citep{TheAstropyCollaboration2018}.
This research made use of Numba \citep{10.1145/2833157.2833162}.

\bibliographystyle{mnras}
\bibliography{references}

\appendix
\section{Model parameter priors}\label{sec:appendix_priors}

In this section we derive formulae that we use to compute and update the priors for our likelihood model's volunteer skill and subject difficulty parameters. Crucially for the efficiency of our framework, these formulae can all be factored into terms that depend only on the current working batch and terms that depend only on prior information. This allows us to straightforwardly update the skill parameters of returning volunteers without having to reconsider the annotations they contributed to previous working batches. 

\subsection{Beta priors for $p^{\mathrm{fp}}$ and $p^{\mathrm{fn}}$}

Our model for volunteer skill assumes that the event in which volunteer $j$ incorrectly provides a false positive clump annotation is Bernoulli $\mathrm{Bern}(p_{j}^{\mathrm{fp}})$ and similarly that the event that a volunteer misses a real clump is $\mathrm{Bern}(p_{j}^{\mathrm{fn}})$. Note that in general $p_{j}^{\mathrm{fp}} \neq 1 - p_{j}^{\mathrm{fn}}$. 

The probabilities $p_{j}^{k},\;k\in[\mathrm{fp},\mathrm{fn}]$ for a particular volunteer are unknown \textit{a-priori} and must be estimated using that volunteer's annotations $Z_{j}$. Let $n_{j}^{\mathrm{fp}}$ be the number of annotations in $Z_{j}$ that are determined to be false positives and $n_{j}^{\mathrm{fn}}$ be the number of real clumps that the volunteer missed. The \textit{posterior} distribution for $p_{j}^{k}$ is
\begin{align}
p(p_{j}^{k}|Z_{j}) &= \pi(p_{j}^{k})p(Z_{j}|p^{k}) = \pi(p_{j}^{k})\prod\limits_{r=1}^{\vert Z_{j}\vert}\mathrm{Bern}(p_{j}^{k}) \notag\\
&=  \pi(p_{j}^{k})\cdot(p_{j}^{k})^{n_{j}^{k}}(1-p_{j}^{k})^{(\vert Z_{j}\vert - n_{j}^{k})}
\end{align}
We place beta distribution priors on the values of $p_{j}^{\mathrm{fp}}$ and $p_{j}^{\mathrm{fn}}$, such that:
\begin{equation}
\pi(p_{j}^{k}) \sim \mathrm{Beta}(p_{j}^{k};n_{\beta}p_{0}^{k}, n_{\beta}(1-p_{0}^{k}))
\end{equation}
where the beta distribution
\begin{equation}
\mathrm{Beta}(\theta;a,b)\propto\theta^{1-a}(1-\theta)^{1-b}
\end{equation}
This approach simulates having previously performed $n_{\beta}$ Bernoulli trials with success probability $p_{0}^{k}$.
The Beta distribution is the \textit{conjugate prior} of the Bernoulli distribution, so using a beta prior on a Bernoulli distribution yields a posterior that is also Beta distributed. Expanding the \textit{posterior} distribution for $p^{k}$ yields
\begin{align}
p(p^{k}|Z_{j}) &\propto (p_{j}^{k})^{(1-n_{\beta}p_{0}^{k})}(p_{j}^{k})^{n_{j}^{k}}\notag\\
&\qquad\cdot(1-p_{j}^{k})^{(1-n_{\beta}(1-p_{0}^{k}))}(1-p_{j}^{k})^{(\vert Z_{j}\vert - n_{j}^{k})}\notag\\
&\propto (p_{j}^{k})^{(1-n_{\beta}p_{0}^{k})}\cdot(p_{j}^{k})^{n_{j}^{k}}\notag\\
&\qquad\cdot(1-p_{j}^{k})^{(1-n_{\beta}(1-p_{0}^{k}))}\cdot(1-p_{j}^{k})^{(\vert Z_{j}\vert - n_{j}^{k})}\notag\\
&\propto (p_{j}^{k})^{(1-n_{\beta}p_{0}^{k} + n_{j}^{k})}\notag\\
&\qquad\cdot(1-p_{j}^{k})^{(1-n_{\beta}(1-p_{0}^{k}) + \vert Z_{j}\vert - n_{j}^{k})}\notag\\
&\propto \mathrm{Beta}\left(p_{j}^{k};n_{\beta}p_{0}^{k}+n_{j}^{k}, n_{\beta}(1-p_{0}^{k}) + \vert Z_{j}\vert - n_{j}^{k}\right)
\end{align}
The expected value of the beta distribution $\mathrm{Beta}(\theta|a,b)$ is
\begin{equation}
\mathbb{E}(\theta) = \frac{a}{a+b}
\end{equation}
Adopting $\mathbb{E}(p_{j}^{k})$ as an estimator for the volunteer skills $p_{j}^{k}$ yields
\begin{equation}\label{eq:skill_prob_estimate}
\begin{aligned}
p_{j}^{k}&\approx\mathbb{E}(p_{j}^{k}) \\
&= \frac{n_{\beta}p_{0}^{k}+n_{j}^{k}}{n_{\beta}p_{0}^{k}+n_{j}^{k} + n_{\beta}(1-p_{0}^{k}) + \vert Z_{j}\vert - n_{j}^{k}}\\
&= \frac{n_{\beta}p_{0}^{k}+n_{j}^{k}}{n_{\beta}(p_{0}^{k}+1-p_{0}^{k}) + n_{j}^{k} + \vert Z_{j}\vert - n_{j}^{k}} \\
&= \frac{n_{\beta}p_{0}^{k}+n_{j}^{k}}{n_{\beta} + \vert Z_{j}\vert } \end{aligned}
\end{equation}

\subsection{Scaled inverse $\chi^{2}$ priors for $\sigma^{2}$}

We use a scaled inverse $\chi^{2}$ priors to compute a posterior distribution over variance parameters of our Gaussian models for  $d_{j}$ and $d_{i,l}$.

We specify a prior that simulates having drawn $n_{\chi}$ samples from a Gaussian distribution with zero mean and variance $\sigma_{0}^{2}$. This implies the following prior density function
\begin{align}
     \pi(\sigma^{2}; n_{\chi}, \sigma_{0}^{2})&= \mathrm{Scale-inv-}\chi^{2}(\sigma^{2};  n_{\chi}, \sigma_{0}^{2})\notag\\ &=\frac{(\sigma_{0}^{2}n_{\chi}/2)^{ n_{\chi}/2}}{\Gamma( n_{\chi}/2)}\cdot\frac{1}{(\sigma^{2})^{1+ n_{\chi}/2}}\cdot\exp\left[ \frac{- n_{\chi} \sigma_{0}^{2}}{2 \sigma^{2}}\right] 
\end{align}
We multiply by the likelihood of the data that were observed. To estimate $\sigma^{2}$ we consider the Jaccard distances $\Delta$ between all \textit{true positive} volunteer boxes and the locations of the established clumps to which they are associated.
\begin{align}
     P(Z|\sigma^{2}) \propto \frac{1}{\sigma^{n}}\exp\left[-\frac{1}{2\sigma^{2}}\displaystyle\sum_{i=1}^{n_{\mathrm{tp}}}\vert\Delta\vert^{2}\right] 
\end{align}
 This yields the posterior density
\begin{align}
     P(\sigma^{2}
     ) 
     &\propto \frac{1}{\sigma^{n_{\mathrm{tp}}}}\cdot\frac{1}{(\sigma^{2})^{1+ n_{\chi}/2}}\notag\\
     &\qquad\cdot\exp\left[-\frac{1}{2\sigma^{2}}\displaystyle\sum_{i=1}^{n_{\mathrm{tp}}}\vert\Delta\vert^{2}\right]\cdot\exp\left[ \frac{- n_{\chi} \sigma_{0}^{2}}{2 \sigma^{2}}\right]\notag\\
     &\propto \frac{1}{\sigma^{n_{\mathrm{tp}} +  n_{\chi}+2}}\cdot
     \exp\left[-\frac{1}{2\sigma^{2}}\left( n_{\chi} \sigma_{0}^{2}+\displaystyle\sum_{i=1}^{n}\vert\Delta\vert^{2}\right)\right]\notag\\
     &\propto \mathrm{Scale-inv-}\chi^{2}\left(\sigma^{2}; n_{\mathrm{tp}}+ n_{\chi}, \frac{n_{\chi} \sigma_{0}^{2}+\displaystyle\sum_{i=1}^{n_{\mathrm{tp}}}\vert\Delta\vert^{2}}{n_{\mathrm{tp}} + n_{\chi}}\right)
\end{align}
The expected value of the scaled inverse $\chi^{2}$ distribution is undefined, so we compute the mode of the posterior as our estimate for $\sigma^{2}$. The mode of the scaled inverse $\chi^{2}$ distribution $\mathrm{Scale-inv-}\chi^{2}(x; \nu, \tau^{2})$ is
\begin{equation}
    \mathrm{Mode}(x)=\frac{\nu\tau^{2}}{\nu+2}
\end{equation}
Substituting for the parameters of our posterior density function yields
\begin{equation}\label{eq:variance_param_estimate}
    \sigma^{2}\approx\frac{n_{\chi}\sigma_{0}^{2}}{n_{\mathrm{tp}}+ n_{\chi}+2}\displaystyle\sum_{i=1}^{n_{\mathrm{tp}}}\vert\Delta\vert^{2}.
\end{equation}

\section{Domain-specific terms}\label{sec:terms}
In this section we provide short definitions for some of the potentially unfamiliar terms that are used in this paper.

\paragraph*{Subject}\label{subsec:subject_defn}
A single image of a galaxy that volunteers are shown via the Zooniverse platform and that they inspect to search for clumps.
\paragraph*{Volunteer}\label{subsec:volunteer_defn}
A member of the public who participated in the \gzcs project by inspecting one or more subjects and used Zooniverse platform interface to search for and annotate the locations on clumps. 
\paragraph*{Annotation}\label{subsec:annotation_defn}
A set of click locations provided by a single volunteer as they inspect a single subject. The click locations are later expanded into a set of square boxes as explained in \autoref{subsec:agg_annots}.
\paragraph*{Label}\label{subsec:label_defn}
A set of zero or more rectangular bounding boxes, derived by our aggregation framework for a single subject image, that estimates the locations of any clumps it contains. 
\paragraph*{Skill}\label{subsec:skill_defn}
A compound metric, describing a particular volunteer, that estimates the probability  will mark a spurious clump, the probability that they will miss a real clump, and the accuracy of the locations they provide for any real clumps they mark.
\paragraph*{Difficulty}\label{subsec:difficulty_defn}
A quantitative metric for the degree to which the properties of a single subject image affect the ability of volunteers to perceive and accurately label any clumps it contains.
\paragraph*{Risk}\label{subsec:risk_defn}
A metric that is designed to quantify the reliability and scientific utility of a single subject's consensus label. 
\paragraph*{Retire}\label{subsec:retire_defn}
Stop collecting annotations for a subject.
\section{Table of symbols}\label{sec:symbols}
In this section we provide a reference table for symbols that recur in multiple sections of this paper.
\begin{table*}
    \centering
    \caption{Table of the most commonly recurring symbols used in this paper. We divide the symbols into categories and provide a brief description of how they should be interpreted. Complete descriptions of each symbol are provided in the main text at the point they are first introduced.}
    \label{tab:table_of_symbols}
    \begin{tabular}{l|c|p{0.6\textwidth}}
         Category& Symbol & Description \\
         \hline
         \multirow[t]{4}{*}{Object indices}&$i$& Index over subjects. \\
         &$j$& Index over volunteers. \\
         &$k$& Index over a volunteer $j$'s clump identifications for a single subject. \\
         &$l$& Index over aggregated clump locations.\\
         \hline
         \multirow[t]{4}{*}{Subjects}&$S$& The global set of subject images.\\
         &$S_{j}$&The set of subject images inspected by volunteer $j$.\\
         &$s_{i}$&A single subject image in $S$.\\
         &$\mathcal{R}_{i}$&The risk for subject $s_{i}$.\\
         &$N_{i}^{\mathrm{fp}}$&The expected number of spurious clump locations (false positives) in the label for subject $i$.\\
         &$N_{i}^{\mathrm{fn}}$&The expected number of missed clumps (false negatives) in the label for subject $i$.\\
         &$N_{i}^{\sigma}$&The expected number nominally true positive clump locations in the label for subject $i$ that differ from the (unknown) true clump location by a Jaccard distance greater than 0.5.\\
         \hline
         \multirow[t]{4}{*}{Subject difficulties}&${\sigma_{i}^{l}}^{2}$&The variance of a Gaussian model for the Jaccard distance offset between the estimated location of the $l$th detected clump for subject $i$ and its corresponding (unknown) true location.\\
         &$\mathcal{D}_{i}$&The difficulty of subject $i$ defined the set of ${\sigma_{i}^{l}}^{2}$ values for all detected clumps in the image.\\
         \hline
         \multirow[t]{2}{*}{Volunteers}&$V$&The global set of volunteers.\\
         &$V_{i}$&The subset of volunteers who inspected subject $i$.\\
         \hline
         \multirow[t]{4}{*}{Volunteer skills}&$p_{j}^{\mathrm{fp}}$&The probability that volunteer $j$ will click on a spurious clump.\\
         &$p_{j}^{\mathrm{fn}}$&The probability that volunteer $j$ will miss a real clump.\\
         &$\sigma_{j}^{2}$&The variance of a Gaussian model for the Jaccard distance offset between volunteer $j$'s true positive click locations and the corresponding (unknown) true clump locations, independent of subject. \\
         &$\mathcal{S}_{j}$&The skill of volunteer $j$ defined as the set $\{p_{j}^{\mathrm{fp}}, p_{j}^{\mathrm{fn}}, \sigma_{j}^{2}\}$.\\
         \hline
         \multirow[t]{6}{*}{Annotations}&$Z$&The global set of volunteer annotations.\\
         &$Z_{i}$&The set of annotations for a single subject image provided by all the volunteers who inspected it.\\
         &$\tilde{Z}_{n}$&A randomly selected subset of $Z$ containing exactly $n$ annotations per subject.\\        
         &$z_{ij}$&A single annotation provided by volunteer $j$ after inspecting subject $i$.\\
         &$B_{ij}$&The set of boxes, corresponding to click locations provided by volunteer $j$ for subject $i$.\\
         &$B_{i}$&The set of all boxes, corresponding to click locations provided for subject $i$ by all volunteers who inspected it.\\
         &$b_{ij}^{k}$&A single box, corresponding to the location of  a single click provided by volunteer $j$ for subject $i$.\\
         &${\sigma_{ij}^{k}}^{2}$&The variance of a Gaussian model for the Jaccard distance offset between volunteer $j$'s $k$th true positive click location for subject $i$ and its corresponding (unknown) true clump location.\\
         &$a_{ij}^{k}$&An integer value that maps the $k$th click in volunteer $j$'s annotation of subject $i$ to a specific clump in that subject's estimated label (or to the dummy facility if it is deemed to be a false positive).\\
         \hline
         \multirow[t]{4}{*}{Labels}&$Y$&The global set of subject labels.\\
         &$y_{i}$&The unknown true label for subject $i$.\\
         &$b_{i}^{l}$&A single box comprising part of the unknown true label for subject $i$.\\
         &$\hat{y}_{i}$&The \textit{estimated} label for subject $i$ that is computed by our framework.\\
         &$\hat{b}_{i}^{l}$&A single box comprising part of the \textit{estimated} label for subject $i$.\\
         &$p^{\mathrm{fp}}_{l}$&The probability that the $l$th clump in the \textit{estimated} label for a subject is a false positive.\\
         &$p^{\sigma}_{l}$&The probability that the Jaccard distance between the $l$th clump in the \textit{estimated} label and the corresponding (unknown) true clump location exceeds 0.5.
    \end{tabular}
\end{table*}

\section{Comparison with \textsc{Scikit Learn} \texttt{MeanShift} clustering algorithm}\label{sec:mean_shift_comparison}

We emphasise that the aim of this paper is not to present a novel and very complicated clustering algorithm. Indeed, our focus is the likelihood model that we use to estimate the \gzcs volunteers' skills, the difficulty of the subjects that they inspect, and the reliability of the consensus labels that we derive. Nonetheless, we recognise that there are many well established clustering algorithms in the literature and that some of them may outperform our framework's ability to actually \textit{detect} clumps, even if they cannot provide the same auxiliary information about the final subject labels. Presenting exhaustive comparison between our framework and every alternative algorithm is beyond the scope of this paper. However, we have tested several of the methods available from the \textsc{Scikit Learn} Python package \citep{scikit-learn}. In this section we present a representative comparison between our framework and the \textsc{Scikit Learn} \texttt{MeanShift} clustering algorithm. We set the \texttt{MeanShift} algorithm's \texttt{bandwidth} parameter set equal to the size of the SDSS imaging PSF for each subject image and all other parameters were left set to their default values\footnote{See \url{https://scikit-learn.org/stable/modules/generated/sklearn.cluster.MeanShift.html}}. 

\begin{figure}
    \centering
    \includegraphics[width=0.48\textwidth]{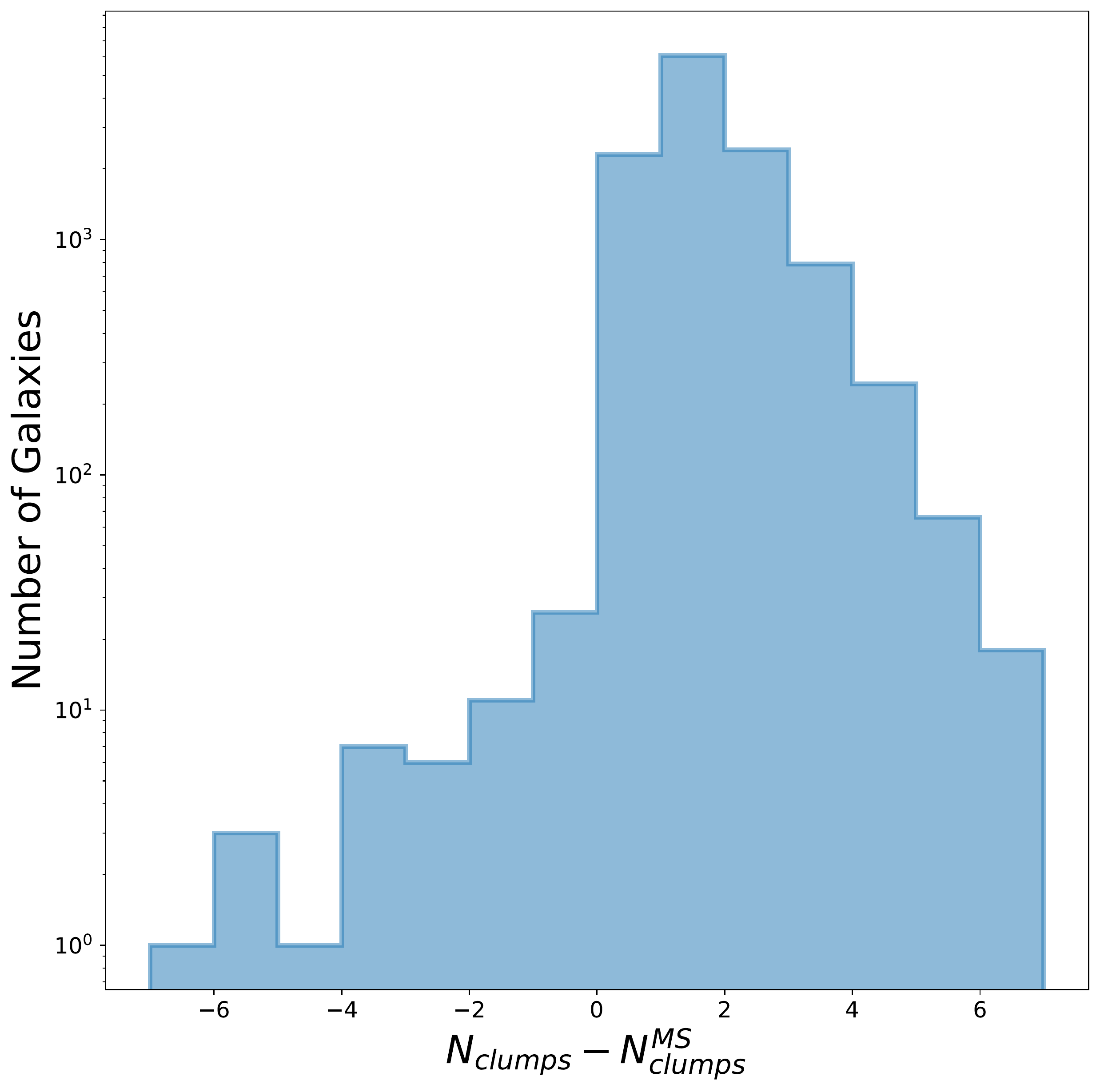}
    \caption{Distribution of the distribution of the difference between the number of clumps detected by our framework and the number detected using the \texttt{MeanShift} algorithm for each subject in the \gzcs subject set.}
    \label{fig:agg_v_ms_dist}
\end{figure}

\begin{figure*}
    \centering
    \includegraphics[width=\textwidth]{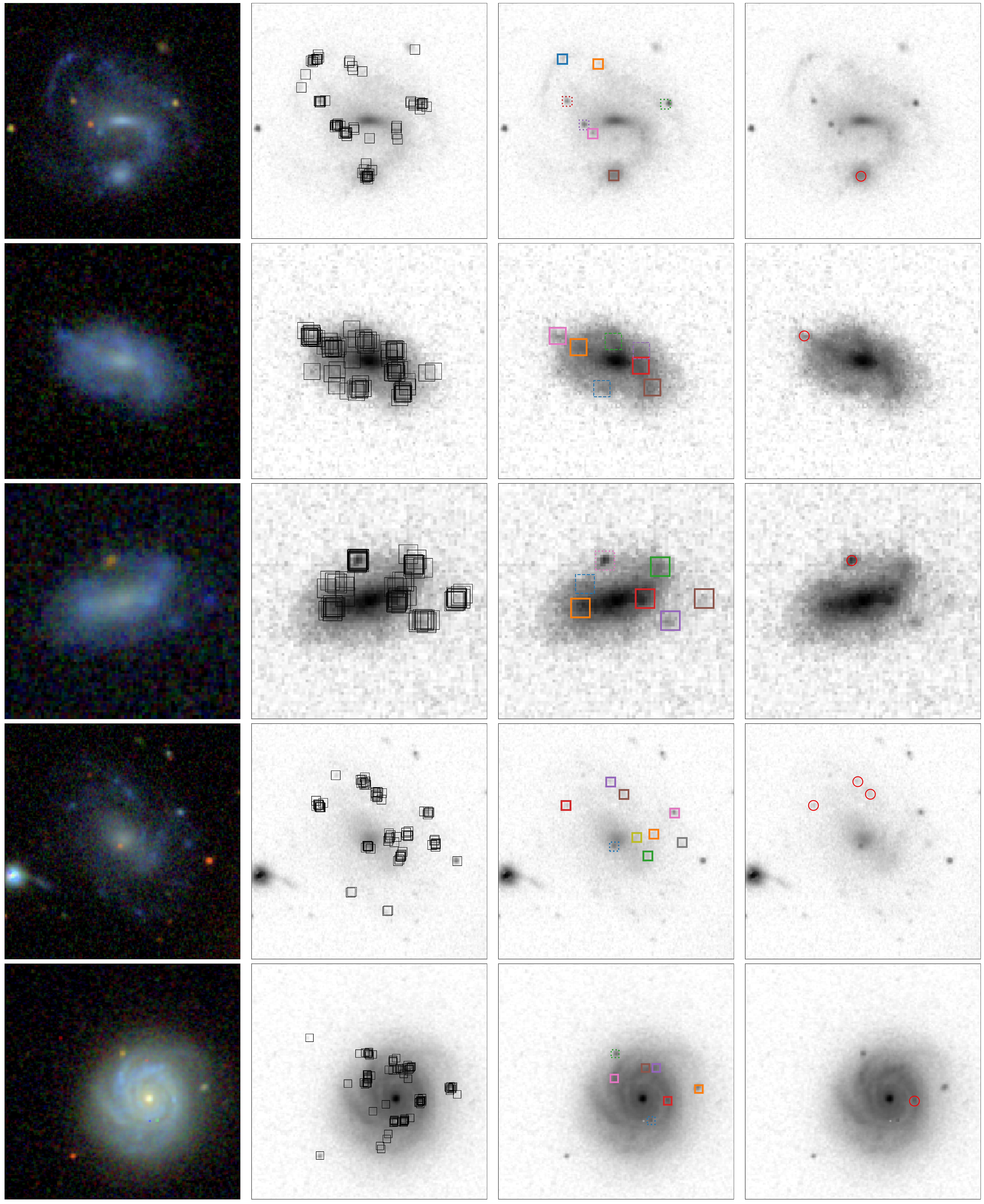}
    \caption{Examples of clump-hosting galaxies, for which our framework detects more clumps than the \textsc{Scikit Learn} \texttt{MeanShift} algorithm. The \textit{first column} shows galaxy images as they were seen by volunteers. The \textit{second column} overlays all volunteer annotations on a grey-scale image of the same galaxy. The coloured boxes in the \textit{third column} show the clump locations that out framework identifies. Dashed boxes indicate clumps with false positive probabilities $p_{l}^{\mathrm{fp}}>0.8$ assigned by framework label. Finally, the red circles in the \textit{fourth column} show the clumps detected by the \texttt{MeanShift} algorithm.}
    \label{fig:agg_v_ms_more_examples_0-5}
\end{figure*}

\begin{figure*}
    \centering
    \includegraphics[width=\textwidth]{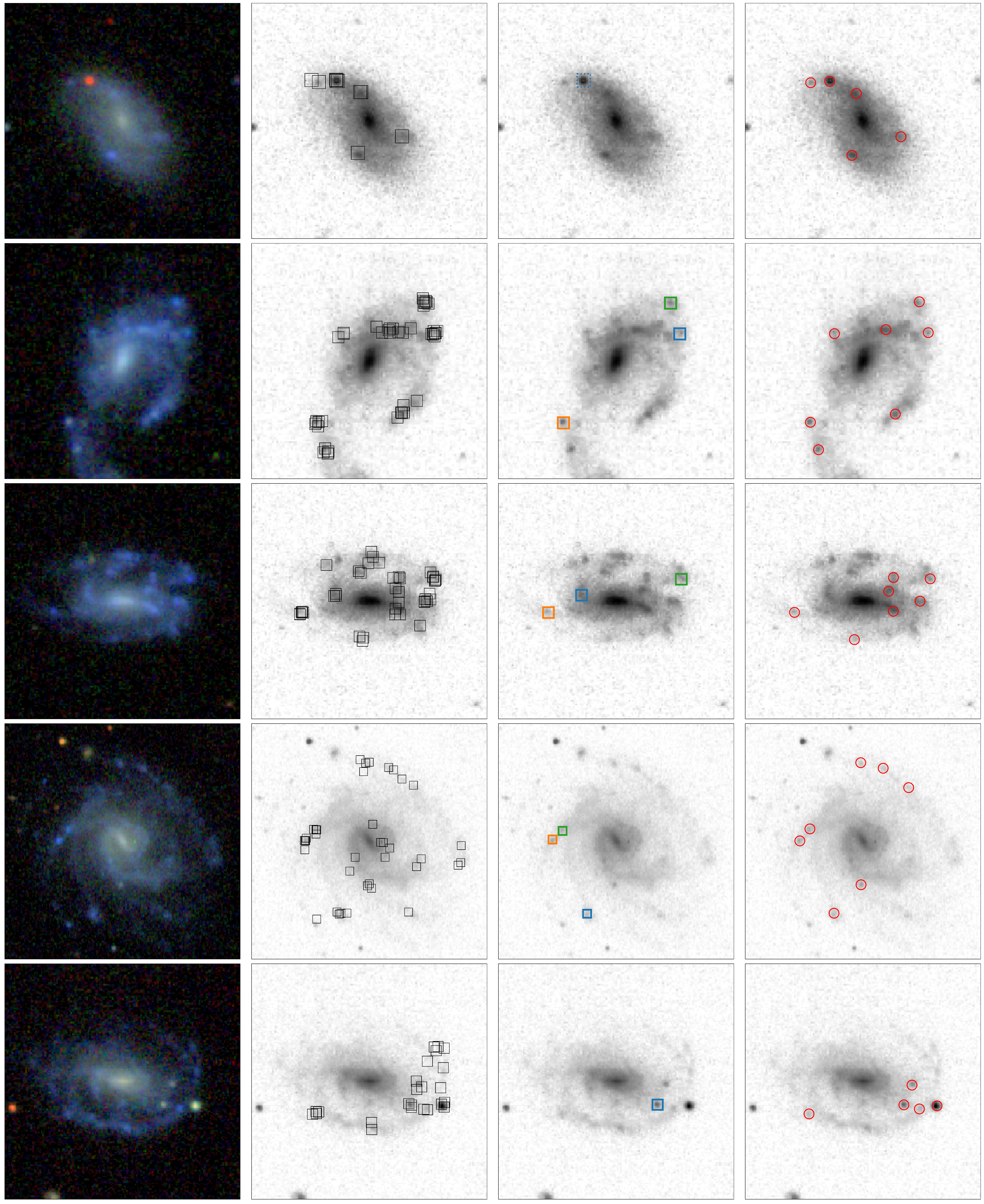}
    \caption{Examples of clump-hosting galaxies, for which our framework detects fewer clumps than the \textsc{Scikit Learn} \texttt{MeanShift} algorithm. The images, boxes and circles shown in the various columns have the same meaning as in \autoref{fig:agg_v_ms_more_examples_0-5}.
    }
    \label{fig:agg_v_ms_less_examples_0-5}
\end{figure*}

\autoref{fig:agg_v_ms_dist} shows the distribution of the difference between the number of clumps detected by our framework and the number detected using the \texttt{MeanShift} algorithm for each subject in the \gzcs subject set. For the majority of subjects, our framework detects more clumps than the \texttt{MeanShift} algorithm. In \autoref{fig:agg_v_ms_more_examples_0-5} we show some representative subjects for which our framework detects more clumps than the \texttt{MeanShift} algorithm and in \autoref{fig:agg_v_ms_less_examples_0-5}, we show subjects for which the reverse is true. It is not obvious from these figure that either algorithm is particularly biased towards detecting clumps with specific properties. There is some evidence that our algorithm detects fainter potential clumps than the \texttt{MeanShift} algorithm, and seems less vulnerable to misidentifying objects like stars and background galaxies as clumps. Even when such objects are detected by our framework, they tend to be assigned false positive probabilities greater than 0.8. In some cases, our framework fails to detect clumps that many volunteers identify. We speculate that this is a result of a small number of volunteers with very high $p_{j}^{\mathrm{fp}}$ identifying the clump, which causes our framework to deem other volunteers' clicks as false positives as well.

\end{document}